\documentclass[12pt,openany]{book}

\usepackage[a4paper,margin=1in]{geometry}
\usepackage{titlesec}
\usepackage{fancyhdr}
\usepackage{tikz}
\usetikzlibrary{arrows.meta}
\usepackage{textgreek}
\usepackage{hyperref}
\usepackage{balance}
\usepackage[greek,english]{babel}
\usepackage[centertags]{amsmath}
\usepackage{amsfonts}
\usepackage{amssymb}
\usepackage{amsthm}
\usepackage{newlfont}
\usepackage{stmaryrd}
\usepackage{mathrsfs}
\usepackage{euscript}
\usepackage{siunitx}
\usepackage{physics}
\usepackage{algorithm2e}
\usepackage{graphicx,subcaption}
\usepackage{enumitem}
\usepackage{natbib}
\usepackage{graphicx}
\usepackage{color}
\usepackage[english,greek]{babel}
\usepackage{floatrow}
\usepackage{caption}
\usepackage{hyperref}
\usepackage{hhline}
\usepackage{cleveref}
\usepackage{import}

\theoremstyle{plain}
  \newtheorem{theorem}{Theorem}[section]

\theoremstyle{definition}

\theoremstyle{remark}

  \newtheorem{example}[theorem]{Example}

\numberwithin{equation}{section}

\let\al=\alpha \let\be=\beta \let\de=\delta 
\let\ve=\varepsilon  \let\ga=\gamma 
  \let\om=\omega 
\let\si=\sigma

\newcommand{\caD}{{\mathcal D}}

\newcommand{\caL}{{\mathcal L}}

\newcommand{\caO}{{\mathcal O}}

\newcommand{\caQ}{{\mathcal Q}}

\newcommand{\caT}{{\mathcal T}}

\newcommand{\bbC}{{\mathbb C}}

\newcommand{\bbG}{{\mathbb G}}

\newcommand{\bbL}{{\mathbb L}}

\newcommand{\bbN}{{\mathbb N}}

\newcommand{\bbR}{{\mathbb R}}
\newcommand{\bbS}{{\mathbb S}}

\newcommand{\bbZ}{{\mathbb Z}}

\newcommand{\opunit}{\text{1}\kern-0.22em\text{l}}

\newcommand{\bsP}{{\boldsymbol P}}

\DeclareMathAlphabet{\mathpzc}{OT1}{pzc}{m}{it}

\newcommand{\ra}{{\rangle}}

\newcommand{\fig}{Fig.\;}

\newcommand{\id}{\textrm{d}}

\usepackage{floatrow}
\usepackage{fancyhdr}
\usepackage{color,soul}

\DeclareCaptionJustification{justified}{\leftskip=0pt \rightskip=0pt \parfillskip=0pt plus 1fil}
\begin{document}
\begin{titlepage}
\centering
{\Huge\bfseries\color{red}What is Nonequilibrium?\par}
\vspace{2cm}
\includegraphics[width=0.5\textwidth]{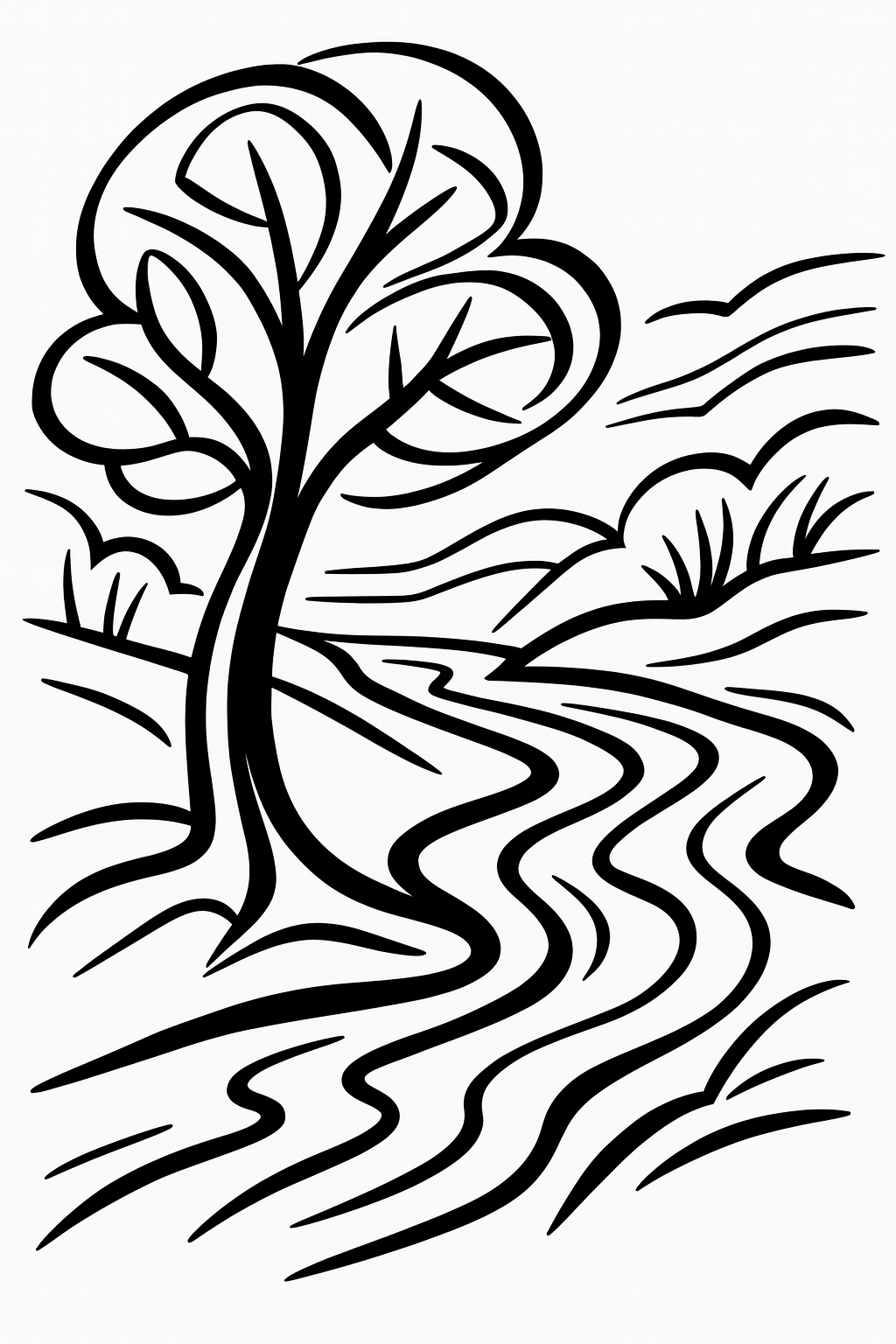}\par
\vspace{1.5cm}
{\Large Christian Maes\par}
{\large Department of Physics and Astronomy, KU Leuven\par}
\vspace{1.5cm}
{\textbf{Introduction to nonequilibrium physics}\par}
\vspace{0.5cm}
{19--20 May 2025\;\;\;\;\;\; ---\;\;\;\; Leuven School on Noneq Stat Mech\par}
\vspace{2cm} \hspace{12cm}
\small{{\it version 2.7}}
\end{titlepage}
\addtolength{\topmargin}{-0.5cm}

\vspace{2cm}

\lhead{C. Maes}
\rhead{}


\pagenumbering{roman}
\chapter*{Preface}
\lhead{}
\rhead{Preface}
\pagenumbering{roman}
Nonequilibrium physics is a rapidly evolving research field, both experimentally and theoretically. Without a doubt, nonequilibrium phenomena represent most fascinating and relevant areas in the natural sciences.  The scope is vast, encompassing a growing number of `temporal' questions across a wide range of scientific disciplines.  Nonequilibrium physics is at the same time foundational, addressing problems concerning chance and evolution, self-organization, and the origin of dissipation, inspiring  in fact new directions in physics and cosmology, and providing more unifying approaches to the science of complexity, as well as playing a central role in various interdisciplinary efforts aimed at understanding pattern formation, stabilization, selection processes, natural intelligence, and control theory. It goes much beyond (even irreversible and/or stochastic) thermodynamics while maintaining interesting and even important connections with traditional branches of physics, including soft-matter physics, rheology, solid-state physics, astrophysics and fluid dynamics.\\
Birds and planes are able to fly thanks to dissipative processes\footnote{d'Alembert (1752) proved that for potential flow there would be neither drag nor lift on a body moving with constant velocity in an incompressible and inviscid fluid.}, and the nonequilibrium domain continues to produce unifying models and  applications, ranging from the life sciences and micro-bioengineering, to the development of new materials and advances in robotics, all the way to climate science and sustainability.  
Classic problems such as stabilization, homochirality, baryogenesis, homeostasis,  self-assembly, ``unusual'' response, the naturalness of Finsler geometry, and (why not) the emergence of time may now be revisited from fresh nonequilibrium perspectives.\\

While a unified theoretical framework for nonequilibrium physics remains elusive, systematic progress was made over recent decades, particularly in systems that satisfy the condition of local detailed balance. In this context, pathspace probabilities (the plausibility of system trajectories, encapsulated through dynamical ensembles and fluctuation theory) have emerged as central elements in a Gibbs-like formalism for states and configurations on spacetime.\\

Here, in these introductory notes, our goal is more modest: to explain what nonequilibrium physics is about, and why it matters: {\it what indeed entails studying nonequilibrium?}\\
Of course, many excellent books and reviews contribute to the field in diverse and valuable ways, too many to cite exhaustively. They are essential companions for reference and study.\\  To help define the scope of nonequilibrium physics and to provide useful background, we include a (surely incomplete) selection of books and notes that, in one way or another, address aspects or introduce tools for the study of nonequilibrium phenomena:

\vspace{0.3cm}

\begin{itemize}

\item 
 C. Mejía-Monasterio and L. Rondoni,
{\it Nonequilibrium Statistical Mechanics I, 
Foundations and Modern Applications}. TMP, Springer Cham, 2025.
   
\item
A. Sarracino, A. Puglisi and A. Vulpiani, {\it Nonequilibrium Statistical Mechanics --Basic concepts, models and applications}, IOP Publishing, Bristol, 2025.

\item
U. Seifert, {\it Stochastic Thermodynamics}.  Cambridge University Press, 2025.

\item
L.F. Cugliandolo, {\it Introduction to Langevin Stochastic Processes}. arXiv:2504.11994v1 [cond-mat.stat-mech] 16 Apr 2025.

\item 
D.T. Limmer, {\it Statistical Mechanics and Stochastic Thermodynamics ---
A Textbook on Modern Approaches in and out of Equilibrium}, Oxford Graduate Texts, 2024.

\item
N. Shiraishi, {\it An Introduction to Stochastic Thermodynamics: From Basic to Advanced}. Fundamental Theories of Physics 212, Springer, 2023.

\item 
H. Tasaki, Online course ``A Modern Introduction to Nonequilibrium Statistical Mechanics'' (https://haltasaki.github.io/OL/noneq/e/index.html), 2023.

\item
P. Gaspard, {\it The Statistical Mechanics of Irreversible Phenomena}.  Cambridge University Press, 2022.

\item
P. Strasberg, {\it Quantum Stochastic Thermodynamics: Foundations and Selected Applications}. Oxford Graduate Texts, Oxford University Press, 2022.

\item J.~Bricmont, {\it Making Sense of Statistical Mechanics}, Springer (Undergraduate Lecture Notes in Physics), 2022.

\item
L. Peliti and S. Pigolotti, {\it Stochastic Thermodynamics: An Introduction}. Princeton University Press, 2021.

\item
R. Livi and P. Politi, {\it Nonequilibrium Statistical Physics: A Modern Perspective}. Cambridge University Press, 2017.

\item
G. Gallavotti, {\it Nonequilibrium and Irreversibility}.
Springer (Theoretical and Mathematical Physics series), 2014.

\item D. Kondepudi and I. Prigogine, {\sl Modern Thermodynamics: From Heat Engines to Dissipative Structures}. John Wiley \& Sons, 2014.  

\item
A. Kamenev, {\it Field Theory of Non-Equilibrium Systems}.
Cambridge University Press, 2011.

\item
P.~L.~Krapivsky, S.~Redner, and  Eli Ben-Naim, {\sl
A Kinetic View of Statistical Physics}, Cambridge University Press, 2010. 

\item 
K. Sekimoto, {\it Stochastic Energetics}. Lecture Notes in Physics, Springer-Verlag Berlin Heidelberg, 2010.

\item
N.~Pottier, {\sl Nonequilibrium Statistical Physics: Linear Irreversible Processes},
OUP Oxford, 2010. 

\item
C. W. Gardiner, {\it Handbook of Stochastic Methods}. Springer, 4th ed., 2009.

\item
M.~Henkel, H.~Hinrichsen and S.~L\"ubeck,
{\sl Non-Equilibrium Phase Transitions: Volume 1: Absorbing Phase Transitions (Theoretical and Mathematical Physics)}, Springer, 2009.

\item
E. Presutti, {\it Scaling Limits in Statistical Mechanics and Microstructures in Continuum Mechanics}. Springer, 2008/2009.

\item
V.~Balakrishnan, {\it Elements of Nonequilibrium Statistical
Mechanics}, Ane Books India (2008).

\item
C.~M.~Van Vliet, {\sl Equilibrium and Non-equilibrium Statistical Mechanics},
World Scientific, 2008.

\item
J.~Cardy,
{\sl Non-equilibrium Statistical Mechanics and Turbulence},
Cambridge University Press, 2008.

\item
G.~Lebon, D.~Jou, and J.~Casas--V\'azquez,
{\sl Understanding Non-equilibrium Thermodynamics -- Foundations, Applications, Frontiers}, Springer-Verlag Berlin Heidelberg, 2008.

\item
G.~Mazenko, {\sl Nonequlibrium Statistical Mechanics}, Wiley-VCH, 2006.

\item 
C.~Cercignani, {\sl Transport Phenomena and Kinetic Theory: Applications to Gases, Semiconductors, Photons, and Biological Systems.}  Birkhäuser, 2006.

\item
W.~Ebeling and I.M.~Sokolov, {\sl Statistical thermodynamics and
stochastic theory of non-equilibrium systems}, Singapore:World
Scientific, 2005.

\item
V.~Privman, {\sl Nonequilibrium Statistical Mechanics in One Dimension},
Cambridge University Press, 2005. 

\item 
M.~Le Bellac, F.~Mortessagne, and G.~G.~Batrouni, {\sl Equilibrium and
Non-Equilibrium Statistical Thermodynamics}, Cambridge, 2004.

\item
P.~A.~Martin: {\it Physique statistique des processus
irreversibles}, Lecture Notes of the DEA de Physique Th\'eorique,
notes by F.Coppex, ENS Lyon, Fall 2001--Spring 2004.

\item
Da-Quan Jiang-Min Qian, Min-Ping Qian, {\sl Mathematical Theory of Nonequilibrium Steady States
On the Frontier of Probability and Dynamical Systems}, Lecture Notes in Mathematics,
Volume 1833, 2004. 

\item
Tian-Quan Chen, {\sl A Non-equilibrium Statistical Mechanics: Without the Assumption of Molecular Chaos},
World Scientific, 2003.

\item
G.~Mazenko, {\sl Brownian Motion, Fluctuations, Dynamics and Applications.}, Clarendon Press Oxford, 2002.

\item
R.~Zwanzig, {\sl Nonequilibrium statistical mechanics}, Oxford, 2001.

\item
J.~Marro and R.~Dickman, {\sl Nonequilibrium Phase Transitions in Lattice Models}, Cambridge University Press, 1999.

\item
P.~Gaspard, {\sl Chaos, Scattering, and Statistical Mechanics},
Cambridge University Press, Nonlinear Science Series No. 9, 1998.

\item 
C. Cercignani, {\it Ludwig Boltzmann: The Man Who Trusted Atoms}. Oxford University Press, 1998.

\item
R.~Balescu, {\sl Statistical dynamics, matter out of equilibrium}, Imperial College Press, 1997.

\item
L.~Schimansky-Geier, {\sl Stochastic dynamics},
Springer, 1997. 

\item
D. Zubarev, V. Morozov, and G. R\"opke, {\it Statistical Mechanics of Nonequilibrium Processes}. Akademie Verlag, 1996-1997.

\item
H. Risken, {\it The Fokker–Planck Equation}. Springer, 1996.

\item
U.~Frisch, {\sl Turbulence: The Legacy of A. N. Kolmogorov}.
Cambridge University Press, 1995.

\item
N.V.~Krylov, {\it Introduction to the Theory of  Diffusion Processes}. Translations of Mathematical Monographs,
 American Mathematical Society (1995).

\item
B.~Schmittmann and R.~K.~P.~Zia, {\sl Statistical Mechanics of Driven Diffusive Systems},
Vol. 17 of Phase Transitions and Critical Phenomena, eds. C. Domb and J.L. Lebowitz,
Academic Press, N.Y., 1995.


\item
H.~Spohn, {\sl Large Scale Dynamics of Interacting Particles},
Springer--Verlag, Heidelberg, 1991.

\item 
D.~J.~Evans and G.~P,~Morriss, {\sl Statistical Mechanics of Nonequilibrium Liquids}, Academic Press, London 1990, Theoretical Chemistry Monograph Series.

\item 
C.~Cercignani, {\sl The Boltzmann Equation and Its Applications}, Springer, 1988.

\item
D.~Chandler, {\sl Introduction to Modern Statistical Mechanics}, Oxford University Press, 1987.

\item
J.~Keizer, {\sl Statistical Thermodynamics of Nonequilibrium
Processes}, Spinger--Verlag, 1987.

\item 
J.L.~ Lebowitz and E.W.~Montroll, Eds, {\sl Nonequilibrium phenomena. II - From stochastics to hydrodynamics}, North-Holland Physics Publishing (Studies in Statistical Mechanics. Volume 11), 1984.

\item Y.~Kuramoto, {\it Chemical Oscillations, Waves, and Turbulence}, Springer Series in Synergetics (SSSYN, volume 19), 1984.

\item
H.~Risken, {\sl The Fokker-Planck Equation}, Springer, Berlin
1984.

\item 
E. W.~Montroll, J.L.~Lebowitz, Eds., {\it Nonequilibrium Phenomena I: The Boltzmann Equation}.
North Holland Pub., 1983. 

\item
H.~Grabert, {\sl Projection Operator Techniques in Nonequilibrium Statistical Mechanics}, Springer--Verlag, Berlin, 1982.

\item
 N.G.~van Kampen, {\sl Stochastic Processes in Physics and
Chemistry}, North--Holland, 1981.

\item
O.~E.~Lanford III, Time evolution of large classical systems,
\emph{Lecture Notes in Phys.} \textbf{38}, 1--111 (1975); The hard
sphere gas in the Boltzmann-Grad limit, \emph{Physica}
\textbf{106A}, 70--76 (1981).

\item I. Prigogine, {\sl From Being to Becoming: Time and Complexity in the Physical Sciences}. W H Freeman \& Co, 1981.

\item
 R.~Kubo, M.~Toda, and N.~Hashitsume, {\sl Statistical Physics 2: Nonequilibrium
Statistical Mechanics}, Springer 1978.

\item
G.~Nicolis and I.~Prigogine, {\sl Self-organization in nonequilibrium systems: from dissipative structures to order through fluctuations}, A Wiley-Interscience Publication, 1977.

\item
R.~Balescu, {\sl Equilibrium and nonequilibrium statistical mechanics}, Wiley--Interscience publication, 1975.

\item I. Prigogine and P. Glansdorff, {\sl Thermodynamic Theory of Structure, Stability and Fluctuations}. John Wiley \& Sons, 1971.


\item
I.~Prigogine, {\sl Non-Equilibrium Statistical Mechanis}, Interscience Publishers --- Wiley, 1962.

\item
S.~R.~de Groot and P.~Mazur, {\sl Non-equilibrium Thermodynamics},
North Holland Publishing Company, 1962.

\item
E.~W.~Montroll, {\sl Lectures on nonequilibrium statistical mechanics},
IBM Research Center, 1960.

\item
M.~Kac. {\sl Probability and Related Topics in Physical Sciences},
Interscience Publishers Inc., New York, 1959.
\end{itemize}
The present text aims to stimulate study and a basic understanding of what nonequilibrium physics may be about; yet, probably lacking in examples and nontrivial model calculations. For a more comprehensive and balanced treatment of the subject, we may refer to the references and books cited above.\\

From the table of contents, the reader will identify three main parts: (1) a characterization of the nonequilibrium condition, largely by contrast to equilibrium; (2) a retelling of some of the great performances of the more distant past, presenting the perspectives of Boltzmann and Onsager; and (3) a focus on more recent methods and concepts, from local detailed balance and the identification of entropy fluxes to dynamical fluctuation theory, which also incorporates the frenetic contribution.\\

There is significant overlap with published work, with the unfortunate consequence that the text has become somewhat idiosyncratic. Regrettably, there are an awful lot of omissions and major oversights. Also for that reason, comments and suggestions are very welcome; updates and newer versions will appear as teaching material on my webpage\\
\url{https://fys.kuleuven.be/english/staff/christ}.   

\vspace{0.5cm}
\noindent  Christian Maes\hspace{8cm} \;\;\;\;\; \;Leuven, 31/01/2026

\tableofcontents  

\chapter{In contrast with equilibrium}
\lhead{C. Maes}
\rhead{Contrast with equilibrium}
\pagenumbering{arabic}

\section{Differences with equilibrium}
Whether you call it disequilibrium, unequilibrium, out-of-equilibrium, or nonequilibrium, there remains the loud impression that nonequilibrium statistical mechanics can only be characterized as the excessively
large complement of equilibrium physics.
Still, does understanding the ocean not begin with our perception of it from the shore? Similarly, a good starting point for grasping the challenge posed by nonequilibrium systems is to compare with thermodynamic equilibrium, if only for stressing differences. {\it E.g.}, a common confusion is to assume that equilibrium simply means ``some well-behaved stationarity,'' {\it i.e.}, statistical time-invariance with some stability. It obviously boils down to terminology but we want equilibrium to mean more than time-invariance; the question is not just whether a system’s macroscopic properties change over time or not. In fact, many nonequilibrium systems are stationary and stable — just like us..., and their statistical properties stay happily constant over time, at least for suitable  observables and under the appropriate time-scales.  Rather, steady nonequilibrium refers to driven or fueled systems, possibly agitated in time by memory and external fields, component-wise or globally, via bulk rotational forces or via thermodynamically-frustrating boundary conditions.  A typical consequence is the maintenance of temporal cycles and currents of all sorts where particles, mass, energy or momentum get and keep getting transported. \\

To illustrate, take this metaphor: an equilibrium system resembles a still lake, whereas a stationary nonequilibrium system mirrors a flowing river. Every part of the lake is just another part of the same lake, uniform in temperature, composed of the same substance concentrations, and governed by the same interactions (aside from boundary effects). The lake dynamics is symmetric under time-reversal, at least on the meaningful time scales. If a tree falls into the lake, it might alter the lake constraints or boundary conditions, but the system eventually returns to a quiescent passive state, restoring the very character of the lake.

In contrast, in its essence, a flowing river cannot be fully captured in a single snapshot. Its currents may be smooth or turbulent, but the water is always in motion. When a tree falls into the river, it disrupts the flow, potentially redirecting currents and reshaping the riverbed. The river or pipeline is constantly driven by gravity or pumps, sustaining a continuous transport of matter and energy that leads to erosion and change, 
\begin{otherlanguage}{greek}
πάντα ῥεῖ .
\end{otherlanguage} 
Unlike the tranquil and time-reversible dynamics of a lake, the behavior of the river reflects a persistent directional flux, a hallmark of nonequilibrium. Although both systems may appear stationary from afar, their underlying physical nature is profoundly different.  The steady far-from-equilibrium condition is nonperturbative with respect to equilibrium; it cannot be created by a spatiotemporally local stimulus.  Rather, nonequilibrium can only be inherited and cannot be created.\\

The image above can be generalized to mesoscopic systems where noise invariably enters. Indeed, when moving intelligently between scales of length, time, or energy (as we do in statistical mechanics), the study of fluctuations enters.  Again, there are important differences between equilibrium and nonequilibrium ensembles.\\  
In equilibrium, the thermodynamic potentials (energy, entropy, free energies {\it etc}) are governing the static fluctuations: we can estimate the likelihood or implausibility of a static fluctuation from free energy estimates, and that gets an operational meaning in terms of work and suitable response coefficients.  For nonequilibrium systems, so far there is no operational meaning that is (easily) attached to the static fluctuation functionals, and a zoo of kinetic effects enters: there is no Gibbs-like formula for the probability of some (fixed-time) density profile that is comparable in power and generality with equilibrium.\\
Nonequilibrium is not answering Hamlet's question; rather it wants to be characterized by a process or dynamical ensemble, which means deriving probabilities on the space of trajectories, {\it i.e.}, distributions on spacetime configurations.  The projection of that distribution on an equal-time layer may create nonlocal effects; only in the case where past and future are interchangeable can we identify well-defined (energy-entropy) landscapes governing the probabilities of configurations.\\

Compared to equilibrium processes, the breaking of (statistical) time-reversal invariance for typical trajectories of steady nonequilibria obviously comes with many drastic changes in phenomenology, and it has interesting and important effects.  Time-symmetry breaking\footnote{which is sometimes realized in terms of memory effects.} is, in fact, the main characteristic of nonequilibrium and manifests itself in the violation of standard response and fluctuation-dissipation relations. However, paradoxically, the nature and the role of the time-symmetric fluctuation sector in the dynamical ensemble give sensational opportunities for selecting stationary nonequilibrium conditions in fundamentally different ways than via potential differences.  We are no longer ``landscaping,'' {\it i.e.}, directing the dynamics of a state or of a collective coordinate to mimic the mechanical motion of a particle on a manifold.  Nonequilibrium has a vastly dissimilar phenomenology\footnote{Useful visualizations are provided by the prints of M. C. Escher, such as `Ascending and Descending' (1960) and `Waterfall' (1961), which  call for a Finsler-like geometry rather than a Riemannian one.} and allows totally different selection, stabilization, and steering mechanisms compared to equilibrium processes. Uncovering these mechanisms and thereby unifying the associated many-body phenomenology constitutes one of the main challenges today, with the potential to revolutionize physics.  That is true  for steady regimes (as emphasized in the above metaphor), but it also holds for transient behavior.\\

From here, the reader can immediately jump to the examples in Chapter III.  The generalities that follow here and in Chapter II can easily be skipped when in more of a hurry.

\section{Dissipation}
There is a common, almost ``folk'' idea about how nonequilibrium systems behave: they are thought to organize themselves in such a way as to (try to) relax back toward equilibrium.\footnote{Some would add, ``as fast as possible.'' That intuition is sometimes invoked as a motivation for the so-called maximum entropy production principle \cite{Paltridge1979}, proposed as a way to characterize nonequilibrium steady states; see \cite{Bruers_2007,minep} for critical discussions.}
This view may have been inspired by the stability of equilibrium itself --- famously formulated in the Le Chatelier–Braun principle (1884–1887): when an equilibrium system is disturbed by a change in temperature, pressure, or concentration, it responds in a way that counteracts the disturbance, settling into a new equilibrium.\\
In that way, because relaxation toward equilibrium is accompanied by an increase in entropy (for a closed, isolated system), one may think of nonequilibrium systems as dissipative by nature: they continually release heat into their surroundings, with which they exchange energy and matter.  Locally, all seems to be put in place for fast return to equilibrium.\\

However, that view is incomplete. One cannot adhere solely to Clausius' adage ``The energy of the universe is constant; the entropy of the universe tends to a maximum.'' While sufficient for Willard Gibbs in the foundation of equilibrium thermodynamics, nonequilibrium studies cannot be based exclusively on the same energy-entropy considerations that still appear so prominently in chemical physics and that explain equilibrium phase diagrams. Dissipation and efficiency, entropy (production) and power are no longer alone.  Rather, (thermo)kinetic effects become central and dissipation is just one essential aspect in the richness and stability of the nonequilibrium condition.  Not only understanding the origin of dissipation, but also, equally important, the constructive role it plays in organizing physical systems, is one of the main challenges of the field.\\

Dissipation measures the breaking of time-reversal symmetry, \cite{maes1999fluctuation,maes2000entropy}.  Indeed, steady nonequilibrium states typically sustain continuous currents and exhibit time-asymmetric behavior in their trajectories or time series. However, this picture is not universally valid. For instance, when a system possesses memory, it may remain perfectly symmetric under time-reversal and yet still be dissipative --- effectively ``spilling'' heat. The key lies then in its coupling to additional, often hidden, degrees of freedom through which currents actually flow.\footnote{A simple illustration is the position dynamics of one-dimensional run-and-tumble particles; see \cite{Demaerel2018}.}\\
Similarly, nature also provides remarkable counterexamples to the opposite, such as superconductors and superfluids, where persistent, dissipationless currents can exist. In such systems, currents continue to flow even after the external driving -- the battery, so to speak --has been turned off.\\

In other words, even without prior exposure to the subject, the reader should appreciate that the behavior of nonequilibrium systems --- especially living or complex ones ---  can be intricate. The distinction between equilibrium and nonequilibrium is not absolute: it depends on the relevant scales of time, energy, and length that one chooses to examine.\\
Under suitably simplified conditions\footnote{{\it E.g.}, within the setup of classical mechanics of particles with short-range and bounded interactions; see for instance \cite{Maes:2002:NCH,local}.}, we can state two fundamental observations. First, a steady current cannot exist without some form of heat dissipation. Second, the presence of heat flow almost always signals the existence of underlying currents --- of matter, charge, or another conserved quantity. These intimate links among currents, fluctuations, and dissipation lie at the very heart of nonequilibrium statistical mechanics.\\
Addressing these deeper questions requires concepts that go beyond the scope of the present notes. In particular, the thermodynamic limit and time-scale separation are essential ingredients in constructing and interpreting thermal reservoirs and open-system behavior.

\section{On three levels}
It gets time for a first classification. While relevant for all of physics, it is a central aspect of statistical mechanics and even more for nonequilibrium purposes.\\ 

From a methodological point of view, statistical mechanics gives us tools to connect different levels of physical description. The Gibbs formalism was used throughout the 20th century as workhorse for explaining macroscopic behavior in terms of microscopic laws and constituents. Of course, it has its own content, dealing with the origin and nature of phase transitions, material properties, critical phenomena, and so on. Many of these effects are understood through statistical analysis and fluctuation theory.\\
It is useful then to distinguish, roughly, three levels of physical description:

\begin{enumerate}
    \item \underline{Microscopic level}:    This is where we find the fundamental laws and basic constituents. Depending on what we are trying to describe, this could mean classical or quantum mechanics, Maxwell theory or quantum electrodynamics {\it etc}. The point is simply to find a  starting point in a (well-established) physical theory. For our purposes, the microscopic level is best thought of as the one of classical mechanics. That includes stellar dynamics or planetary motion as far as the internal mechanisms and degrees of freedom may be ignored. It is the world of billiard balls but also of single small molecules and hydrogen atoms in some decoherent approximation.

    \item \underline{Macroscopic level}:    Here we deal with systems that have very many degrees of freedom, systems with particle numbers on the scale of Avogadro’s number,  on a physically coarse-grained level of description. That is where we talk about thermal baths or reservoirs, introduce temperature and entropy, study hydrodynamic equations, and see the emergence of thermodynamic behavior. New law-like behavior appears, such as the Second Law or the ideal gas law. This is also the regime in which we study phase transitions and spontaneous symmetry breaking.  The large-scale structure of the universe, the phases of water, Ohm's law,... belong to macroscopic physics.  Interestingly, one can often forget details of the microscopic world; as long as certain symmetries and conservation principles are obeyed, statistical limit laws ensure that the macroscopic description appears quite independent from the microscopic ingredients. 
    
    \item \underline{Mesoscopic level}: This covers small systems or a few particles interacting with large reservoirs. In that sense, the mesoscopic description depends both on the microscopic and on the macroscopic levels: the reservoirs are described macroscopically, while the system of interest is small. One can also reach the mesoscopic level through reduced or effective descriptions—by going to a low-energy regime, writing down an effective action, or integrating out fast or hidden degrees of freedom. This is where fluctuations become important and where stochastic processes emerge. Brownian motion is the classic example. Typical length scales are between the atomic (microscopic) and bulk (macroscopic) regimes—roughly from nanometers to micrometers.  Here we do not mean so much quantum dots or carbon nanotubes, but focus on small systems such as colloids or molecular motors that are weakly coupled to a thermal environment.
    \end{enumerate}
However, we should nuance this (admittedly oversimplified) classification. Fluctuation theory already appears at the microscopic level whenever there is uncertainty in initial or boundary conditions, as is common already in (few-body) classical dynamical systems. Likewise, the intrinsic uncertainty of quantum mechanics leads to the Born rule and probabilistic descriptions even for single-particle trajectories. Fluctuation theory should not be confused with mesoscopic physics, when the latter is meant to describe physics on nano-to-micrometer scales. Similarly, macroscopic physics may be infiltrated by fluctuations, thinking of disorder or of turbulence, macroscopic chaos etc.\\  Secondly, not all of physics has followed the same steps and distinctions. Think about gravity theory; it is not clear to what extent Einstein's field theory is macroscopic or microscopic.  What is microscopic about spacetime geometry etc? On the side of quantum mechanics, some of us would probably claim that it is (only or still) an effective theory, {\it for practical purposes only}, waiting to be completed.\\

Nevertheless, as good as we can, and given the previous provisos, by way of statistical mechanics, it is important to distinguish these three levels.  Derivations from micro to meso, or from meso to macro all have their specific challenges.  It is good to know on what scale a particular phenomenon plays and what can be the input for its description and explanation.  That remains of course true for nonequilibrium phenomena, and with additional and specific difficulties. For example, far from equilibrium, the Gibbs formalism largely fails for describing static fluctuations, temperature and entropy become problematic, long-range correlations appear and locality becomes an issue.  Let us illustrate that indeed with two quantities that we know and love from equilibrium statistical mechanics.  In a way, they still make the decomposition in Sections \ref{pif}--\ref{strdy} of the action governing path probabilities for nonequilibrium systems:\\

{\bf Temperature:}   There is no mentioning of temperature in Newton's {\it Principia}, as it is solely concerned with mechanics. It took some time indeed before the project of deriving thermodynamics from classical mechanics got going. The first steps in statistical mechanics were made exactly with the statistical interpretation of temperature. As an early precursor, Daniel Bernoulli already mentions that faster motion means hotter gas. Next, there were various stages and various people. For example, Clausius claims that temperature is linked to {\it vis viva} (twice the kinetic energy, introduced by Leibniz), but he did not treat the molecular velocities statistically.  The first genuinely statistical interpretation of temperature is usually credited to Maxwell, in ``On the Dynamical Theory of Gases,'' Philosophical Magazine (1859).  In that paper\footnote{Thinking perhaps of the statistical analysis in Quetelet's {\it Physique Social} (1842) that was partially translated in English by Herschel and influenced Darwin's writing of ``The Origin of Species'' and Maxwell's conceptual understanding alike.}, he introduces the Maxwellian velocity distribution for molecules in a gas, and clearly states that
temperature measures the average kinetic energy of molecules. Today, we still speak about a kinetic temperature, and we characterize the temperature of interstellar space (cosmic microwave background) from the Planck law.\\
Kelvin (1848) defines absolute temperature independently of any particular substance, based on Carnot’s theory of heat engines.
With the statistical foundation of thermodynamics and after Clausius, foremost in the work of Gibbs' 1875–1878 papers ``On the Equilibrium of Heterogeneous Substances,'' absolute temperature $T$ appears as $T^{-1}=\id S/\id E$ for the derivative of the thermodynamic entropy $S$ with respect to the energy $E$.  In that sense, temperature is an equilibrium concept.\\ While, for classical thermodynamics, temperature assumes local equilibrium and large numbers of degrees of freedom, people do speak about temperature for systems composed of only a few degrees of freedom, or far from equilibrium.  However, in that case, things get messy and one must specify the measurement process, which often leads to various definitions of {\it effective} temperature.  It can be defined from response theory or from the statistics in the occupation of energy-levels, and it reflects a more conservative point of view of keeping to older concepts. The introduction of dynamical activity and frenesy \cite{fren}, {\it e.g.} in Section \ref{pif},  may be viewed as an attempt to break away from that tradition to reach a fresher and more powerful perspective for nonequilibrium physics.\\

{\bf Entropy}:  Entropy $S$ is a protean concept in equilibrium physics, where that same quantity is related to heat (via Clausius relation $\id S = \delta Q^\text{rev}/T$), to fluctuations (via Boltzmann relation $S = k_B \log W$) and to statistical/thermodynamic forces (via Onsager's current-force relation $J = \bbL \cdot \nabla S$).  Moreover, entropy plays a central role in linear response theory around equilibrium (in the fluctuation-dissipation relations) and in the gradient-flow aspect of kinetic and macroscopic equations where the entropy is monotone increasing in time. We visit Boltzmann's and Onsager's picture in Chapters \ref{bpic}--\ref{onsp}; and a version of Clausius heat theorem comes in Section \ref{clas}.\\
For steady out-of-equilibrium systems, there are traditionally only two entries for the concept of entropy.  The first and widely-used instance appears in irreversible thermodynamics and is directly related to the assumption of local equilibrium, \cite{gas2,dGM}.  One divides a macroscopic system in regions which are much smaller than the size of the (macroscopic) system but still much larger than microscopic distances (like interaction range).  It remains (locally) an equilibrium concept.  The second widespread use of entropy for nonequilibrium systems is for an ensemble of identical and independent (small and open and driven) systems.  We call it a nonequilibrium or active (dilute) gas, basically following then Boltzmann's entropy (or, H-functional) for a dilute gas.  That is what appears in Section \ref{therk} for a gas of driven colloids (and in much of the literature today).  All the other instances of entropy in this text are related to thermodynamic equilibrium entropy, where always, miraculously, Shannon, Gibbs, Clausius and Boltzmann entropies coincide.  That is not to be expected for macroscopic nonequilibrium system; see also \cite{lebowitz_goldstein2004boltzmann,Goldstein2019,Maes_2012}.  In Sections \ref{h1}-- \ref{limita} we discuss the Boltzmann picture in a wider setting, and we give a more general definition \eqref{be} of a nonequilibrium entropy for a macroscopic system with nontrivial interactions; of course, it is compatible with the Second Law when adding the change of entropy in the environment but to have a consistent picture, we need the condition of local detailed balance, explained in Section \ref{dlbb}.\\

In what follows, we shall restrict ourselves to simple, illustrative ``toy'' models and theoretical frameworks. These, we hope, may serve as stepping stones toward a broader and more systematic understanding of dissipative phenomena.  David Ruelle asserts in \cite{Ruelletoday} that this approach, while not answering all questions, would bring nonequilibrium statistical mechanics to a similar level of understanding that equilibrium statistical mechanics had reached after Gibbs' work:
\begin{quote}
{\it My inclination is to postpone the study of the large-system limit: Since it is feasible to analyze the nonequilibrium properties of finite systems — as Gibbs did for their equilibrium properties — it seems a good idea to start there.}\\
(David Ruelle, 2004)
\end{quote}

However, before, we end this Chapter with some more material on the mesoscale for equilibrium models.

\section{Static equilibrium fluctuations}\label{gii}

The usual and more standard description of
\ thermodynamic equilibrium is concerned with a condition of a many-body system where a collection of \ macroscopic observables no longer changes with time, and all driving forces have vanished. There are no temperature gradients, no net mechanical force, and all chemical potentials are equal for species that can be exchanged.  As a result, no net macroscopic flows of energy, particles, mass or volume exist.  The intensive properties (temperature $T$, chemical potential $\mu$) are uniform throughout the system\footnote{For a fluid under gravity $g$, mechanical equilibrium requires a balance between pressure gradients and body forces. The presence of a pressure gradient $\nabla P\neq 0$ is not a sign of disequilibrium, but rather a necessary condition for equilibrium in a non-uniform potential field.  In that way, the hydrostatic equilibrium condition $\nabla P = -\rho g$ (where $\rho$ is the density) remains consistent with thermodynamic equilibrium: the entropy is maximized subject to the external constraints.  In general relativity, Tolman (1930) showed that in a gravitational field, the condition for thermal equilibrium is actually that $T(r)\sqrt{g_{00}(r)}$ is constant, where $g_{00}(r)$ is the time-component of the spacetime metric. Under weak gravity (on Earth) a sealed column of gas in a gravitational field does not show thermal gradients in thermal equilibrium.}.  There is no net entropy production and the condition is invariant under time-reversal. \\

On an even more fundamental level, the equilibrium condition can be characterized as a condition of maximum entropy given the relevant set of macroscopic quantities and given the constraints on the system.  For example, for a system at fixed volume and particle number in thermal contact ({\it i.e.}, allowing energy exchanges) with a large heat bath fixed at a certain temperature, equilibrium is characterized from minimizing the Helmholtz free energy, a functional of the relevant state characteristics such as density or energy profile.\\
Return to equilibrium is a gradient flow in which steepest descent to the minimum (equilibrium) free energy is sought. Then, back in equilibrium, the macroscopic quantities do no longer change their value and the system appears at rest.  There are no macroscopic currents and all statistical features are symmetric under time-reversal. R.I.P.\\

The probability distributions characterizing the static fluctuations in thermodynamic equilibrium are known as Gibbs distributions\footnote{The notion of Gibbs distribution is in many ways subtle, and it is built around the idea of well-defined relative energies. In probabilistic terms, they are described by their finite conditional distributions, in terms of a quasilocal interaction. Not all (even very regular) probability distributions are Gibbsian and, needless to say, one cannot decide equilibrium {\it versus} nonequilibrium to characterize a process by only looking at the stationary distribution (static fluctuations).}. They give probabilities of states that follow the Boltzmann-Gibbs weight $\rho_\text{eq} \sim \exp[-\beta E]$ for a system in contact with a heat bath at inverse temperature $\beta$ and in a state with energy $E$.   Energy is understood in the mechanical sense, as a function of the microscopic state.  Recall here the ideas of pioneers like Boltzmann and Einstein, who saw the Gibbs distributions in the light of macroscopic (static) fluctuation theory for equilibrium systems.  For that purpose, it remains very instructive to read the first pages of \cite{Einstein1910Opalescence}.  General references about the Gibbs formalism and large deviations include \cite{vanEnterFernandezSokal1993,Varadhan1984,MartinLof1979,denH,Dorlas1999,LanfordLargeDeviations,Ellis1985}.\\
We take the idea in its simplest form where we could abbreviate it as ``equilibrium distributions maximize the entropy given constraints.''  We come back to that picture in Chapter \ref{bpic}.
Here, we already illustrate that for the celebrated Maxwell distribution describing the velocity histogram for a dilute gas (ignoring potential energy).\\
  Consider a gas of $N$ particles of mass $m$.  We also fix the total energy denoting it by a constant $E$ (which is kinetic). At any moment, every particle has a certain velocity $\vec{v}_i$, while there is a constant kinetic energy $E$.  We ask how many particles that velocity $\vec{v}_i \simeq \vec{u} \pm \vec{\delta}$ approximately with small tolerance $\delta$, is equal to a given $\vec{u}$; we denote that number by  $N_{\vec{u}}(\vec{v}_1,\ldots,\vec{v}_N)$.  The answer is
\begin{equation}\label{tt}
\frac{N_{\vec{u}}(\vec{v}_1,\ldots,\vec{v}_N)}{N} \simeq
\frac{e^{-\beta m |\vec{u}|^2/2}}{(2\pi m k_B T)^{3/2}}
\end{equation} 
{\it i.e.}, given by the Maxwellian at inverse temperature $\beta = 1/(k_B T)$) where the temperature $T$ is taken from $E = 3N k_B
T/2$.  For very large $N$, the (near) equality \eqref{tt}
is verified for almost all $(\vec{v}_1,\ldots,\vec{v}_N)$ on the constant $E$-surface. It is in that sense we say that the Maxwell--distribution is the {\it typical} velocity distribution. \\
In other words, when we randomly throw noninteracting particles in a volume for a given fixed total energy, we expect to find the homogeneous Maxwell distribution for a temperature that corresponds to that energy.  That is typical.  Fluctuations refer to deviations from this law.  Here is how Boltzmann wrote about that case (to be read slowly and with care):
\begin{quote}
{\it One should not forget that the Maxwell distribution is not a state
in which each molecule has a definite position and velocity, and
which is thereby attained when the position and velocity of each
molecule approach these definite values asymptotically.... It is
in no way a special singular distribution which is to be contrasted
to infinitely many more non-Maxwellian distributions; rather it is
characterized by the fact that by far the largest number of possible
velocity distributions have the characteristic properties of the
Maxwell distribution, and compared to these there are only a relatively
small number of possible distributions that deviate significantly
from Maxwell’s.}\\ (Ludwig Boltzmann, 1896)
\end{quote}

 Let us look closer at the Gibbs probability distribution $\rho_\text{eq}(x) = e^{-\beta E(x)}/Z$, say for the ensemble population on a finite number of states $x$ for potential\footnote{For classical systems, the kinetic energy is easily integrated out.  In what follows, we often write $E$ for the energy function of a multilevel system resembling the possible energies of a quantum system.} $E$, at inverse temperature $\beta=(k_BT)^{-1}$.  We still avoid the most interesting and important case of spatially extended systems containing many interacting atoms, molecules or particles of some sort. For that finite state space, the relative entropy\footnote{For translation-invariant equilibrium systems with a quasilocal interaction, the Shannon entropy and the Gibbs entropy coincide, and also equal the Boltzmann and Clausius entropies.  For nonequilibrium entropies, and how they separate, we refer to \cite{Maes2012NonequilibriumEntropies,negcap,lebowitz_goldstein2004boltzmann}.  The coincidence for macroscopic systems in thermodynamic equilibrium of Clausius, Boltzmann and Gibbs entropies should of course not be taken for granted in the nonequilibrium world.  Another possible pitfall is to identify probabilities with (macroscopic) densities, and expected values with typical values.  For a steady and open macroscopic nonequilibrium system, it does make sense to speak about its (analogue to Boltzmann) entropy, but only for a dilute gas is there a chance to coincide with the Shannon entropy of its density.  See more in Section \ref{bent}.} of a probability distribution $\mu$ with respect to $\rho_\text{eq}$ is
\begin{eqnarray}
S(\mu|\rho_\text{eq}) = \sum_x \mu(x)\log\frac{\mu(x)}{\rho_\text{eq}(x)} = \beta \sum_x E(x) \mu(x) - S(\mu) + \log Z \geq 0 \label{rele}\\
\text{with Shannon entropy }\; S(\mu) = - \sum_x \mu(x) \log\mu(x) \geq 0\notag
\end{eqnarray}
In other words, the relative entropy contains a difference of the free energy functional ${\cal F}(\mu) := \sum E(x) \mu(x) -k_BT\,S(\mu)$, evaluated in $\mu$ and $\rho_\text{eq}$ respectively:
\[
S(\mu|\rho_\text{eq}) = \beta\left({\cal F}[\mu] - {\cal F}[\rho_\text{eq}]\right), \quad  {\cal F}[\rho_\text{eq}] = -k_BT\log Z
\]
The relative entropy plays a role in finding the probability of a deviation in the histogram, say $\mu$ instead of $\rho_\text{eq}$.  We explain.\\
Take $N$ independent copies of the canonical equilibrium distribution, {\it i.e.}, draw $N$ copies $X_i$ from $\rho_\text{eq}$. The $X_{i}$'s are then independent and identically distributed. Next, look at the fraction of copies where $X_{i}=x$. Formally, we have the following random empirical distribution 
\begin{equation}\label{empd}
\nu(x)=\frac{1}{N}\sum_{i=1}^{N}\delta_{X_{i},x}
\end{equation}
where $\delta$ indicates the Kronecker delta-function (zero or one).
When $N$ grows very large, this fraction approaches the equilibrium distribution, {\it i.e.}, $\nu(x)\rightarrow \rho_\text{eq}$ for $N\rightarrow\infty$, with probability one with respect to $\rho_\text{eq}$. That is the law of large numbers for $\rho_\text{eq}$.

\begin{figure}[H]
\includegraphics[scale=0.8]{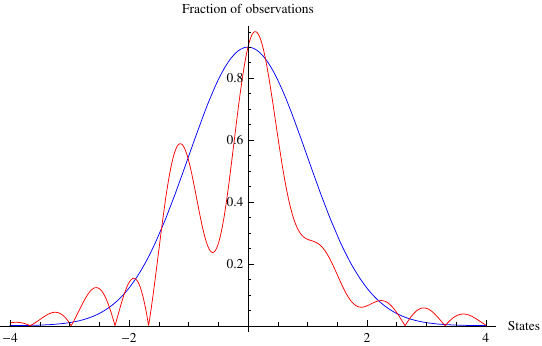}
\caption{The red curve is the distribution for a finite (small) $N$ while the blue curve is the distribution for $N\uparrow +\infty$ corresponding to the equilibrium distribution for some energy function.}
\end{figure}

Now we can ask about corrections to that limit, {\it i.e.}, what is the probability to find a certain distribution $\mu$ for this fraction? How plausible is it to find the histogram $\nu$ in \eqref{empd} given by $\mu$?  We then evaluate static fluctuations, and the result is
\[
\mbox{Prob}_{\text{eq}}[\frac 1{N} \sum_{i=1}^N \delta_{X_i,x} \simeq \mu(x) \text{ for all } x] \sim e^{-N\,S(\mu|\rho_{\text{eq}})},\quad N\uparrow +\infty
\]
in a logarithmic asymptotic sense, where the rate is exactly given by the relative entropy \eqref{rele}.\\
As a consequence, 
\begin{equation}\label{freevar}
\mbox{Prob}_\rho[\frac 1{N} \sum_j \delta_{X_j,\cdot} \simeq \mu] \simeq e^{-N\beta\,[{\cal F}[\mu] - {\cal F}[\rho_\text{eq}]]}
\end{equation}
By construction, the equilibrium distribution $\rho_\text{eq}$ (by the Law of Large Numbers)  maximizes the left-hand side
and hence, looking at the right-hand side, we recover the equilibrium distribution from minimizing $\cal F(\mu)$ over all probability distributions $\mu$. That explains the Gibbs variational principle (at least in this gassy context): the equilibrium distribution $\rho_{\text{eq}}$ is the minimizer of the free energy:
\[
{\cal F}[\mu] \geq -k_BT\log Z = {\cal F}_{\text{eq}}
\]
In other words, $\rho_{\text{eq}}$ maximizes the Shannon entropy (to become the Gibbs entropy)
$S(\mu) = -\sum_x\mu(x) \log\mu(x)$ when keeping the energy $\sum_x \mu(x) E(x)$ fixed at some value.\\
There is a second good reason to appreciate that formula \eqref{freevar}, since we know that the free energy has a good operational meaning, for instance related to isothermal work. Free energies govern the static fluctuations \eqref{freevar}, and they have a clear physical meaning. In other words, the static fluctuation in and around thermodynamic equilibrium are characterized by operationally or experimentally accessible thermodynamic potentials.  Explanations and details can be taken from the literature, including \cite{Einstein1910Opalescence,vanEnterFernandezSokal1993,LanfordLargeDeviations,MartinLof1979,Ellis1985}.\\

That is no longer true ({\it i.e.}, has no operational version) for nonequilibrium systems\footnote{The close-to-equilibrium situation still enjoys variational principles such as minimum and maximum entropy production principles, \cite{minep}.}, and we must therefore invent and try new methods of analysis: there does not appear to exist in general a powerful formula characterizing the static fluctuations of a macroscopic nonequilibrium system in terms of operationally accessible physics. One needs to {\it derive} the fixed time macroscopic behavior including fluctuations from dynamical ensembles.  The good news is there indeed; at least for a broad and interesting class of nonequilibrium processes, we can characterize the {\it dynamical} fluctuations in terms of a number of more accessible quantities, albeit that some are nonthermodynamic.  We
learn about these in Sections \ref{pif} and \ref{maft}.  When speaking about the probability of dynamical fluctuations, we also refer to trajectory- or path-ensembles.  In any event, we need to turn to dynamical considerations.

\section{Equilibrium jumping}

\subsection{Under detailed balance}\label{relax}

We continue by adding elementary points of recognition for, in a long string of words,  ``equilibrium mesoscopic stochastic dynamics.''
For simplicity, we continue to focus mostly on systems with a finite number of states.  We can imagine that they give a coarse-grained description of a chemo-mechanical condition, or that they correspond to the relevant and possible configurations. We always work for the case that the system is in thermal contact with a (unique) equilibrium heat bath, with constant finite volume and particle number.  That is the context of the so-called canonical ensemble where the Helmholtz free energy is the natural thermodynamic potential (as in the previous section).  Again, statements about probabilities must be regarded in the ensemble-sense, like for a gas of these systems, to make physically meaningful statements.\\
 
\begin{figure}[H]
\centering
\includegraphics[scale=0.5]{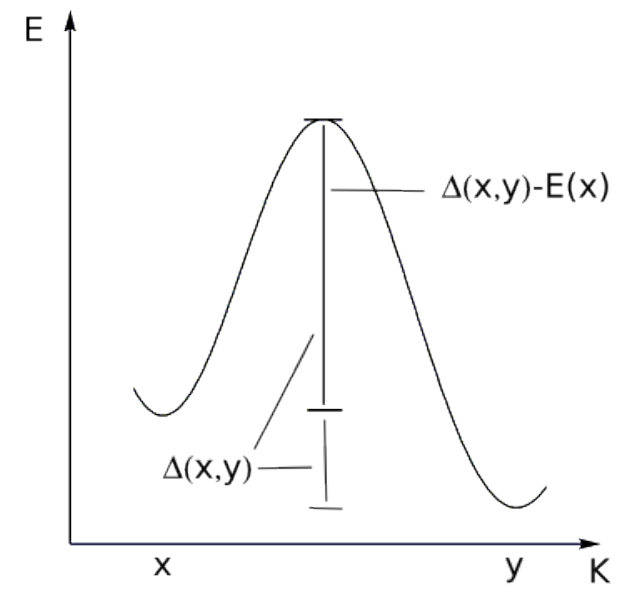}
\caption{$k(x,y)=a(x,y)e^{-\beta\left[\Delta(x,y)-E(x)\right]}$ with prefactor $a(x,y)=a(y,x)$ that depends on the specific path in the energy landscape. The $\Delta(x,y)=\Delta(y,x)$ measures the height of the barrier between states $x$ and $y$.  E.g. $\Delta(x,y) = [E(x) + E(y)]/2$ as in \eqref{arh}.}\label{figa}
\end{figure}

\begin{figure}[H]
\includegraphics[width=0.5\linewidth]{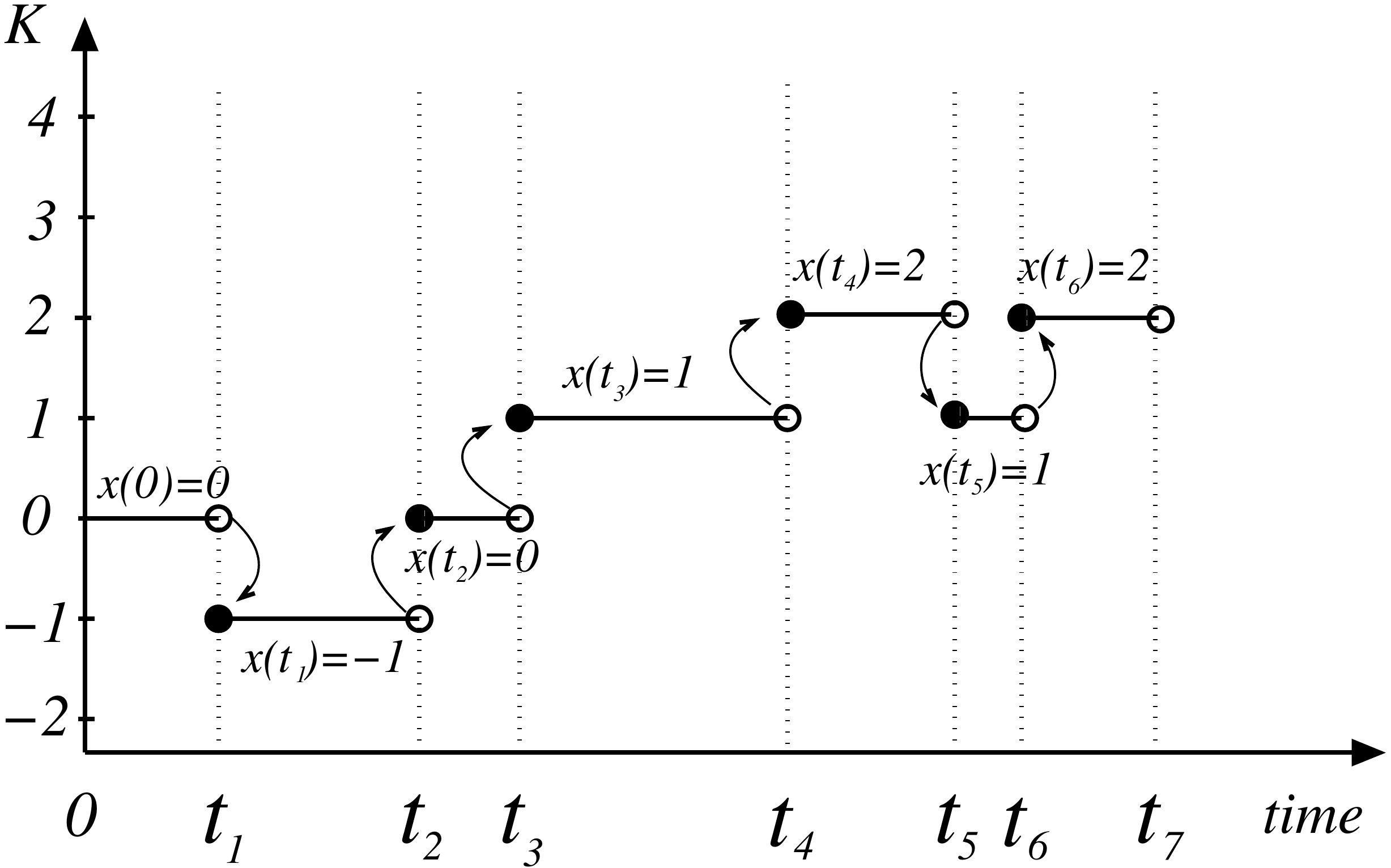}
\caption{A continuous time random walk on $\bbZ$.}
\label{ju}
\end{figure}

Let us consider a finite state space $K$ with states $x\in K$.  We have in mind many ($N$) independent particles that each can have energies $E(x)$, using a set-up that goes back to Boltzmann and his treatment of multilevel systems.  In that state of mind, we focus on energy (ex)changes. More specifically, we consider a Markov jump process with transition rates for the jump $x\rightarrow y$ given by
\begin{equation}\label{arh}
k(x,y) = \psi_\beta(x,y)\,\exp -\frac{\beta}{2}[E(y)-E(x)],\quad \beta = \frac 1{k_B T}
\end{equation}
with symmetric $\psi_\beta(x,y)=\psi_\beta(y,x)$ in general also depending on the inverse temperature $\beta$.  It may be given by an Arrhenius formula where we associate $x,y$ to local minima of a more continuous energy profile; see Fig.~\ref{figa}.
Remember that trajectories consist here of piecewise constant functions in time as illustrated in Fig.~\ref{ju}.

Most important about the above transition rates \eqref{arh} is that they satisfy detailed balance:
\[
\frac{k(x,y)}{k(y,x)} = \exp -\beta [E(y)-E(x)]
\] 
It implies that the process is invariant under time-reversal, 
\begin{equation}\label{detb}
\rho_\text{eq}(x)k(x,y) = \rho_\text{eq}(y)k(y,x)
\end{equation}
when running stationary with (single time) equilibrium distribution
\[
\rho_{\text{eq}}(x) = \frac 1{Z}\,e^{-\beta E(x)}, \quad Z = \sum_x e^{-\beta E(x)}
\]
It means that the equilibrium process does not show an arrow of time: for instance, the time-sequence $(x,y)$ is equally probable as $(y,x)$.\\

Furthermore, the free energy functional
\[
{\cal F}[\mu] := \sum_x \mu(x) E(x) + k_BT\,\sum_x\mu(x)\log\mu(x)
\]
defined on general probability distributions $\mu$ on $K$ is monotonically decreasing in time, {\it i.e.}, $${\cal F}[\nu_{t}]\searrow  {\cal F}_{\text{eq}}$$ where
${\cal F}_{\text{eq}} := {\cal F}[\rho_{\text{eq}}] = -k_BT \log Z$, and $\nu_t$ is the probability distribution at time $t$ as solution of the Master equation\footnote{Named, respectfully, after Pauli, \cite{Pauli1927}. Its origin is Einstein's radiation theory \cite{Einstein1917} and the application of the Dirac-Fermi Golden Rule approximation.}
\[
\frac{\id\nu_{t}(x)}{\id t} = \sum_{y}\left[k(y,x)\nu_{t}(y) -k(x,y)\nu_{t}(x)\right],\qquad x\in K
\]
The resulting evolution, in the case of detailed balance, is an example of nonlinear gradient flow.

\subsection{Clausius heat theorem}\label{clas}

The Clausius heat theorem (1865) uses the idea of a quasistatic process.  Such processes are central in the discussion of thermodynamic transformations.  One can think of changing the volume of the system very slowly, perhaps combined with very slow changes of the temperature in the heat bath.  The relevant time-scale (to compare what is slow or fast) is the relevant relaxation time of the system.\\
For illustration, look at a system dynamics having time-dependent transition rates for $x\rightarrow y$,
 $$k_{t}(x,y) = \exp\{-\frac{\beta_{t}}{2}\left[E(y,\alpha_{t})-E(x,\alpha_{t})\right]\}$$
  Here, $\beta_{t}$ is the time--dependent inverse temperature of the environment and $\alpha_{t}$ is a time--dependent external parameter (volume perhaps). For each fixed time $t$ the dynamics is detailed balance with corresponding stationary distribution $\rho_{\text{eq}(t)}(x)\propto \exp[-\beta_t E(x,\alpha_t)]$; for a fixed time $t$, one can think of $\rho_{\text{eq}(t)}(x)$ as an instantaneous equilibrium distribution. In reality, for the time-dependent process, the system would be constantly frustrated, lagging behind, depending on the time-scales of relaxation to equilibrium on the one hand and the rate of change in the time--dependent parameter on the other hand. Now imagine the case of extremely slow changes in time of the parameters. In that quasistatic limit the solution to the time--dependent Master equation  converges at each time to the instantaneous $\rho_{\text{eq}(t)}(x)$. Since on top of that we have also assumed detailed balance, we get full reversibility.  Those are the reversible evolutions of thermodynamics; they can be discussed as curves in $PV-$diagrams etc.\\
  
  It is indeed easy enough to check that, in that limit, time disappears and we get the ``geometric'' relation,
  $$\beta\left[\id\langle E\rangle - \langle\frac{\partial}{\partial\alpha}E\rangle\,\id \alpha\right] = \id\left[\beta\langle E\rangle +\log Z_{\alpha,\beta}\right]$$ where 
  the '$\id$' makes a differential in the $(\alpha,\beta)-$plane,  and the averages $\langle\cdot\rangle$ are taken with respect to $\rho_{\text{eq}(\alpha,\beta)}(x)=e^{-\beta E(x,\alpha)}/Z_{\alpha,\beta}$. That equality confirms the Clausius heat theorem for quasistatic processes: the right-hand side, and hence also the left-hand $\delta Q^\text{rev}/(k_BT) $ (energy change minus work, divided by $k_BT$), is an exact differential.  It defines the state function entropy $S$ via $\delta Q^\text{rev}/T = \id S$.\\

More in general, if the parameters do not change slow enough with respect to the relaxation of the system, we get entropy production.  In the given case where we imagine many independent copies (dilute gas of such finite systems) we can see how the Shannon entropy of the resulting density is changing.  For the probability distribution $\nu_t$ at time $t$: $$S(\nu_{t})=-\sum_{x}\nu_{t}(x)\log\nu_{t}(x)$$ 
Taking the time derivative leads to a  balance equation,
$$\frac{\id}{\id t}S(\nu_{t})=-\sum_{x}\frac{\id\nu_{t}(x)}{\id t}\log\nu_{t}(x)-\sum_{x}\nu_{t}(x)\frac{1}{\nu_{t}(x)}\frac{\id\nu_{t}(x)}{\id t}$$ The second term on the right-hand side is zero because the normalization of probabilities  $\sum_{x}\nu_{t}(x)=1$ holds for all time $t$. We insert the Master equation
$$
\frac{d\nu_{t}(x)}{dt} = \sum_{y}\left[k_{t}(y,x)\nu_{t}(y) -k_{t}(x,y)\nu_{t}(x)\right]$$ so that the change in Shannon entropy can be rewritten as
 \begin{eqnarray}
\frac{\id}{\id t}S(\nu_{t})&=& \frac{1}{2} \sum_{x,y}\left[k_{t}(x,y)\nu_{t}(x) -k_{t}(y,x)\nu_{t}(y) \right]\log\frac{\nu_{t}(x)}{\nu_{t}(y)}\nonumber\\
&=&\frac{1}{2} \sum_{x,y}\left[...\right]\log\frac{\nu_{t}(x)k_{t}(x,y)}{\nu_{t}(y)k_{t}(y,x)} - \frac{1}{2} \sum_{x,y}\left[...\right]\log\frac{k_{t}(x,y)}{k_{t}(y,x)}\nonumber
\end{eqnarray}
The first term is always positive because it takes the form $(a-b)\log\frac{a}{b}\geqslant 0$. Using detailed balance,
\[
\log \frac{k_{t}(x,y)}{k_{t}(y,x)} = \beta_t\left[E(x,\alpha_t)-E(y,\alpha_t)\right]
\]
which is the change of energy in the heat bath during the system jump $x\rightarrow y$ at time $t$, times the inverse temperature. Hence, writing $\frac{\delta{Q}}{\id t}$ for the incoming heat flux at time $t$, we end up with 
$$\frac{\id}{\id t}S(\nu_{t}) =  \mbox{ a positive\ term} + \beta_t\frac{\delta Q}{\id t}$$
or$$\frac{\id}{\id t}S(\nu_{t}) - \beta_t\frac{\delta Q}{\id t} \geqslant 0.$$
It is a toy-version of the Second Law for transformations of parameters in a jump process that runs under detailed balance, presenting the physics of the linear Boltzmann equation.

\subsection{Statistical forces}\label{staf}

One of the most powerful images throughout science is that of mechanical motion in a potential landscape.
The potential changes along some coordinate that we can imagine as the position of a particle on a toboggan to visualize the future motion of the coordinate.  Let us see how that works for a finite-dimensional degree of freedom $x \in \bbR^n$ which is coupled to an equilibrium bath.  For that we have in mind a joint energy function (or potential) $E(x,\eta)$ where the $\eta = (\eta_i)_{i=1}^N$ stand for the degrees of freedom in the bath.  The mechanical force on $x $ is then
\[
-\nabla_x E(x,\eta)
\]
depending on the bath (or hidden) degrees of freedom $\eta$.
The statistical force arises by integrating out the $\eta$ for an equilibrium ensemble.  We take the canonical ensemble at inverse temperature $\beta = (k_BT)^{-1}$,
\[
\langle\nabla_x E(x,\eta) \rangle := \frac 1{Z(x)}\, \sum_{\eta} e^{-\beta E(x,\eta)} \nabla_x E(x,\eta) = \nabla_x {\cal F}
\]
for Helmholtz free energy  
\[
{\cal F}(x) := -k_BT\,\log Z(x), \quad Z(x):=\sum_{\eta} e^{-\beta E(x,\eta)}
\]
In other words, $x$ undergoes a mean force that is minus the gradient of the free energy of the bath.  That free energy of the bath changes as $x$ is displaced.  The equilibrium position of $x$ will correspond to the minimum of $\cal F$ with gradient flow towards it.\\

We have not written in detail about the exact coupling between $x$ and $\eta$; see \cite{MaesNetočný2019}.  In many cases, we think of $x$ as more macroscopic and slower than the fast and more microscopic degrees of freedom $\eta$.  In fact, we can also think of $x$ as representing collective or macroscopic variables.  Then $x$ could be the local density, in which case we better change the notation $x\rightarrow \rho$, where $\rho$ is now a function of a spatial coordinate $r$ throughout the volume.  Yet, similarly, we continue to think of the (functional) derivative of the free energy as a force.  
  That also enters the balance equations of irreversible thermodynamics, and a slight generalization leads to the notion of stress tensor.

\section{Kinetic Ising model}
Suppose we have a lattice cube $V$ where a ``spin'' $\sigma_i$ is assigned to each site $i\in V$.  The spin configuration is $x = (\sigma_i, i\in V)$.  We also specify boundary conditions by fixing all the spins outside $V$ (while only the spins at the boundary $\partial V$ really matter).  A spinflip dynamics on the $x$ is specified when we give the rate $c(i,x)$ at which a site $i \in V$ undergoes a spinflip $x\rightarrow x^i$.  Here $x^i$ is the new configuration where all spins are equal except for the one at site $i$.  We write the backward generator as a sum
\begin{equation}\label{defLi}
L = \sum_{i\in V} L_i, \qquad L_i f(x) = c(i,x) [f(x^i) - f(x)]
\end{equation}
where the flip rate $c(i,x)$ is assumed to depend only on the spins $\sigma_j,j\in \cal N_i$, in some neighborhood $\cal N_i$ of $i$.  For simplicity, just take the set of nearest neighbors $j\sim i$ plus the site $i$ itself.  In this way we generate a Markov dynamics on the spin configurations in $V$.  We can make the process detailed balance with respect to the energy (in $V$),
\[
U(x) = -J\sum_{i \in V, j\sim i} \sigma_i\sigma_j - b\sum_{i\in V} \sigma_i
\]
where the first sum is over nearest neighbor pairs (so we need the boundary conditions).  That energy corresponds to the standard Ising model with coupling $J$ and magnetic field $b$.  Detailed balance for inverse temperature $\beta$ requires
\begin{equation}\label{li}
c(i,x) = c(i,x^i) \exp\{ -\beta [U(x^i)-U(x)]\}
\end{equation}
where, most importantly, the relative energies $U(x^i)-U(x)$ are well defined.  There is then a unique equilibrium process with the Gibbs-Boltzmann weight making the reversible distribution $\rho(x) = Z^{-1}\exp(-\beta U(x))$.\\

It is important to note that $\rho$ is also reversible with respect to each $L_i$, defined in \eqref{defLi}--\eqref{li}.  We have indeed that for all functions $f,h$,
\begin{eqnarray}
\sum_x \rho(x) f(x^i)\, h(x) &=&  \sum_x Z^{-1}\exp(-\beta U(x^i))\, f(x)\,h(x^i) \notag\\\ 
&=& \sum_x \rho(x) f(x) \exp(\beta [U(x)- U(x^i)])\,h(x^i)\notag\\
&=& \sum_x \rho(x) f(x) \frac{c(i,x)}{c(i,x^i)}\,h(x^i)
\end{eqnarray}
We can take $h(x) = c(i,x)\,g(x)$ and the above equality reads, for all $i\in V$,
\[
\sum_x \rho(x) f(x^i)\, c(i,x)\,g(x) = \sum_x \rho(x) f(x) c(i,x) \,g(x^i)
\]
which also implies the local invariance $\sum_x \rho(x) L_if(x) =0$ for all $i\in V$.  That is called local reversibility (not to be confused with local detailed balance).  As a matter of fact, that local reversibility truly characterizes the distribution $\rho$ as a Gibbs distribution\footnote{Called the Dobrushin-Lanford-Ruelle characterization.}.  It is related to the spatial Markov property of Gibbs distributions, which is crucial for many applications such as local restoration and error correction\footnote{That typically gets much more complicated for stationary nonequilibrium distributions, where one needs restoring on the spacetime level.}.

\chapter{Nonequilibrium subjects}
\lhead{C. Maes}
\rhead{Nonequilibrium subjects}

To introduce structure into the vast domain of nonequilibrium physics, we can divide it --- somewhat arbitrarily --- into a few distinct, though still extensive, categories. At present, it is useful to distinguish at least three main types of nonequilibrium situations.

\begin{enumerate}
\item
{\bf Relaxation behavior.}  
A macroscopic system, whether closed or open, generally evolves in time on macroscopic scales, even when left undisturbed and for most initial conditions. During this interval, before the system reaches its time-asymptotic condition, its collective properties and macroscopic appearance continue to change. This period is called the \underline{transient regime}.  For example, we may prepare a system with an atypical energy distribution and then allow it to evolve freely. The motion of its many microscopic constituents—molecules, atoms, and so forth—can lead to changes in macroscopic quantities such as density or energy profiles.\\
A regime of particular interest is that in which the relaxation time is much larger than the microscopic timescale of molecular collisions. 
In this regime, civilizations rise and decay, currents may appear or decay, phase separation can occur, and order parameters may relax toward stationary values. The best-studied example is the relaxation (or return) to thermodynamic equilibrium, where unitarity or detailed balance plays a fundamental role. Understanding this return process has long been a central subject of nonequilibrium physics, particularly because it raises the question of how dissipation emerges despite the microscopic reversibility of the underlying dynamics.
This topic connects to many important concepts, such as the nature of thermal baths and the formulation of response and fluctuation relations for systems—large or small—coupled to such environments.
Kinetic effects that hinder or complicate the smooth relaxation to equilibrium include glassy dynamics, jamming transitions, constraints, ageing, memory effects, and localization.\\
A more general form of relaxation is the convergence to a steady, possibly nonequilibrium, condition that occurs in open systems, not necessarily macroscopic, and/or where numerous conserved quantities may exist.  Phenomena of metastability and prethermalization belong here as well.
\item 
{\bf Steady-state behavior in contact with spacetime well-separated equilibrium baths.} 
When a system reaches a stationary condition, its observable properties no longer change with time on the scale of observation. Thermodynamic equilibrium is one example of such a stationary state. However, open and driven systems can also maintain stationary profiles of physical quantities --- such as densities --- while supporting nonzero currents.

These complexions are known as \underline{nonequilibrium steady states} (NESS). Only in the last several decades has the study of steady-state nonequilibrium systems become a mainstream topic in physics.
Typically, we consider an open system in contact with multiple large reservoirs, each weakly coupled to the system but maintained at distinct macroscopic conditions. A familiar example is a metal rod placed between two thermal baths held at different temperatures. The reservoirs may also be chemical or mechanical in nature, and external driving may arise from rotational forces, applied shear, or other means. In this regime, the principle of local detailed balance is often invoked (more in Section \ref{dlbb}).\\
Steady-state behavior can also arise when \underline{time-dependent external forces} act on a system, leading to retardation and dissipation effects. For instance, a ferromagnet subjected to a periodically varying magnetic field provides a relevant example. In such cases, the system continually attempts to relax toward the instantaneous equilibrium corresponding to the current external parameters, yet it always lags behind them. Two limiting cases can be distinguished: the quasistatic limit, in which external parameters change very slowly compared to the system’s relaxation time, and the fast (or fluid) limit, in which the parameters vary very rapidly. Both limits can, in certain contexts, be described as adiabatic\footnote{``Quasistatic processes'' are very slow compared to relevant relaxation times; ``adiabatic processes''  proceed without exchange of energy through molecular processes, {\it i.e.}, no heat, not fordable.  Reversible processes  are quasistatic and without driving (allowing instantaneous time-reversal symmetry) and, following Clausius in Section \ref{clas}, the entropy is invariant when the process is adiabatic as well.}.
\item 
{\bf Systems coupled to nonequilibrium media or non-separated baths.}
A third class of nonequilibrium phenomena arises when a system interacts with one or more nonequilibrium reservoirs or with baths that are not well-separated in space or time.
In such situations, the system may be embedded in or \underline{coupled to a complex } nonequilibrium environment. This is particularly relevant in biological contexts, where heterogeneity and multiple nonequilibrium sources coexist and interact.\\
These setups give rise to new types of ``engines,'' in which energy or matter flows between nonequilibrium reservoirs --- such as biological baths --- while performing work. In other cases, the separation between system and bath becomes blurred: for example, when media are either quantum entangled (for small devices) or when they are not time-separable and showing many time-scales ({\it e.g.}, for climate problems), we enter that third type of nonequilibrium as well.
\end{enumerate}

The source of nonequilibrium can thus be three--fold (at least): (1) relaxation problems: the system is {\it on its way} to some equilibrium or to some more exotic stationary condition, (2) steady-state problems: the system is open and undergoes driving via external reservoirs, rotational forces or time-periodic external fields, but is in a steady condition, (3) absence of spatial locality or temporal resolution: the system is coupled to various nonequilibrium media, or the different media are no longer spatially or temporally separable.\\
Of course, many intermediate and hybrid situations exist, with rich overlaps among these fundamental types.

\section{Phenomenology}\label{gena}

Nonequilibrium is ubiquitous. The most amazing
examples are probably found in life processes, or perhaps in the
origin of life itself.  There we see an open system fully engaged in transport processes, with little engines, pumps and cycles, and the
emergence of order on diverse spatio-temporal scales out of molecular complexity. Molecular motors give mesoscopic realizations of chemical engines and of transport on  molecular scales.
  The cell itself, how it moves and how its membrane fluctuates on $\mu m$-scales, brings nonequilibrium statistical mechanics to life.    Chemical reactions are traditionally also sources of nonequilibrium changes.  There, the relation between biochemical functioning and thermo-mechanical properties is of central interest today.
\\ However, the variety of phenomena where
nonequilibrium considerations are essential is much larger.  We find them at cosmological scales of the observable
 universe and beyond the smallest sizes of nanotechnology.  Subatomic processes, the creation and annihilation of particles, and spontaneous fluctuations at the smallest dimensions produce sources of noise and matter, from which our world is finally made.  On more earthly scales, evolution and the fight for low entropy create cycles of life and destruction that we recognize in all of nature. 
   Turbulent flow and similar nonlinear evolutions give other macroscopic realizations of nonequilibrium effects.
  Climatology and ecology ask questions about the atmosphere and ocean dynamics and about foodweb chains as
 nonequilibrium systems.\\
 Above all that, we see the Second Law of thermodynamics at work in all processes, putting fundamental limitations on all theories and models of this universe.\\
We list some standard examples referring to nonequilibrium phenomenology:

\begin{itemize}
\item Consider a material ({\it e.g.} a metal rod) connected to a hot environment at one end and to a cold environment at the other end. Energy flows from the hot to the cold side, at least usually. The difference in temperatures at the left and right ends is ``frustrating'' the system and, on the appropriate time-scale where the conflicting reservoirs remain each at equilibrium keeping their temperatures fixed, a stationary heat current will be maintained in the system.  Heat will be flowing in/out of these reservoirs changing their energy (without changing their temperature).  That gives rise to a stationary entropy flux, applying the Clausius relation $\id S = \delta Q^\text{rev}/ T_\text{high}$ in the bath at temperature $T_\text{high}$.  Statistical mechanics wants to derive the transport equations (such as the Fourier law with possible modifications), and to understand the more microscopic nature of thermal conductivity. Obviously, we are also interested in nonlinear regimes and fluctuations.  {\it E.g.}, what does it take to essentially stop or even to reverse the heat current?  Furthermore, what is the role of the environment reservoirs and how does the coupling with the rod influence the heat and energy transport?  How small can we take this setup?

\begin{figure}[H]
\centering
\includegraphics[scale=0.8]{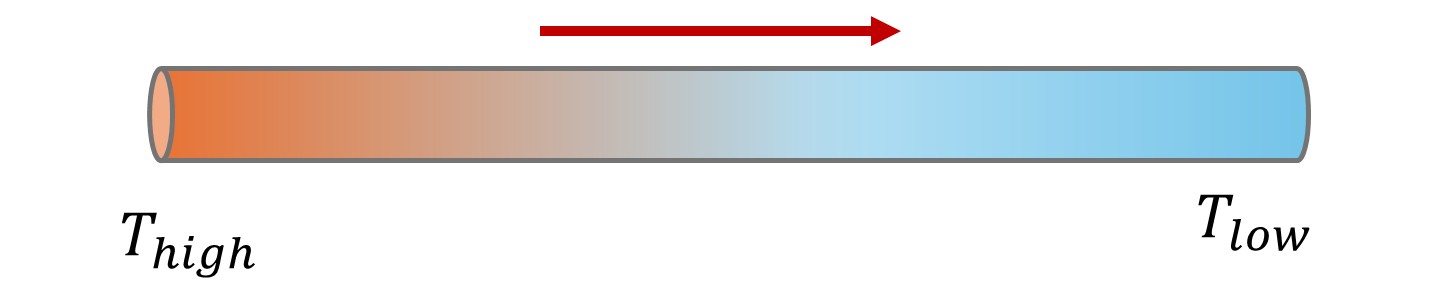}
\end{figure}

\item In an electric circuit, charges are displaced and these currents are usually created by an external source such as a battery. The battery works to displace electric charges maintaining a current which itself causes Joule heating in any resistor. That is a way of making hot water for tea, by the heat dissipation into the environment.  There will also be fluctuations and noise on the potential and current over and through a resistor.  Depending on the type of circuit and on its elements, negative differential conductivities could be observed.  How is all that linked together from more microscopic modeling and analysis?

\item At the membrane of a biological cell, currents of ions are passing through pores due to gradients in electrical and/or chemical potential.  These are influenced by the environment and are subject to fluctuations.  Also within the cell there is transport and dissipation driven {\it e.g.}, by the ATP-concentration. As we are dealing with active matter ({\it e.g.}, molecular motors), there may be features in the relation between fluctuations and response of the currents that differ significantly from the passive case introduced in the previous two examples.

\begin{figure}[H]
\centering
\includegraphics[scale=0.8]{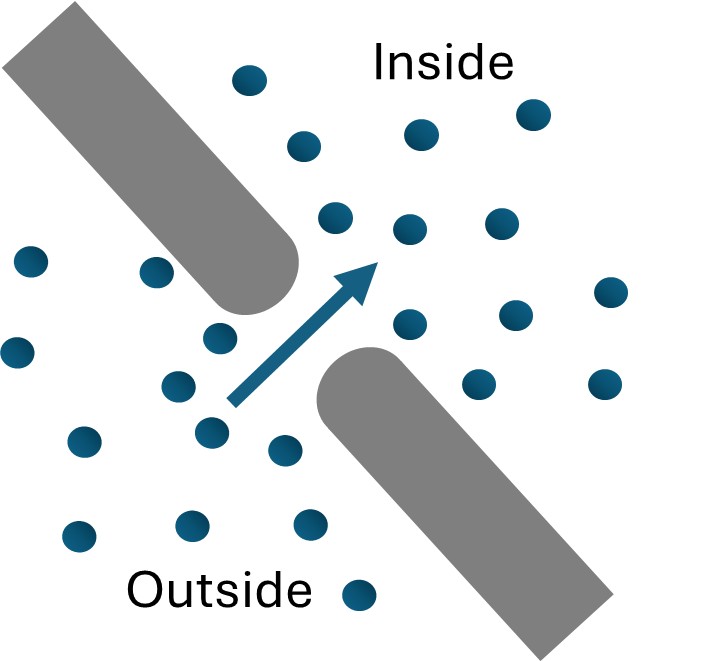}
\end{figure}
For all these transport examples, there appear spontaneous current (of energy, charge and particle) fluctuations that differ from equilibrium (where the first moment equals zero).

\item Thermoelectric effects (and other effects that couple currents) involve coupled transport of heat and charge, driven by temperature gradients and electric fields, giving a textbook example of a nonequilibrium steady state. The Peltier effect (1834) states that an electric current between different metallic conductors will cause production of absorption of heat at the junctions.  The Seebeck effect (1826) is the appearance of an emf in a circuit if the junctions are maintained at different temperatures [thermocouple].
 It was inspiring for the understanding of reciprocity and its development into linear response theory (Onsager relations). Classic demonstrations (like the Seebeck or Peltier effect) opened the understanding of close-to-equilibrium systems. Modern research into thermoelectricity in nanoscale systems and under large gradients pushes well beyond linear response, making it very much a frontier of nonequilibrium physics.

\item Rayleigh–Bénard convection (1916–1917): a fluid heated from below develops a regular pattern of convection rolls above a certain temperature gradient.  It demonstrates spontaneous symmetry breaking and pattern formation in a nonequilibrium steady state.  Energy flow is crucial here for the development of the type of macroscopic order we see in these Rayleigh–Bénard rolls.\\
To characterize the transition between the conductive and the convective regime, it is useful to introduce the Rayleigh number Ra for a fluid column of height $h$ and with upper and lower temperatures $T_u, T_b$,
\[
\text{Ra} = \frac{g\beta}{\nu \alpha}\,(T_b-T_u) \,h^3
\]
where $\nu$ is the kinematic viscosity, $\alpha$ is the thermal diffusivity, and $\beta$ is the thermal expansion.  The Rayleigh number measures the ratio  between characteristic time scales for thermal diffusion $h^2/\alpha$ {\it versus} thermal convection $\nu/(\delta \rho h g$ with $\Delta \rho = \rho\beta \Delta T$.  The transition to convection happens around Ra $ \simeq 1708$.  It plays an important role in atmosphere dynamics.  In the case the fluid is active (like composed of bacteria), the nature of the transitions from conduction over convection to turbulence remains less clear. 

\item The Belousov–Zhabotinsky reaction (1950s–1960s) is an example of a  (driven) chemical oscillator exhibiting periodic concentration (of different color!) changes.  It can be understood from coupling nonlinear reaction equations.  It is now a classic example of chemical nonequilibrium dynamics and temporal pattern formation.  It indeed demonstrates that nonequilibrium systems can exhibit sustained oscillations.   A related phenomenon is the (possible) occurrence of time-crystals.  It relates to the understanding of synchronization and the organization of sustained nonlinear oscillations in dissipative systems.

\item Granular materials and vibrated granular layers (1990s): when we shake grains, there may develop stripes, hexagons, or localized excitations.  They form examples of collective behavior in athermal systems far from equilibrium. Here is also the land of sandpile and earthquake models, realistic and less realistic, with  a possible time-scale separation between diffusive and reactive scales that they share with some reaction-diffusion models.

\item Cold atom experiments and quantum quenches (2000s–present) apply sudden changes (quenches) to parameters in ultra-cold atom systems and study their subsequent evolution.  They are quantum analogues of nonequilibrium transient dynamics in the study of thermalization, many-body localization, and prethermalization, challenging traditional ideas of relaxation to thermal equilibrium.

\item Turbulence in fluids and plasmas occurs at high-Reynolds-number where flows develop chaotic, multiscale velocity fields.  It is a paradigmatic nonequilibrium problem with enormous practical and theoretical significance.  In particular, the phenomenon of  energy cascades and scaling laws (partially studied in the Kolmogorov theory) illustrate universal behavior far from equilibrium.

\item Dynamical fluctuation behavior (1990s–present): also the trajectories of very small systems from nano to mesoscales can be measured and used to define notions of heat, work and entropy production.  Moreover, dynamical activity appears to play a crucial role in the response theory when the system is driven away form equilibrium. These days, optical manipulations and fast-camera observations allow detailed visualization and study of these systems via statistical properties of their trajectories.  We are talking then about technological and experimental advances such as optical tweezers 
and optical traps (Nobel Prize in Physics, 2018) and the broader use of fluorescent dyes and lasers to track biological systems, including quantum dots (Nobel Prize in Chemistry, 2023).
The ultra-fast camera (Nobel Prize in Physics, 2023) and key developments in colloidal physics and soft condensed matter have also played a major role. Super-resolution optical microscopy highlights individual molecules. We now have direct measurements of molecular forces, also driving molecular machines (Nobel Prize in 
Chemistry, 2016), which mimic biological motors in an artificial way. New measuring 
instruments, such as the atomic force microscope (AFM), invented by IBM scientists around 1985, have made it possible to scan the mechanical properties of molecules and cells—similar to how a record player’s needle reads sound. Before that, the scanning tunnelling microscope (Nobel Prize in Physics, 1986) had already been a breakthrough, but the AFM removed the limitation that objects must be electrically conductive.

\item Active Matter Systems (2000s–present): concerning systems like self-propelled particles (bacteria, synthetic swimmers) exhibiting flocking, clustering, etc.
The nonequilibrium is due to constant energy consumption at the microscale. Novel phases are revealed and showing transitions without equilibrium analogs.

\item Fermi Acceleration (Plasmas, Cosmic Rays): a stunning example of energy gain through stochastic interactions in a driven environment, central in astrophysical plasma physics and nonthermal energy distributions. The mechanisms ({\it e.g.}, particle energization, shock acceleration) embody far-from-equilibrium dynamics and the coupling between waves and matter.  Related instabilities due to wave-particle energy transfers are well documented, \textit{e.g.} the origin of Langmuir waves \cite{langmuir} (and the ``negative mass effect'' \cite{negativemass, negativemass2}), Faraday waves \cite{waveparticleexperiment, waveparticleexperiment2} (where a flat hydrostatic surface becomes unstable by a vibrating receptacle), plasma heating and acceleration more generally, \cite{plasmaphysicsbook}, and in the interaction of waves with active matter \cite{beyen2025couplingelasticstringactive}.

\item Physics of Life: life itself is arguably the ultimate nonequilibrium phenomenon. Life exists and persists far from equilibrium, constantly consuming energy to maintain order, process information, and reproduce.  It also includes the understanding of the conversion of energy ({\it e.g.}, from sunlight or food) into chemical work — an ongoing nonequilibrium process that sustains life's low entropy state.  Biological engines allow DNA replication, transcription, and translation, all are driven by chemical gradients (ATP, GTP hydrolysis), all irreversible and far from equilibrium.

\item Climate science: Earth continuously receives energy from the Sun and radiates it back into space. This constant energy flow keeps the atmosphere, oceans, and biosphere in nonequilibrium states.  Jet streams, Hadley cells, ocean currents, El Niño events — all are emergent structures in a far-from-equilibrium fluid system.
 Melting ice sheets, tipping points, and ocean circulation shifts involve irreversible dynamics, central to nonequilibrium processes.  Climate feedbacks (like water vapor amplification or albedo loss) are fundamentally nonlinear and lead to complex, possibly chaotic dynamics. The climate is a multiscale, coupled system: atmosphere, ocean, biosphere, cryosphere — each with its own nonequilibrium behavior. Climate science is one of the grandest and most urgent nonequilibrium challenges in contemporary physics.
 \end{itemize}

\section{Theoretical strides}
The first steps in  nonequilibrium statistical mechanics have been made in kinetic theory, in particular of gases. The formulation of the Second Law of thermodynamics has various authors.  Rudolf Clausius, \cite{clausius1865} in 1850-1865, is often credited with the first clear formulation, ``Heat cannot of itself pass from a colder to a hotter body.''
He introduced the concept of entropy in 1865 and gave a general mathematical formulation.\\ William Thomson (Lord Kelvin) around 1851 independently formulated the Second Law, with a version that focused on the impossibility of converting all heat into work, ``It is impossible to construct a heat engine which does nothing but convert heat into work.''  However, there was the precursor Sadi Carnot who in 1824 wrote his ``“Reflections on the Motive Power of Fire.'' In more abstract ways, he studied heat engines and introduced the idea of efficiency limits.  After these ``thermodynamic beginnings'' fluctuation theory, correcting equilibrium thermodynamics, was soon added. Ludwig Boltzmann was the one to introduce the most famous kinetic equation, proving a so-called H-theorem.  More foundational even were Boltzmann's microscopic derivations of the Second Law, and the thoughts for deriving macroscopic irreversibility from microscopic laws that are time-reversible, \cite{Boltzmann_2012cam}.\\
In more modern times, there is the work on kinetic theory and diffusive scaling limits from Newtonian particle systems \cite{Spohn1991}, and the derivation by Oscar Lanford, in the Boltzmann-Grad limit, of the Boltzmann equation from particle dynamics, \cite{lan,lan2}.\\

A second major breakthrough started with the 1931 work of Lars Onsager, \cite{onsager19311,onsager19312}, in particular for establishing reciprocity as a consequence of microscopic reversibility. He also pioneered linear response theory. The main idea, continued by Onsager and Machlup  \cite{onsagermachlup,Onsager1953}, was to connect response and relaxation with dynamical fluctuations.  That was taken up later by John Gamble Kirkwood (1946), \cite{kirkwood1946}.  Kirkwood formalized the idea that macroscopic transport properties arise from microscopic fluctuations.\\
With a similar goal, but via other methods, the latter was finally formulated around equilibrium in the 1950-60s by various groups including Welton, Green, Kubo and others.  It concerned the establishment of linear response theory and the fluctuation-dissipation relations close to equilibrium; see {\it e.g.} \cite{Kubo1957,Kubo1966,Callen1951}.\\ In more recent years, nonequilibrium response theory has been formulated in various ways, \cite{upd,resp}, with important operational relations such as summarized in the Harada-Sasa equality that connects energy dissipation in a system to violations of the fluctuation--dissipation theorem, \cite{HaradaSasa2005,Harada2006}. \\

In his landmark 1940 paper \cite{Kramers1940}, Kramers provided a quantitative framework for escape rates over energy barriers due to thermal noise. Freidlin-Wentzel theory (1970), \cite{FW,fen} develops the theory of random perturbations around deterministic dynamics.  It provides a rigorous framework for studying the behavior of systems experiencing weak noise and rare transitions.  It is foundational for deriving reaction rate or transition rate theory.\\

Around the middle of the 20th century was also the development, mostly by the Brussels school around Prigogine, of irreversible thermodynamics, \cite{dGM}.
It is a theory of nonequilibrium systems which presents the basic phenomenology of many irreversible processes.  It considers spatially extended systems consisting of several subsystems, each of them in thermal equilibrium but where ``global'' thermodynamic equilibrium is broken. Within that framework, the ``total entropy`` is defined as the sum of equilibrium entropies over all subsystems, including the environment (usually represented in terms of boundary conditions) in case the system is still not isolated. There can even be a continuum of subsystems in which case we speak about ``local'' equilibrium,  with local temperatures, pressures or chemical potentials, and we introduce the entropy ``density''. 
Due to gradients of thermodynamic parameters (like temperature or chemical potential) called ``thermodynamic forces`` $F_\alpha$ there exist flows between the subsystems (like those of energy or mass) called ``thermodynamic fluxes`` $J_\alpha$. Forces and fluxes are naturally paired with each other so that the ``entropy production`` measuring the increase rate of total entropy has the form
\[
  \sigma = \sum_\alpha J_\alpha F_\alpha
\]
The Second Law of thermodynamics requires $\sigma \geq 0$ in the steady state.
Linear irreversible thermodynamics assumes
\[
  J_\alpha = \sum_\gamma L_{\alpha\gamma} F_\gamma
\]
with ``linear response coefficients'' $L_{\alpha\gamma}$ satisfying the ``Onsager-Casimir reciprocity relations'' $L_{\alpha\gamma} = L^*_{\gamma\alpha}$. The star indicates time-reversal of the microscopic dynamics, {\it e.g.}, it inverts the sign of velocities and magnetic fields. In ``even'' systems $L_{\alpha\gamma} = L^*_{\alpha\gamma}$ and then the matrix $[L_{\alpha\gamma}]$ is ``symmetric'' and ``positive-definite''.  A more recent summary (with many other insightful discussions) is found in the book by Pierre Gaspard, \cite{gas2}.\\

Of course, we know about the understanding of Brownian motion by Albert Einstein in one of his 1905 papers.  Much has been added to that, in various styles and extensions.  Skipping more pioneering work by Langevin and Smoluchowski, we refer to the work of Robert Zwanzig for projection-operator methods, using coarse-graining from classical Hamiltonian mechanics. In precise terms, Brownian motion was shown to emerge from Hamiltonian systems using quantum statistical mechanics in 
\cite{DeRoeck2010quantum}.\\

Joel Lebowitz’s work from the 1950s–70s established rigorous and physically insightful connections between microscopic reversible dynamics and macroscopic irreversible behavior, laying the groundwork for modern nonequilibrium statistical mechanics. Already starting in the 1950's, Lebowitz formalized ensembles for nonequilibrium situations, including systems in contact with multiple reservoirs, ({\it e.g.}, heat baths at different temperatures); \cite{leb55,LebowitzFrisch1957Knudsen,Lebowitz1978ExactResults,Lebowitz1956Thesis}.  He emphasized the importance of stationary nonequilibrium states as key objects of study.  It was then understood in various works that the condition of local detailed balance gives a good recipe for constructing physically motivated nonequilibrium models; see \cite{kls,leb55,leb57,xing2025,derrida,hal,chi,time,ldb}.  They were used for studies of nonequilibrium phase transitions ({\it e.g.}, for driven lattice gases \cite{KatzLebowitzSpohn1984}) and for a program on deriving hydrodynamic equations \cite{presut2009,KL,sasa2014}. An important reference book here is the one by Herbert Spohn on large scale dynamics of interacting particle systems, \cite{Spohn1991}.  Over the years, more exact solutions have been found, and macroscopic dynamical fluctuation theory for diffusive particle systems got a powerful and interesting framework; \cite{derrida,Derrida1998,M,F,T,sdiv}. Dynamical fluctuation theory for jump processes is included in \cite{maes2007entropy}.\\

From around 1960, various models started to appear to get some understanding of far-from-equilibrium systems. Computational methods started to enter, including molecular dynamics simulations (1967 by Alder and Wainwright), thermostating models (by Evans and Morriss, 1984), and stochastic chemical kinetics (by Gillespie in the 1970s).  Probability theory got much stimulation from the emerging theory of stochastic particle dynamics, \cite{Liggett1985,Spitzer1970}.\\
Toward the end of the 20th century, it became clear that fluctuation theory for steady nonequilibrium systems requires a theory on the level of trajectories, \cite{gibbsian}.  Various quantities such as heat, work and entropy fluxes from the theory of irreversible thermodynamics got translated in variables depending on the trajectory, \cite{ken,vulpi,seif,Shiraishi2023}. It became known as ``stochastic thermodynamics,'' already in the title of \cite{VandenBroeck1986StochasticThermodynamics} from 1986.  Some importance was given to fluctuation theorems for the entropy fluxes, and they were proven in various contexts and for different types of situations, starting with the Gallavotti-Cohen type fluctuation relations and the free energy equalities of Chris Jarzynski, \cite{GC,gc2,Jarzynski, crooks,time,2000}.\\
Re-discovering the importance of nondissipative aspects is a more recent tendency, \cite{nondiss,frenesy}. Kinetic time-symmetric aspects complement in many ways the considerations from stochastic thermodynamics.  Dynamical activity plays a crucial role in stabilization and self-assembly, steering and control of nonequilibrium systems. We also mention the work on ratchets, {\it e.g.}, in \cite{AstumianBier1994,Astumian1997BrownianMotor,Parrondo1998ReversibleRatchets,MeursVanDenBroeckGarcia2004} stimulating research on current generation and selection. Active matter is a more recent addition to nonequilibrium stat mech with exciting insights for biology and new materials. The basic models were already introduced in the 1990's (or earlier) though, {\it e.g.}, \cite{Vicsek1995,TonerTu1995,Schweitzer1998,SimhaRamaswamy2002}. A seminal review article is \cite{Marchetti2013}.\\

In the mean time, important work continues on steady-state thermodynamics.  Pioneering work includes \cite{oono,sasa_tasaki2006,hatano_sasa2001} with more recent results in \cite{sasa_nakagawa2025,nakagawa_sasa2019,nakagawa2024_arxiv}.  Here, questions of stabilization and phase transformations get formulated (and answered) for out-of-equilibrium spatially extended systems.  Perhaps, progress and studies of nonequilibrium calorimetry should also be included in that project; {\it cf}.\cite{epl,cejp,jir,calo,negcap,fiori2024specificheatdrivencurieweiss,pritha,jchemphys}.\\

The list above is far from complete.  
 Even then, many interesting and important performances have not been cited, and we must refer to more specialized books and reviews as for instance mentioned in the Preface of the present notes. 

\section{Nonequilibrium (open) problems}

\begin{quote}
{\it Ah, but a person's reach should exceed their grasp,
Or what's a heaven for?}\\
Free after Robert Browning’s poem ``Andrea del Sarto'' (1855).
\end{quote}

As a way of challenging the reader, we mention several types of problem related to the (greater) understanding of nonequilibrium processes.  Of course, posing difficult or endless questions requires little insight.  Take it then as an encouragement to at least keep some of the holy grails alive and present. Here is a selection.

\begin{enumerate}
\item
\underline{Phase transitions}:  in what sense do equilibrium phase transitions survive nonequilibrium driving?  How are equilibrium phase diagrams affected and what are the main qualitative changes?  Are there universality classes; are there new ones? When and how do critical exponents change?  Do there appear new (types of) phase transitions as the driving gets larger? \\ 

Needless to say that we are very far from having a nonequilibrium equivalent or even an analogue to Gibbs' monumental work ``On the Equilibrium of Heterogeneous Substances'' (1876-78).
There, Gibbs introduced various key-concepts from which we understand the equilibrium state of various substances including issues of solubility and phase change.  Among many other things, we find there the Gibbs phase rule, which predicts the number of independent degrees of freedom $F$ that can be changed without altering the number of phases $P$ in equilibrium, $F = C - P + 2$, where $C$ is the number of components.
This rule is fundamental for analyzing and predicting the behavior of heterogeneous systems, but we have no clue what to make of that for steady nonequilibrium systems.  Can it be that what was unstable (or metastable) for equilibrium, becomes stable under nonequilibrium conditions?\\
There does not appear to exist even a powerful Landau theory.  What is the power of mean-field treatments under driving, \cite{beyen2024phase,fiori2024specificheatdrivencurieweiss}?  How to (even) define universality?  \\

We also miss a nonequilibrium theory of nucleation and metastability.  That is meant to be in or around steady nonequilibria.  We can imagine for example that a driven or active system resides for a long time in a passive zero-current condition, and then, by a fluctuation, relaxes to a stable nonequilibrium which is active and/or current carrying.  Such resurrections are seen in nature but hardly in any theory of nature. Are instantons, are nucleation and crystal formation subject to different {\it nonequilibrium} laws and reaction rates compared with the situation under detailed balance?

Recent strides in those directions have been made in various directions.  There are for example the revolutions with reference to the Kardar-Parisi-Zhang equation, \cite{kardar1986dynamic,takeuchi2018appetizer}.  It offers a specific study of nonequilibrium growth and interface height fluctuations.  Interesting connections have been made with stochastic Burgers equation and directed polymers, with random matrix theory and nonGaussian rescaling and limit theorems, and encompassing a rich phenomenology .  At the same time, it has opened a new discussion on the meaning of unversaility classes in nonequilibrium physics, {\it e.g.}, how they are related to the same critical exponents (which ones) and/or limiting distributions.     In that respect, we may add that there has been no or little consideration of nonequilibrium models within the framework of conformal field theory.  It appears that nonequilibrium systems are very often without spatial correlation length, over much or all of the phase diagram and {\it on top of that}, phase transitions and critical behavior can show up.

\item \underline{New materials and devices}: There are serious gaps in our understanding of nonequilibrium assembly and of the nonequilibrium synthesis of molecules.  
Can one control response in terms of conductivity, elasticity, chirality and viscosity via the kinetics of nonequilibrium driving?  As a special case, how to mimic biological materials, working multicellularity, macroscopic synchronization, and how to organize analogues of biological functioning such as memory and intelligence?  How to manufacture self-reproduction and selction mechanisms?  Is the signal--to--noise ratio typically getting smaller when far from equilibrium?  That is important in the construction of biosensing devices.

\item
\underline{Turbulence}: Turbulence refers to a macroscopic nonlinear regime with possibly strong collective and dissipative effects, \cite{kolmogorov1941local,kolmogorov1941dissipation,frisch1995turbulence}.  What is the status of Kolmogorov's theory (1941) describing an ``energy cascade'' where large eddies break into smaller ones, transferring energy until viscosity dissipates it, characterized by a universal \(-5/3\) power law for energy distribution in the inertial range, sometimes called  ``the last great unsolved problem of classical physics?''    We believe to see reproducible effects and related equations \cite{zaki2025turbulence}, but so far no convincing and unifying underlying theory\footnote{An interesting toy-scenario is given by Burgers turbulence, \cite{bec2007burgulence}, failing however to reproduce the phenomena of real turbulence, where pressure induces long-range effects in reciprocal space.}.\\

The question is then to get a unified vision on turbulence that is compatible and guides experimental observation, \cite{bec2024statistical}.  Key-words here are the so-called dissipative anomaly (meaning the breakdown of Eulerian time-reversibility\footnote{Possibly caused by the developments of ever-increasing gradients in the kinetic energy profile as the viscosity gets smaller.}), the emergence of spontaneous stochasticity (with stochastic Lagrange trajectories and the phenomenon of Richardson dispersion \cite{richardson1926atmospheric,salazar2009two}), and spatial intermittency which is the breakdown of single scale invariance over distances related to the low-viscosity scales.

\item \underline{Glass transition} Another example is the physics of amorphous and disordered materials, specifically concerning the glass transition.  There a huge change in viscosity is witnessed over a small temperature change, with no or little foundation.  Is the glass transition in the end a (nonequilibrium) transition in dynamical activity?\\
The problem touches deeply on a more general question of how noise on more microscopic levels can be enlarged to macroscopic scales.  In other words, is the origin of macroscopic chaos to be found in the emergent macroscopic dynamics, or how is all that the natural result of an amplification (with corresponding synchronization) of more microscopic fluctuations.

\item 
\underline{Stat mech of gravity and geometric degrees of freedom:} With lots of interest into the problem of quantum gravity, one easily overlooks the fact that a satisfactory fluctuation theory for gravity is still lacking, even in the classical regime. Already Newtonian gravity presents unusual and hard problems for the Gibbs formalism, and it is not clear how to deal with its Second Law behavior, clearly relevant for the history of the universe and for galaxy formation.  With its (very) long-range attraction and without screening, it is quite different from the usual situation in statistical mechanics.\\
The source of all nonequilibrium in our universe is probably to be found in the low-entropy condition of gravitational / geometric degrees of freedom in what is called the very early universe\footnote{The meaning is not entirely clear, and it is useful to make a  difference between the big bang model (empirically established for the history of our universe over the last 13.8 billion years) and {\it some} big bang (containing heavily popularized ideas about {\it some} beginning of the universe but without a shred of evidence --- somewhat similar to theories of inflation).  There is no support for the scientific hypothesis of Lemaître that there was a physical beginning of space, time, and matter.}.  But we do not really know what we mean by low-entropy geometry, nor how low entropy gets transferred to (or influences) matter degrees of freedom, {\it e.g.} in the problem of baryogenesis.  Perhaps there are even low-frequency corrections (to the Planck law) in the cosmic microwave background that find their origin in gravitational / geometric effects at the time of recombination.\\
And what is the relation between (spacetime) curvature and nonequilibrium effects? Entropy and temperature appear in black hole thermodynamics but it remains unclear how they connect with Clausius heat theorem and with Boltzmann's statistical picture.

\item \underline{Derivation of induced forces or interactions} on and between probes in a fast nonequilibrium medium.  In particular, when do these forces show non-reciprocity or higher-order interactions?  How does friction and noise combine?  Is memory different from activity?\\  Is there at least a heuristic theoretical derivation of physically correct nonequilibrium dynamics, describing the return to steady nonequilibrium, going beyond merely mimicking or even reproducing some of the experimental findings?  {\it E.g.}, do thermostated dynamics, often used in computational approaches similar to a molecular dynamics simulation using a thermostat algorithm to control and maintain the system's temperature or (kinetic) energy at a constant value, have a microscopic derivation? What is the role of chaos? \\
Finally, it is experimentally well-established that nonequilibrium long-range
correlations of concentration fluctuations appear in diffusions of a solute
in a solvent or of colloids.  It remains unknown how such correlations are established
dynamically. 
\end{enumerate}

\chapter{Nonequilibrium model examples}
\lhead{C. Maes}
\rhead{Nonequilibrium examples}

To get started 'in theory,' we select (mathematical) examples of nonequilibrium models. 

\section{Modified Sutherland-Einstein relation}\label{sue}
\lhead{C. Maes}
\rhead{Modified fluctuation-dissipation relation}
How is the relation between diffusion and mobility coefficients changed under nonequilibrium conditions?
In the paper \cite{gle2} we consider a colloid of mass $m$ in $\mathbb{R}^3$, with position $\vec{r}_t$ following the Langevin dynamics,
\begin{eqnarray}
\dot{\vec{r}}_t = \frac{\vec{r}_{t+dt} -\vec{r}_t}{dt} &=& \vec{v}_t \label{general}\\
 m\dot{\vec{v}}_t  = m\frac{\vec{v}_{t+dt} -\vec{v}_t}{dt} &=& \vec{F}(\vec{r}_t,\vec{v}_t)-\gamma m \vec{v}_t + \sqrt{2m\gamma T}\,\vec{\xi}_t \nonumber
\end{eqnarray}
We write $\vec{v}_t$ for the velocity of the colloid. It is passive and subject to the influence of a heat bath in thermal equilibrium at temperature $T$.  The bath is represented by the vector $\vec{\xi}_t$ of standard Gaussian white noises, each of the components having a Gaussian distribution with mean zero $\left<\xi_{t,i}\right> = 0$ and with covariance $\left<\xi_{t,i}\xi_{s,j}\right> = \delta_{i,j}\delta(t-s)$.  We set Boltzmann's constant $k_B=1$. Note that the damping $\gamma$ defines a time-scale and appears in the noise amplitude (part of the so-called Einstein relation).
For external force we take $\vec{F}$, depending periodically on the position $\vec{r}$ so that there is no confinement.

The corresponding Fokker-Planck equation reads
\begin{eqnarray}\label{fokplagen}
 \frac{\partial \mu_t}{\partial t} = -\vec{v}\cdot\vec{\nabla}_r\mu_t-\vec{\nabla}_v\cdot \Big[\Big(\frac{\vec{F}-m\gamma  \vec{v}}{m}\Big)\mu_t - \frac{\gamma T}{m}\vec{\nabla}_v \mu_t\Big]
\end{eqnarray}
for probability density $\mu_t$ at time $t$.
We introduce three important quantities that help to characterize the behavior of such a system.
First, there is the (symmetric) diffusion (matrix) $D(t)$,  defined as
\[
 D_{ij}(t) = \frac{1}{2t}\Big<(\vec{r}_t-\vec{r}_0)_i;(\vec{r}_t-\vec{r}_0)_j\Big>
  \]
where the subscripts $i,j$ denote the components, and the right-hand side is a covariance: for observables $A$ and $B$
\begin{equation}\label{trunc}
 \Big<A;B\Big>=\Big<AB\Big> -\Big<A\Big>\Big< B\Big>
 \end{equation}
 depending on initial conditions, {\it e.g.}, given initial position and velocity of the colloid.
For diffusive systems such as described above, this diffusion function
is expected to have a large time limit, called the diffusion matrix
\[ D_{ij} = \lim_{t\to\infty}D_{ij}(t) \]
meaning that the (co)variance of the displacement of the particle is linear for large times $t\gg 1/\gamma$,
with slope given by the diffusion constant.\\
Secondly, the mobility matrix  $M(t)$ is defined by first adding to (\ref{general}) a constant (but small) force $\vec{f}$, replacing $\vec{F}(\vec{r},\vec{v})
\rightarrow \vec{F}(\vec{r},\vec{v}) + \vec{f}$. Then, the mobility
measures the change in the expected displacement of the colloid:
\[ M_{ij}(t) = \frac{1}{t}\left.\frac{\partial}{\partial f_j}\Big<({\vec r}_t-{\vec r}_0)_i\Big>^{f}\right|_{\vec{f}=0} \]
where the superscript $f$ indicates that the average is to be taken in the dynamics with the extra force $\vec{f}$.
Also that mobility is supposed to have a large-time limit, which makes the mobility matrix
\[ M_{ij} = \lim_{t\to\infty}M_{ij}(t) \]
equal to the linear change in the stationary velocity by the addition of a
small constant force.\\
In the special case of a detailed balance dynamics, we have the Sutherland--Einstein relation,
\begin{equation}\label{einstein} M_{ij} = \frac{1}{T}D_{ij} \end{equation}
which is an instance of the more general fluctuation-dissipation theorem.  It is subject of the discussion on response in Section \ref{pif}; see also \eqref{eq2:dQSD}.\\  However, when rotational forces are present, we need a third player, which is the force-velocity time-correlation to correct the (equilibrium) Sutherland-Einstein relation,
in the form
\begin{eqnarray}
 && M_{ij} = \frac{1}{T}D_{ij} - \lim_{t\to\infty}\frac{1}{2\gamma mT }\,\int_0^t ds\,\Big<\frac{({\vec r}_t - {\vec r}_0)_i}{t};F_j(\vec{r}_s,\vec{v}_s)\Big>\label{genresult5}
\end{eqnarray}
As we see, the correction to the equilibrium mobility--diffusion relation is measured by a spacetime correlation between applied forcing and displacement, \cite{gle2}. One can show that the deviations with respect to the Sutherland-Einstein relation are second order in the nonequilibrium amplitude.\\

The symmetrized mobility matrix is
\begin{equation}\label{symr}
\frac{M_{ij}+M_{ji}}{2} = \frac{D_{ij}}{2T} + \frac{\delta_{i,j}}{2\gamma m} - \lim_{t\to \infty}\frac{1}{4\gamma^2 m^2T t}\Big< {\vec \Phi}_i;{\vec \Phi}_j \Big>
\end{equation}
where each term is symmetric.  The vector $\vec{\Phi}$ is explicitly given by the rotational impulse
\begin{equation}
\vec{\Phi} =  \int_0^t ds\,\vec{F}(\vec{r}_s,\vec{v}_s)\label{phi}
\end{equation}
Under conservative forces (no driving), the sum of the second and the third term in \eqref{symr} give the first term on the right-hand side.  Otherwise, in diffusive nonequilibrium the correction term to the Sutherland--Einstein relation is nonzero with the symmetrized mobility matrix as the explicit difference between a diffusion--related matrix and the force-force covariance matrix.
These results remain essentially unchanged when interactions with other particles are included.  They add, in the sense of Taylor-Aris dispersion and shear-enhanced diffusion, \cite{aris1956dispersion,taylor1953dispersion}, that the nonequilibrium diffusion is in general larger than the mobility and the equilibrium diffusion under (mean zero) nonequilibrium driving.

\section{Agitated molecules}\label{afmp}
\lhead{C. Maes}
\rhead{Agitated molecules}
A traditional way of modeling a small system like a molecule subject to a heat bath or a radiation bath, is to consider energy levels and corresponding transition rates such that the population statistics of the energy levels is given by the Boltzmann distribution.  We can modify that population statistics by pumping or by agitating the molecule. Let us discuss some examples; see also \cite{jchemphys}.

\begin{example}[Molecular switch]\label{ex1}
    Consider an agitated molecular system, where the hierarchy of energy levels randomly switches. Such a molecular switch can be modeled as a Markov jump process on a ladder with (to be specific) three levels; see \fig \ref{barrier}. Each leg $\sigma=\pm $ has three levels $\eta = 1,2,3$, so that the states are of the form $x=(\eta,\sigma)$. Each state located in $\sigma=-$ has an energy $E(\eta,-) =(\eta-1)\ve $  and the states on $\sigma=+$ have an energy  $ E(\eta, +)=(3-\eta)\ve $,  The process is switching legs at rate $\alpha$. The transition rates are chosen as
    \begin{align}\label{rate1}
  & k((1,-),(2,-))=k((2,-),(3,-))=k((3,+),(2,+))=k((2,+),(1,+))= e^{-\beta(\Delta + \ve/2)}\notag \\
  & k((1,+),(2,+))=k((2,+),(3,+))=k((3,-),(2,-))=k((2,-),(1,-))= e^{-\beta(\Delta - \ve/2)} \notag\\
 & k((\eta,-),(\eta,+))=k((\eta,+),(\eta,-))=\alpha 
\end{align}
where $\Delta>0$ is an energy barrier, and  $k((\eta ,\sigma),(\eta',\sigma))$ denotes the transition rate from state $(\eta ,\sigma)$ to state $(\eta',\sigma)$.\\ 

\begin{figure}[H]
\centering
\includegraphics[scale=0.65]{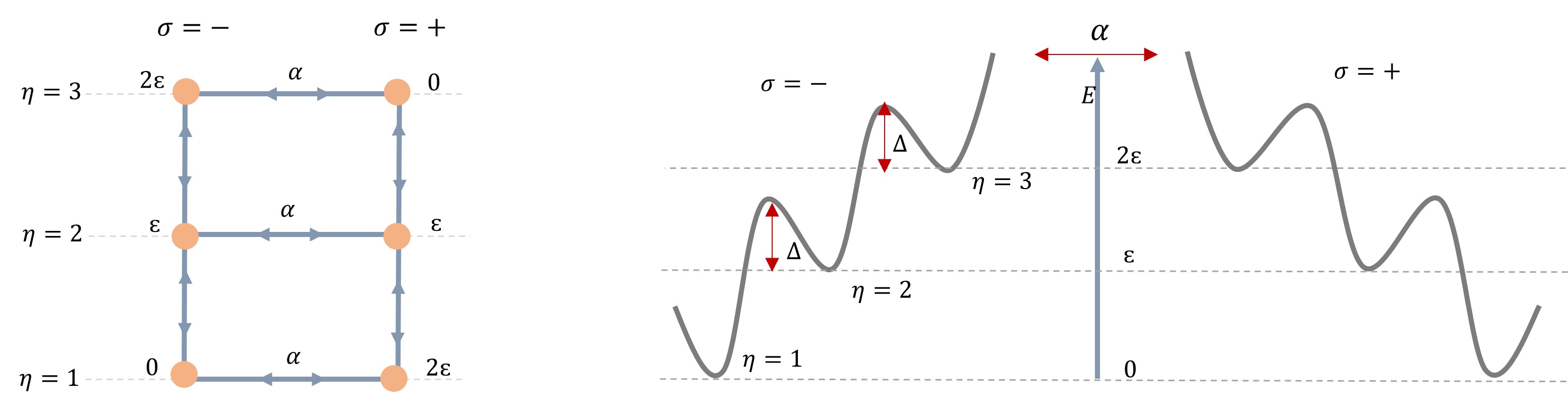}    
\caption{\small{Left:  three-level ladder.  Right: the energy landscape, where $\Delta$ denotes the height of the barrier. A molecular switch can indeed be reproduced experimentally with a colloid moving in an optically simulated and flashing landscape.}}   \label{barrier}
\end{figure}
The heat flowing to the bath at inverse temperature $\beta$ equals $\ve$ at each transition $\eta\rightarrow \eta'$ where the level is changing. The changing of legs is  work done by external sources.\\
Solving the stationary Master Equation for the rates in \eqref{rate1}, the stationary distribution $\rho$ is 
\begin{align*}
\rho(1,-)=\rho(3,+)&=\frac{1}{Z}e^{\frac{\beta  \varepsilon }{2}}  \left(\alpha  e^{\beta  (\Delta +\varepsilon )}+\alpha  e^{\beta  \Delta }+e^{\frac{3 \beta  \varepsilon }{2}}\right)\geq\\
 \rho(2,-)=\rho(2,+)&=\frac{1}{Z}\left(\alpha  e^{\beta  \left(\Delta +\frac{3 \varepsilon }{2}\right)}+\alpha  e^{\beta  \Delta +\frac{\beta  \varepsilon }{2}}+e^{\beta  \varepsilon }\right) \geq\\ 
\rho(3,-)=\rho(1,+)&=\frac{1}{Z}\left(2 \alpha  e^{\beta  (\Delta +\varepsilon )} \cosh \left(\frac{\beta  \varepsilon }{2}\right)+1\right),
\end{align*}
where $Z=2 e^{\frac{\beta  \varepsilon }{2}} \left(e^{\beta  \varepsilon }+1\right) \left(3 \alpha  e^{\beta  \Delta }+e^{\frac{\beta  \varepsilon }{2}}\right)+2.$\\
The stationary distribution of the states on the leg $\sigma=-$ is plotted in \fig \ref{rhoplots} for different values of $\alpha, \Delta$ and temperature. States $(1,-)$ and $(3,+)$ are dominant at low temperatures for $\alpha=0$. For $\alpha\neq 0$, an additional time-scale remains and toward zero temperature for values of $\Delta >\ve/2$ all the transition rates become equal (approaching zero); consequently, all states become equivalent.
\begin{figure}
     \centering
      \begin{subfigure}{0.49\textwidth}
         \centering
         \def\svgwidth{0.8\linewidth}        
        \includegraphics[scale = 0.85]{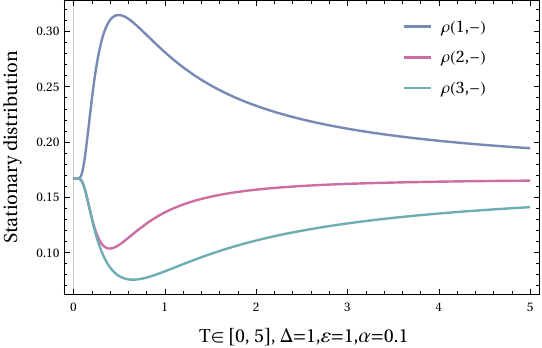}
     \end{subfigure}
     \hfill
     \begin{subfigure}{0.49\textwidth}
         \centering
         \def\svgwidth{0.8\linewidth}        
   \includegraphics[scale = 0.85]{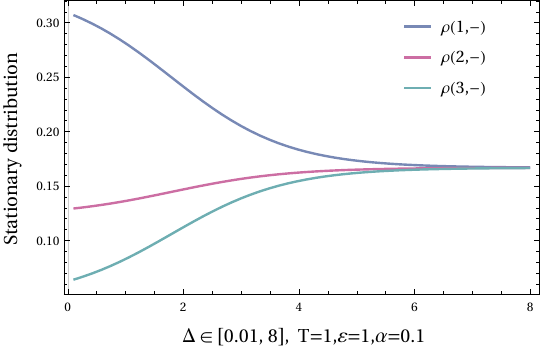}
     \end{subfigure}
      \centering
      \begin{subfigure}{0.49\textwidth}
         \centering
         \def\svgwidth{0.8\linewidth}        
        \includegraphics[scale = 0.85]{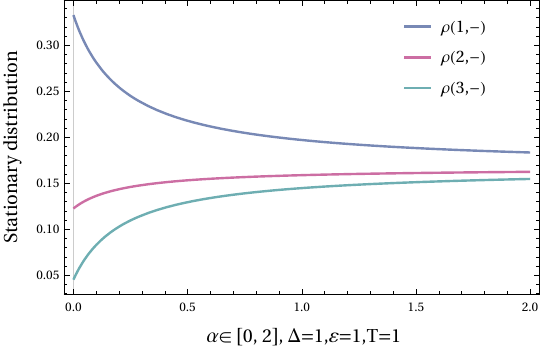}
     \end{subfigure}
\caption{\small{Stationary distributions of the three-level ladder given in  Example \ref{ex1} for different values of $\alpha, \Delta $ and temperature $T$. }}   \label{rhoplots}
\end{figure}

We can follow the kinetics that decides dominant states from nonequilibrium calorimetry; see \cite{calo,negcap,fiori2024specificheatdrivencurieweiss,pritha} for examples.  The heat capacity is plotted in \fig\ref{hc3l}. Taking $\alpha=0$  corresponds to the equilibrium case.  For  $\alpha >0 $, the nonequilibrium heat capacity depends kinetically on the barrier $\Delta$ and may become negative (here, at low temperatures for large enough $\Delta$).
\begin{figure}[H]
     \centering
      \begin{subfigure}{0.49\textwidth}
         \centering
         \def\svgwidth{0.8\linewidth}        
        \includegraphics[scale = 0.85]{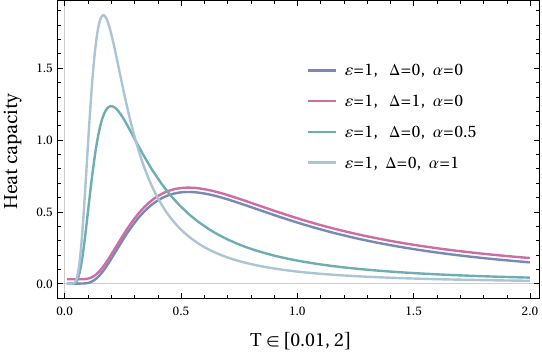}
     \end{subfigure}
     \hfill
     \begin{subfigure}{0.49\textwidth}
         \centering
         \def\svgwidth{0.8\linewidth}        
   \includegraphics[scale = 0.85]{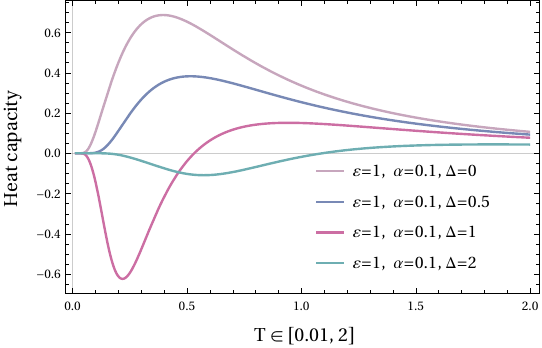}
     \end{subfigure}
\caption{\small{Heat capacity of a $3$-level ladder  as a function of temperature $T$ for different values of  $\alpha, \Delta $ and $\ve$ as defined in the transition rates \eqref{rate1}. We observe an inverted Schottky anomaly for large enough $\Delta$ when $\ve=1$.}}   \label{hc3l}
\end{figure}

\end{example}

\begin{example}[Ratchet]
One can think of this model as a version of what are called ratchets, including models with flashing potentials.  For instance
 a flashing harmonic potential for an overdamped evolution of a position variable in one dimension could look like 
\begin{equation}\label{fh}
	\dot{x}(t) = -U'(x(t),\sigma(t)) + \sqrt{2 T}\, \xi(t)\,,\qquad
	U(x,\sigma) = \frac{\nu}{2}(1 + \lambda\,\sigma) \,x^2
\end{equation}
where $\sigma(t)=\pm 1$ flips at rate $\alpha$, and $\nu$ is a frequency.
The parameter $\lambda\in [0,1]$ couples the position to the dichotomous noise. In the stationary distribution (which is not Gaussian for $\lambda\neq 0)$,
\begin{align}
	\langle x^2\rangle &= \frac{1 + z}{1 + z(1- \lambda^2)} \frac{T}{\nu}\label{va}
\\
    \langle \sigma\, x^2\rangle &= -\frac{z\lambda}{1 + z(1 - \lambda^2)} \frac{T}{\nu}\label{van}
\end{align}
in terms of the dimensionless persistence $z=\nu/\alpha$.  There is heat produced, and if we would make the potential $U$ to flash between a skewed landscape (like a washboard) and a flat landscape, we would also get particle motion.
\end{example}

\begin{example}[Parrondo game]
A more discrete example of the emergence of a particle current, mentioned above, goes under the name of Parrondo game (1996), \cite{Parrondo1996CheatBadMathematician,ParrondoDinis2007RatchetsParadoxicalGames}:\\
The state space is $K= \{1,2,3\}$ and the state at time $n$ is $x_n$.  The Markov chain uses a different rule ($A$ or $B$) at even and at odd times $n$.  Alternating, the following two games are played. Game $A$ is fair coin tossing: we simply move $x\rightarrow x\pm 1 \mod 3$ with equal probability at even times. Game
$B$ is played at odd times and with two biased coins, a good one and a bad one. In
game $B$, the good coin is tossed when $x_n \in \{1,2\}$ and the bad coin is used each time when $x_n=3$. Winning takes $x_{n+1} = x_n + 1$; losing at time $n$ means
$x_{n+1} = x_n - 1$, always modulo 3. The transition
probabilities are then
\begin{eqnarray}
\mbox{Prob}[x_{n+1}=x\pm 1|x_n=x] &=& 1/2,\quad \mbox{ when } n \mbox{ is even}
\nonumber\\
 \mbox{Prob}[x_{n+1} = x + 1|x_n=x] &=& 3/4,\quad \mbox{ when
} n \mbox{ is odd and } x \neq 3\nonumber\\
 \mbox{Prob}[x_{n+1} = x + 1|x_n=x] &=& 1/10,\quad \mbox{ when
} n \mbox{ is odd and } x = 3
\end{eqnarray}
Let us check detailed balance when we would only play game $B$ (at all times):\\
Consider the cycle $3 \rightarrow 1 \rightarrow 2
\rightarrow 3$ .  Its stationary probability (always for game $B$ alone) is Prob$[3
\rightarrow 1 \rightarrow 2 \rightarrow 3] = \rho(3)\times 1/10
\times 3/4 \times 3/4 = 9\rho(3)/160$.  For the reversed
cycle, the probability Prob$[3 \rightarrow 2 \rightarrow 1
\rightarrow 3] = \rho(3)\times 9/10 \times 1/4 \times 1/4=
9\rho(3)/160$ is the same.  The equilibrium distribution for game $B$ is then found to be
$\rho(1) = 2/13, \rho(2)=6/13$ and $\rho(3) = 5/13$. Obviously then,
there is no current  when playing game $B$ and clearly,
the same is trivially verified for game $A$ when tossing with the
fair coin.  Yet, and here is the paradox, when playing periodically game $B$ after game $A$, a current
arises.
\end{example}

\begin{example}[Rower model]
The ``rower model'' is  a two-state oscillator that switches direction upon reaching certain thresholds, used to explore features of ciliary motion, \cite{kotar2010hydrodynamic,gupta2025}.
We consider here a discretized version of a single rower (with $n$ energy levels).\\

Each rower has $2n$ sites $(i,\sigma)$ for $i=0,\ldots,n-1$ and $\sigma=\pm 1$, with associated energies $E(i,\sigma)$ taking values that are multiples of the gap energy $\nu>0$,
\begin{equation}\label{rowE}
 E(i,+1) = i\,\nu, \qquad E(i,-1) = (n-1-i)\nu   
\end{equation}
In Fig.~\ref{row} we connect these sites  for $n=4$ where each edge is a possible directed transition. The transitions $k((i,\sigma);(i\pm 1,\sigma))$ are defined as follows. Along the $\sigma = +1$ branch (left in Fig.~\ref{row}), the upward transitions occur with rate $\tau^{-1} e^{-\beta \nu}$, while the downward transitions have rate $\tau^{-1}$. Along the $\sigma = -1$ branch (to the right in Fig.~\ref{row}), the upward transitions occur with rate $\tau^{-1}$ and the downward transitions with rate $\tau^{-1} e^{-\beta \nu}$.   Here, $\beta$ is the inverse temperature of the heat bath to which the rower is weakly coupled.  Horizontal transitions where an energy $(n-1)\nu$ is transferred to the rower, flipping $\sigma$, are allowed only at the ends of the branches: $k((0,+);(0,-)) = \alpha$ and $k((n-1,-);(n-1,+)) = \alpha$ for rate $\alpha>0$. The opposite transitions have zero rates.
\begin{figure}[H]
    \centering
    \includegraphics[width=0.5\linewidth]{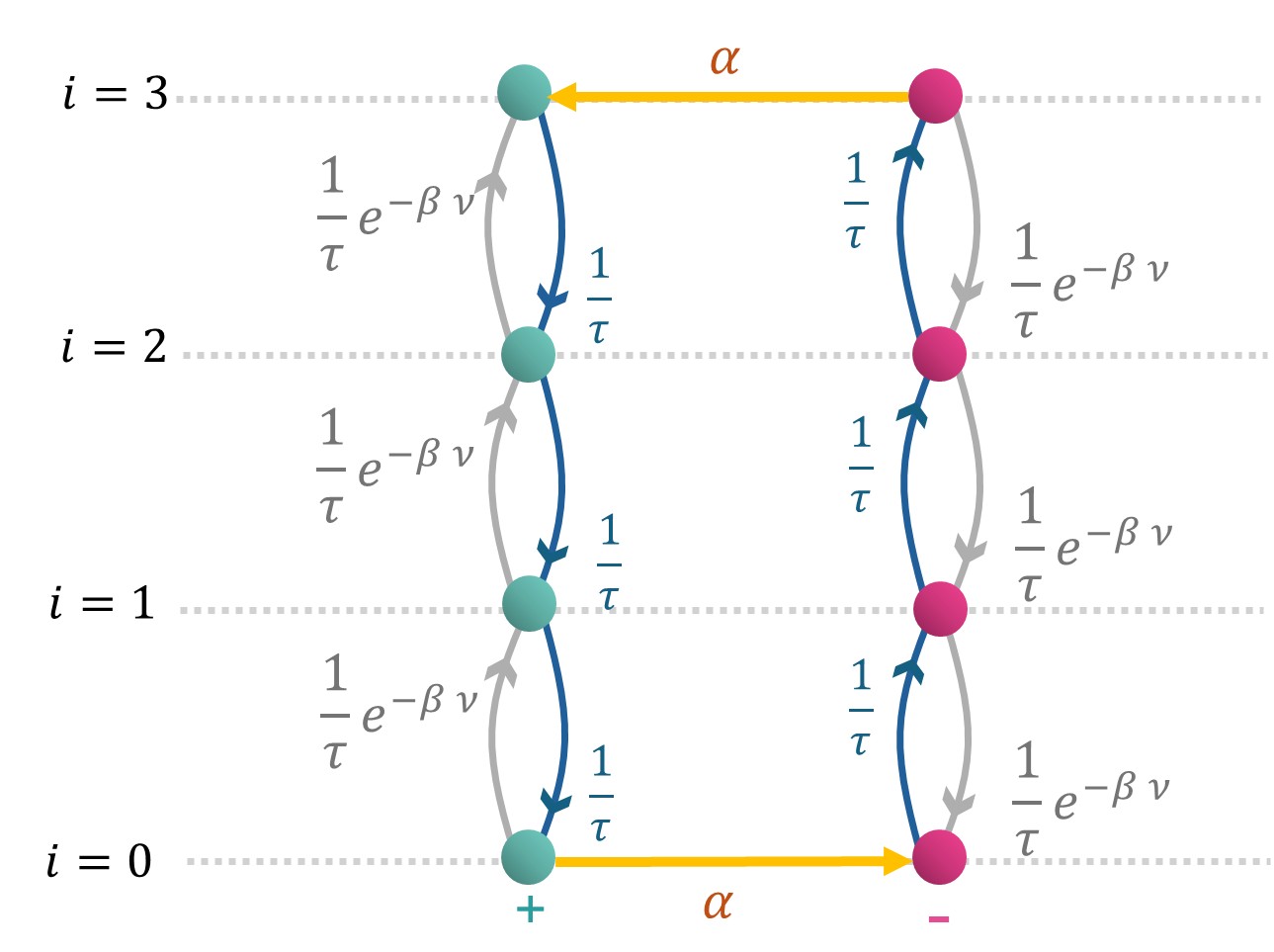}
    \caption{\small{Graph and transition rates for a rower with $n=4$.}}
    \label{row}
\end{figure}
Fig.~\ref{rotr} shows a typical trajectory for different values of the parameters. At the jumps $(0,+) \rightarrow (0,-)$ and $(n-1,-) \rightarrow (n-1,+)$, energy is pumped into the system.

\begin{figure}[H]
    \centering
    \includegraphics[width=0.5\linewidth]{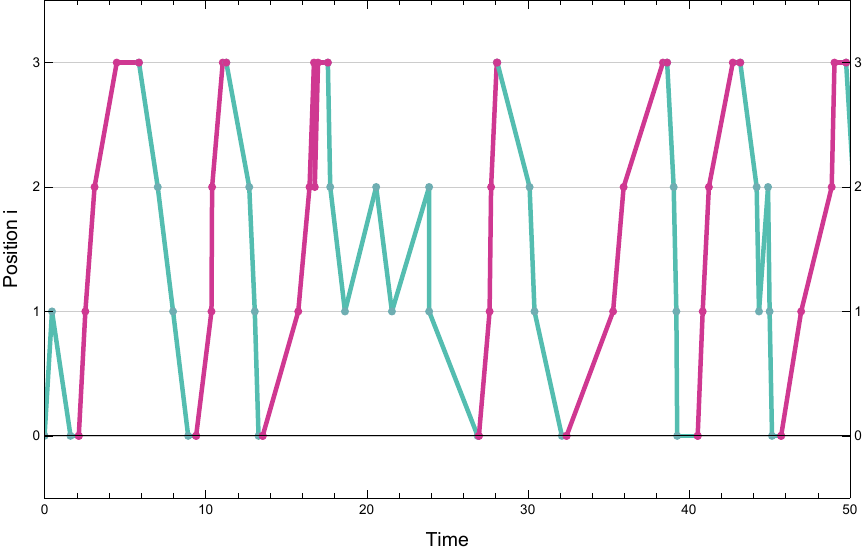}
    \caption{\small{A possible trajectory of the rower described in  \fig \ref{row}, for $\nu=\tau= 1, \beta=2$ and $\alpha =0.5$. The green color indicates that the rower is in states with $\sigma = +1$, and  purple corresponds to $\sigma = -1$. }}
    \label{rotr}
\end{figure}
If we couple such rowers, we may want to ask about the emergence of coordinated motion or wave formation, \cite{kotar2010hydrodynamic}.  Other aspects of interest include the understanding of the heat balances and energy consumption.
\end{example}

Again, lots of variations remain interesting, including the case where we couple these systems for instance via a chain of molecules or a chain of oscillators and study energy transport (between nonequilibrium reservoirs).

\chapter{Boltzmann's picture}\label{bpic}
\lhead{C. Maes}
\rhead{Boltzmann's picture}

A major part of nonequilibrium statistical mechanics has been devoted to the characterization of equilibrium, statics and dynamics.  That got intimately connected with the concept (and the origin) of dissipation, with the derivation of kinetic and hydrodynamic equations, and with the derivation and understanding of the Second Law of Thermodynamics.  The main qualitative idea goes back to Ludwig Boltzmann, the theory of whom Erwin Schr\"odinger wrote,
\begin{quote}
{\it The spontaneous transition from order to disorder is the quintessence
of Boltzmann’s theory... This theory really grants an understanding
and does not reason away the dissymetry of things by means of an
a priori sense of direction of time... No one who has once understood
Boltzmann’s theory will ever again have recourse to such expedients. It
would be a scientific regression beside which a repudiation of Copernicus
in favor of Ptolemy would seem trifling.}
\end{quote}
The present chapter obviously has little original.  There are many excellent expositions, including \cite{bricmont2022making,goldstein2001boltzmann,lebowitz2008timesarrow}.\\

\textbf{Microscopic setup:} We consider a closed, isolated system with conservative Hamiltonian dynamics and flow $\varphi_{t}$. The number of particles $N\gg 1$ (like Avogadro's number), energy $E$, and a regular volume $V$ with smooth elastic boundary conditions are all fixed.  The resulting dynamics develops on a restricted phase space $\Omega_E$ of constant energy that we equip with the Liouville volume element.  Mechanical or microscopic states are made from the generalized
coordinates $q_1,\ldots q_N$ and momenta $ p_1,\ldots,p_N$. We
abbreviate $X=(q_1,\ldots,q_N;p_1,\ldots,p_N)$ for the state of
our universe.   All such states $X$ are considered (microscopically) equivalent\footnote{That is basically the assumption of the microcanonical ensemble, where, following the Laplace indifference principle, we identify the probability of a phase space region as proportional to its Liouville volume.  One motivation is that such a measure is invariant under Hamiltonian flow.  Obviously, it remains a theoretical hypothesis, and empirical data decide how useful it is for a unified understanding.  At any rate, since much is based on the law of large numbers and the corresponding notion of ``typicallity,'' measures having a smooth density with respect to the microcanonical distribution are equally fine.}, possibly taking into account the practical (in)distinguishability of particles. The flow $X(t)=\varphi_{t}(X_{0})$, determining the state at time $t$ from the state at time zero, is volume-preserving and time-reversal invariant. The latter means that there is an involution $\pi$ for which $\pi\varphi_{t}\pi=\varphi_{-t}$. The time reversal $\pi$ reverses all momenta $\vec{p}_{i}\rightarrow-\vec{p}_{i}$; it is called kinematical time-reversal and we assume that the Hamiltonian is invariant for it.  We denote the volume of a phase space region $A\subset
\Omega_E$ by $|A|$ which is the Liouville volume projected on
$\Omega_E$. There is no phase space contraction in the sense that
the image of $A$ under the Hamiltonian flow, $|\varphi_t(A)|=|A|$,
has the same volume as $|A|$.
\begin{figure}[H]
\centering
\includegraphics[scale=1]{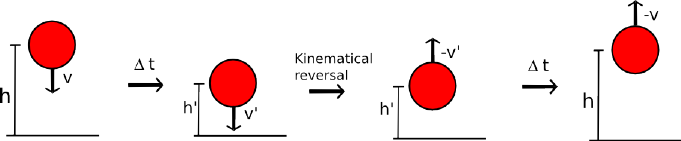}
\caption{Thinking of a ball falling on Earth; we have the same gravity when launching the ball, retracing the spatial trajectory with opposite velocity.}\label{ballf}
\end{figure}
We can lift the time-reversal to the level of trajectories, where we flip both the sign of the momenta and the time-order of states; see Fig.~\ref{ballf}. Each possible trajectory then has a reversed one, equally possible, where all velocities have changed sign.\\

\textbf{Macroscopic variables:} The assignment to microscopic states $X$ of the values of the macroscopic variable is a many-to-one map $X\longmapsto M(X)$. Generally, $M(X)$ can be thought of as a type of spatial profile or histogram, achieved by spatial averaging (as when computing a density) or by counting proportions (such as the fraction of particles having some property somewhere). Whatever the case, the map $M$ induces a (sometimes more fuzzy) partition in the phase space, dividing it into patches of all states $X$ that have about the same macroscopic value $M(X)$. The largest patch is the macroscopic condition called equilibrium.  Limitations to that picture are discussed below in Section
\ref{limita}; assumptions are needed indeed.
\begin{figure}[H]
\centering
\includegraphics[scale=0.5]{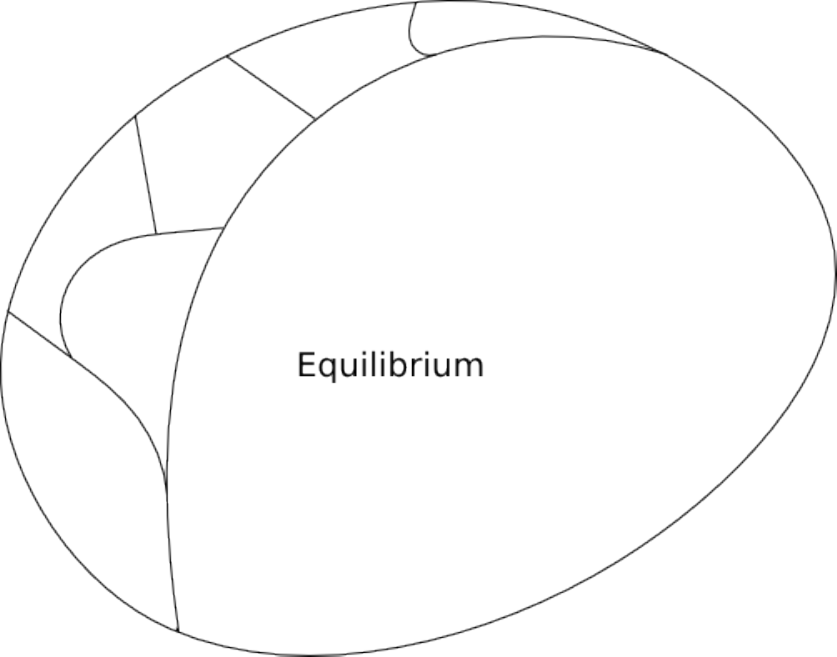}
\caption{Constant energy surface, indicating that `equilibrium' occupies by far the largest volume for reasonable macroscopic variables.  In reality, the figure is very far from doing justice to the real situation at hand: by the law of large numbers as $N\uparrow 10^{23}$, about all phase points belong to `equilibrium.' On that count, nonequilibrium is rare.}
\end{figure}
For many-particle systems the equilibrium region will be overwhelmingly huge compared to the other (nonequilibrium) regions, and in that way the equilibrium values are {\it typical} values just as in the law of large numbers. No details about the system or its dynamics have been specified yet except that we fixed the energy and that we want our macroscopic description to be relevant and (in a sense) complete (see below for macroscopic autonomy). Regardless of that, it tells us that equilibrium is the most probable condition from the macroscopic point of view.  Nonequilibrium conditions are rare, and their presence needs explaining.   All nonequilibria must come from earlier nonequilibria... except when even the notion of time gets lost.\\

Given the above setup, the \textbf{Boltzmann entropy} for a macroscopic condition $m$ is given by 
\begin{equation}
    S(E,V,N;m) = S(m)=k_{B}\log |\{X: M(X)=m\}|, \text{ or } \;\quad S= k_B\log W
\label{starr}    
\end{equation}
where $|A|$ of a phase region $A$ denotes its Liouville volume, always given the constraints on $N$, $E$ and $V$. We hypothesize that it quantifies the plausibility of a macroscopic condition, in the sense that (within the considered microcanonical ensemble),
\[
\frac{\text{Prob}[m]}{\text{Prob}[m']} = e^{[S(m) - S(m')]/k_B}
\]
where $m,m'$ stand for macroscopic conditions, outlook, values,... Taking into account here that $k_B \simeq 1.38 \times 10^{-23}$ J/K,
 a relatively small increase in the entropy of $10^{-3}$ J/K for $m'\rightarrow m$ already gives rise to a factor $\exp 10^{20}$ for the probability ratio.\\
 The breaking of time-reversal symmetry is certainly an important
feature of nonequilibrium systems.  While the underlying microscopic
dynamics is (under usual circumstances) time-reversal symmetric, the
plausibility of the time-reversed history of mesoscopic or even more
macroscopic conditions can greatly differ from that of the original
history. These considerations are very much linked with the concept
of entropy as in the Boltzmann picture of the previous chapter,  and its production.  As written by Max Planck in 1926
\cite{Planck}: {\it "...there is no other general measure for the
irreversibility of a process than the amount of increase of
entropy."}\\

For a microscopic derivation of the Second Law it is useful to define that Boltzmann entropy on microscopic states $X$, via
\begin{equation}\label{enb}
S(X) = S(M(X))=k_{B}\log |\{Y: M(Y)=M(X)\}|
\end{equation}
Now we can follow $S_t = S(\varphi_t(X))$ in time $t$.  
In great generality and excluding major conspiracies, we may expect the microscopic trajectory to wander off towards equilibrium when started off-equilibrium as it is just more plausible to enter more plausible macroscopic conditions.  The result is a monotone increase of Boltzmann entropy, till a maximum is reached.  Such an H-theorem is considered in Section \ref{h1}, and a specific illustration for macroscopic mechanics is discussed in \cite{tas}.\\

Let us give an example of a physically interesting
coarse-graining as initially worked out by Boltzmann in the context of a dilute gas\footnote{Another case, the (modified) Enskog equation for a system of hard spheres, is obtained in \cite{Resibois1978}.}.  Such a gas has essentially no interactions\footnote{At least that is a valid approximation for a static description.  For the dynamics and the relaxation, interactions do matter and the nontrivial collision kernel in the Boltzmann equation depends on them.} between the particles and the energy is determined by
the kinetic energy.  For coarse-graining (and simplifying to positions and velocities), we define the proportion
of particles that have their position and momentum around some point $(r,v)$ in the one-particle phase space. Then, this reduced
state determines both the energy and the particle number. To be specific we take $n$ particles of mass 1 in a finite box $V\subset
\bbR^3$, and we specify small volumes (cells) $\Delta \subset V\times \bbR^3$ of the one-particle phase space. (Think of the
above $N=3n$ for point particles.)  In other words $\Delta =\Delta_{r,v}$ is a little volume around some $(r,v)\in V\times
\bbR^3$. We then denote the density of particles around $(r,v)$ by 
\begin{equation}\label{fbol}
f^{(n)}(r,v) = \frac 1{|\Delta_{r,v}|} \sum_{i=1}^n \chi[
(r_i,v_i) \in \Delta_{r,v}]
\end{equation}
where we sum over all particles and $\chi$ is the indicator function, equal one when the particle's position  $r_i$ and
velocity $v_i$ are around $(r,v)$ and equal zero otherwise. So for every micro-state (positions and velocities of all the $n$
particles) we know the value  of $f^{(n)}$ at each $(r,v)$. All these values together (or, the function itself) correspond to a
particular macroscopic condition.\\
  The entropy counts the degeneracy; in how many ways can we realize such a profile
$f^{(n)}$?  We must calculate the phase-space volume \eqref{starr},
approximately
\[
 W
\equiv\prod_{(r,v)}\frac{{|\Delta_{r,v}|}^{|\Delta_{r,v}|f^{(n)}(r,v)}}{(|\Delta_{r,v}|f^{(n)}(r,v))!}
\]
We can apply Stirling's formula, to find
\begin{equation}\label{h}
S[f] = k_B \log W \simeq -k_B \int \id r\int \id v \,f(r,v) \log f(r,v)
+ \mbox{ constant}
\end{equation}
when $f(r,v)$ is the limiting profile one gets from $f^{(n)}$ in
the limit when the $\Delta$'s shrink to points $(r,v)$ and $n$
goes up. This expression on the right-hand side for the entropy
\eqref{h} is sometimes also called the Boltzmann entropy, but
should not be confused with the more fundamental one \eqref{enb}.

\section{Kac ring model}\label{kmo}
\lhead{C. Maes}
\rhead{Kac ring}
We follow references \cite{pra,DeRoeckJacobsMaesNetocny2003}.\\

\textbf{Model \& dynamics:} The Kac ring model (1958-59) was introduced to simplify the discussion on problems involved in the understanding of the Boltzmann equation, \cite{K,pra,DeRoeckJacobsMaesNetocny2003}.  While the Boltzmann equation is much more
complicated, the Kac model is mathematically simple and free of
those extra technical problems that are not really important for
understanding some crucial aspects of the emergence of macroscopic
irreversibility.  On the other hand, it does not give an understanding of relaxation to thermal equilibrium (as it is restricted to only one type of observable).\\

\begin{figure}[H]
\centering
\includegraphics[scale=0.4]{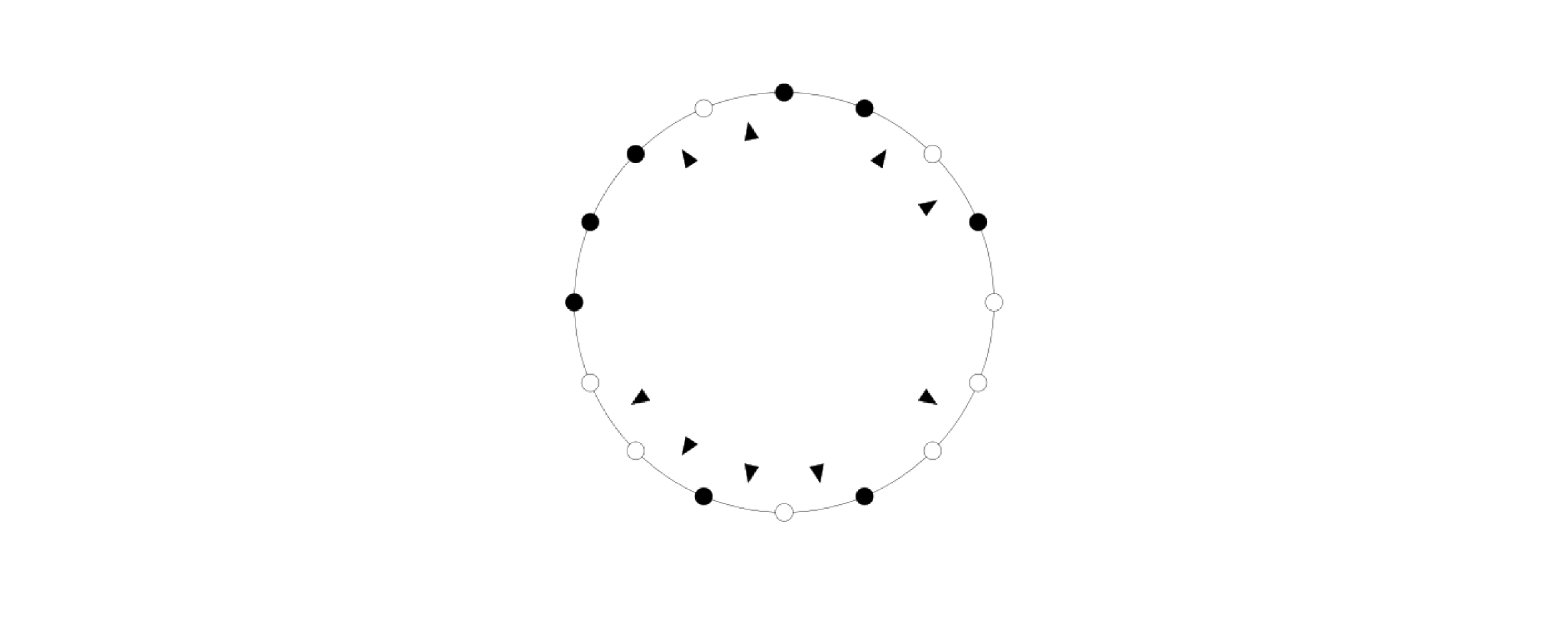}
\caption{\small{Kac ring: black and white balls clip their color when passing a scatterer (black triangle)}}\label{kacf}
\end{figure}

  Consider a ring with $N$ sites ($N+1=1$); see Fig.~\ref{kacf}. There are little particles or balls which each have a color, black or white, assigned to the sites. The color of site $i$ is indicated by
$$\eta(i)=\left\{
\begin{matrix}
+1&(\text{black})\\
-1&(\text{white})
\end{matrix}
\right.$$ 
with $i=1,...,N$. Additionally there are ``scatterers'' in between the sites. The presence of a scatterer between site $i$ and $i+1$ is given by$$g(i)=\left\{
\begin{matrix}
1&(\text{present})\\
0&(\text{absent})
\end{matrix}
\right.$$
and is represented by the black arrows, for $g(i)=1$, in Fig.~\ref{kacf}.
The dynamics consists of a discrete time process where every time step the ring is rotated clockwise, {\it i.e.}, each particle  moves from site $i\rightarrow i+1$, while the scatterers remain unmoved. If a particle passes through a scatterer during the rotation it changes color. Thus, $$\eta_{t}(i+1)=\left[1-2g(i)\right]\eta_{t-1}(i)$$ where $t=1,2,3,\ldots$.  The scatterer configuration is not changing in time. In all, the micro-state $(\eta,g)$ evolves deterministically and is also periodic with a period $\leqslant 2N$. Then each particle will have passed all scatterers exactly twice so it will have retained its original color.  The dynamics is also reversible in the obvious sense, {\it i.e.}, upon rotating the ring counter-clockwise, we recover the original configuration.\\

We can forget about the scatterers as dynamical variables by keeping them unchanged; they are fixed in time with a given average or scatterer density
\[
\rho = \frac{1}{N}\sum_{i=1}^{N}g(i) \in (0,1)
\]
There is an obvious interpretation:
at every time instance $t$, each spin $\eta_t(i)$ jumps to its
successive site, $i+1$, either
 flipping its value if a scatterer is
present, $g(i) = 1$, or keeping its value if $g(i) = 0$.\\
 Sampling the initial colors
$\eta_0$ from a probability law $\mu_0$ on $K$, the probability to
find the color configuration $\eta$ at time $t$ is
\begin{equation}
  \text{Prob}_{\mu_t}[\eta] = \text{Prob}_{\mu_0}[\eta_t]
\end{equation}
That is the present variant of the Liouville equation for
mechanical systems.  Here also the Shannon entropy $S(\mu) =
-\sum_{\eta,g} \mu[(\eta,g)] \log \mu[(\eta,g)]$ is
time-invariant, $S(\mu_t) = S(\mu_0)$. There is just no strictly
increasing Lyapunov function for this dynamical system;
 we repeat that
 the dynamics is $2 N-$periodic. Nevertheless, the coarse-grained color exhibits relaxation to the equilibrium value; to see that, we need to
pass to the macroscopic viewpoint.\\

\noindent\textbf{Macroscopic point of view:}
We consider the macroscopic variable 
\begin{equation}
m^{(N)}(\eta)=\frac{1}{N}\sum_{i=1}^{N}\eta(i)\quad\in[-1,1]\nonumber
\end{equation}
where $m^{(N)}$ is the average color (or ``greyness'') of the particle system.\\
As is easy to verify by an exact numerical calculation, the emergent \emph{macroscopic} dynamics has the form
\begin{equation}\label{maka}
  m^{(N)}_t \mapsto m^{(N)}_{t+1}= \phi (m_t^{(N)})\qquad \text{ with }  \phi (m) = [1 - 2\rho] m,
\end{equation}
at least for very large $N$.  In other words, the macroscopic
greyness $m^{(N)}_t$ evolves autonomously.  
It is important to understand here the status of the so-called {\it Sto{\ss}zahl Ansatz}, a fast but possibly misleading answer also known as the assumption of molecular chaos.  In the present model, it is obtained by assuming that the scattering of the colors should effectively\, amount to a macroscopic evolution obtained by using $\rho$ (the scatterer density) as a probability:
\begin{eqnarray}
\# \text{black}_{t} = (1-\rho)\# \text{black}_{t-1} + \rho\#\, \text{white}_{t-1}\nonumber\\
\# \text{white}_{t} = \rho\#\, \text{black}_{t-1} + (1-\rho)\#\,\text{white}_{t-1}\nonumber
\end{eqnarray}
$$m^{(N)}(\eta_{t})=m^{(N)}_{t}=\frac{1}{N}\left(\# \text{black}_{t} - \# \text{white}_{t}\right)=(1-2\rho)m^{(N)}_{t-1}=(1-2\rho)^{t}m^{(N)}_{0}$$
\begin{figure}[H]
\centering
\includegraphics[scale=1.5]{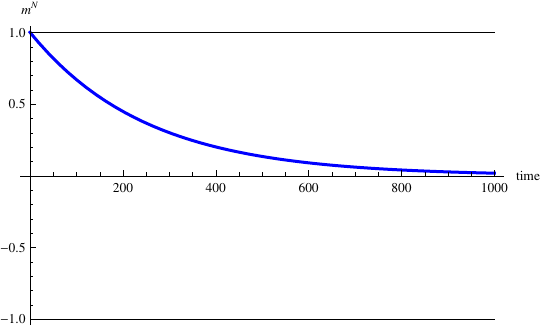}
\caption{\small{Kac ring: the greyness in time $t\in \bbN$ for density $\rho < 1/2$ of scatterers.  When $\rho > 1/2$ there are also points that give a negative 'greyness' but still, as we have discrete time and in particular never actually cross the line of ``zero greyness'', the entropy is always increasing.}}\label{kacff}
\end{figure}
That leads to the correct macroscopic evolution but the equation is certainly not correct as such for all $N$ or for all initial conditions. It is easy to imagine microscopic configurations that
violate \eqref{maka} and the question arises how such a macroscopic
behavior can or must be understood.  It therefore helps here to be mathematically precise.\\

As a reference, we consider a uniform probability distribution ${\mathbb P}^N$ over the color configurations, {\it i.e.}, for each site $\eta(i)=\pm1$ are equally probable (similar to the microcanonical distribution). We also use the notation $a \stackrel{\delta}{=} b$ for $|a - b| \leq
\delta$.  Then the statement \eqref{maka} has the more precise form of a law of large
numbers:
\begin{equation}\label{eq: Kac-autonomy}
  \lim_{N\uparrow\infty}
  {\mathbb P}^N[m^{(N)}(\eta_t)
   \stackrel{\delta}{=} m_0 (1 - 2\rho)^t \,|\, m^{(N)}(\eta_0) =
   m_0;\, \rho^N(g) = \rho]
  = 1
\end{equation}
for all $\delta > 0$. This means there is a set of \emph{typical}
microscopic configurations satisfying the macroscopic law with map
$\phi$; those configurations violating that law make a set of zero
limit measure. \emph{Autonomy} amounts to
\begin{equation}\label{eq: Kac-autonomy1}
  \lim_{N\uparrow\infty}
  {\mathbb P}^N[\forall t \leq \tau:\,
  m^{(N)}(\eta_t) \stackrel{\delta}{=} m_0 (1 - 2\rho)^t \,|\, m^{(N)}(\eta_0) = m_0;\, \rho^N(g) = \rho]
  = 1
\end{equation}
for all $\delta > 0$ and any finite $\tau$.\\

The relaxation along that typical macro-evolution is
obvious by inspection but one can also construct an explicit
witness which is the
 \emph{Boltzmann entropy}
$s(m)$ defined as the large deviation rate function for the
sequence $(m^N)_{N\uparrow +\infty}$ of observables:
\begin{equation}
  {\mathbb P}^N[(m^N(\eta)] \simeq m] \simeq e^{N s(m)}
\end{equation}
This is to be understood in the logarithmic sense after taking the
limit $N \uparrow +\infty$, {\it i.e.}, it is a shorthand for the limit
statement
\begin{equation}
  s(m) = \lim_{\delta\downarrow 0} \lim_{N\uparrow +\infty} \frac{1}{N} \log
  {\mathbb P}^N[m^N(\eta) \stackrel{\delta}{=} m;\,\rho^N(\eta) \stackrel{\delta}{=} \rho]
\end{equation}
This is simply the binomial entropy,
\begin{equation}\label{eq: Kac-entropy}
  s(m) = - \frac{1+m}{2} \log (1+m) - \frac{1-m}{2} \log (1-m), \qquad  -1 < m < 1
  \end{equation}
and one checks that $s(\phi(m)) > s(m)$ whenever $m \neq
0$ (system off equilibrium) and  nonsingular
macrodynamics \eqref{maka}.  Observe that the entropy $s(m)$ is maximal for $m=0$. Therefore, the equilibrium value for the average color, that is overwhelmingly probable, is the grey $m^{(N)}(\eta) =0$.  We can also reasonably expect that a good set of initial conditions will evolve into that grey.

Using the notation $m_t = m_0 (1-2\rho)^t$, it
yields that $s(m_t)$ is a strictly increasing function of
time (for $m_0\neq 0)$. Following Boltzmann's terminology, such a  statement is
called an \emph{H-theorem}; the Boltzmann entropy is a Lyapunov
function.\\
It is essential to understand for the relaxation that we take $N\gg \tau$ (huge number of components), where we refer to macroscopic (or more generally reduced) variables which make an autonomous macro-evolution, and that the statement has a statistical meaning (as relaxation need not be true for {\it all} possible initial microscopic states).\\

\noindent\textbf{Irreversibility and entropy production:}
Consider the time-reversed microscopic dynamics, {\it i.e.}, turning the wheel counter-clockwise and producing the color update
\begin{equation}\label{revd}
  \eta_{t+1}(i) = [1 - 2g(i)]\,\eta_t(i+1)\qquad \mod N
\end{equation}
Observe that the sequence (trajectory)
$(\eta_0,\eta_1,\ldots,\eta_t)$ is allowed (possible) under the
original microscopic dynamics iff
$(\eta_t,\eta_{t-1},\ldots,\eta_0)$ is possible under the time-reversed microscopic dynamics.  That
invertibility is referred to as \emph{dynamical
(time-)reversibility}.  It can be formulated differently by
extending the configuration space $K$ with a `velocity' variable
$v \in \{-1,1\}$ indicating whether we turn the wheel clockwise (+1) or counter-clockwise (-1).  The time-reversed microscopic dynamics is then achieved by
inverting the velocity $v$.\\

The macroscopic time-evolution $\phi$ in which $m \mapsto
(1-2\rho)\, m \in [-1,1]$ is invertible as well (provided that $\rho
\neq \frac{1}{2}$). Yet, the typical macroscopic time-evolution
corresponding to $v=-1$ (counter-clockwise) is not $\phi^{-1}$ but rather $\phi$
again, {\it i.e.}, the law of large numbers~\eqref{eq: Kac-autonomy}
stays true for the time-reversed microscopic dynamics. It simply means that the macroscopic evolution $m
\mapsto m (1 - 2\rho)$ does \emph{not} get inverted by starting
from a \emph{typical} microscopic configuration $\eta$
corresponding to the macroscopic state $m (1-2\rho)$ and by
applying the inverted microscopic dynamics (or by inverting the
velocity). Naturally, there exist microscopic configurations
$\eta$ for which the inverted macro-evolution $m (1-2\rho) \mapsto
m$ along the counter-clockwise dynamics would be observed---this is
precisely what the dynamical reversibility claims---they show up
to be exceedingly \emph{exceptional} under $m (1-2\rho)$, however.
This physical impossibility to invert the macroscopic evolution is
referred to as
\emph{macroscopic irreversibility}.\\

The macroscopic irreversibility in the above sense on the one hand
and the strict increase of the Boltzmann entropy on the other
hand, are often used as synonyms. Let us formulate their relation
a bit more precisely.  Denote by $\bar\eta_t$ the color configuration at time $t$ by the opposite turning (counter-clockwise) of the wheel.  Observe that the two sets of color configurations
\[
  \{\eta_0:\,m^{(N)}(\eta_0) \stackrel{\delta}{=} m_0;\,m^{(N)}(\eta_t)
  \stackrel{\delta}{=} m_t\}
\]
and
\[
  \{\eta_0:\,m^{(N)}(\bar\eta_t) \stackrel{\delta}{=} m_0;\,m^{(N)}(\eta_0)
  \stackrel{\delta}{=} m_t\}
\]
with $m_t = m_0 (1-2\rho)^t$ have the same cardinalities. Hence,
they have the same probabilities under ${\mathbb P}^N$, which we write
as
\begin{equation}\label{eq: inverted1}
\begin{split}
  \log {\mathbb P}^N[m^{(N)}(\eta_t) \stackrel{\delta}{=} m_t \,|\, m^{(N)}(\eta_0)
  &\stackrel{\delta}{=} m_0]
\\
  +\log {\mathbb P}^N[m^{(N)}(\eta_0) &\stackrel{\delta}{=} m_0]
\\
  = \log {\mathbb P}^N[m^{(N)}(\bar\eta_t) \stackrel{\delta}{=} m_0 \,|\, m^{(N)}(\eta_0)
  &\stackrel{\delta}{=} m_t]
\\
  + \log {\mathbb P}^N[m^{(N)}(\eta_0) &\stackrel{\delta}{=} m_t]
\end{split}
\end{equation}
Dividing by $N$, taking the limits $N\uparrow\infty$ and
$\delta\downarrow 0$ in this order, and using the law of large
numbers~\eqref{eq: Kac-autonomy}, we get the large deviation law
\begin{equation}\label{eq: inverted2}
\begin{split}
  \lim_{\delta\downarrow 0} &\lim_{N\uparrow\infty}
  \frac{1}{N} \log{\mathbb P}^N[m^N(\bar\eta_t) \stackrel{\delta}{=} m_0
  \,|\, m^N(\eta_0) \stackrel{\delta}{=} m_t]
\\
  &= s(m_0) - s(m_t)
\end{split}
\end{equation}
or
\begin{equation}\label{eq: main-macro}
  {\mathbb P}^N[m^N(\bar\eta_t) \simeq m_0 \,|\, m^N(\eta_0) \simeq m_t]
  \simeq e^{-N[s(m_t,\rho) - s(m_0,\rho)]}
\end{equation}
which is quite a remarkable relation. Notice first that it
provides another derivation of the H-theorem: since the left-hand
side is less than one, one immediately gets $s(m_t) \geq
s(m_0)$. Further, the left-hand side is nothing but the
probability that a configuration $\eta_0$ sampled from macrostate
$m_t$ and evolved counter clockwise, exhibits the macroscopic evolution from $m_t$ to $m_0$, which is just an inversion of the
typical transition $m_0 \mapsto m_t$.

Macroscopic irreversibility amounts to the statement that such
inverted macroscopic transitions are physically impossible; here
we have a quantitative evaluation how rare they really are: the
large deviation rate function for the backward transition $m_t
\mapsto m_0$ with respect to $v=-1$ just coincides with the
entropy production along the \emph{typical} macroevolution $m_0
\mapsto m_t$ with respect to the microscopic dynamics at $v=1$. Inverting
the logic, this can be read off as a formula for the entropy
production.    That is nothing else than detailed balance, {\it i.e.}, that the transition probabilities  are governed by a change in entropy, so that the change in entropy is a measure of the irreversibility of the transition.\\

We summarize some recurring comments on that type of result.  These are the objections also Boltzmann had to answer when presenting his Boltzmann equation.\\
\begin{itemize}
\item \underline{Bad initial conditions} do exist and are in fact easy to make.  We can let the system run from time zero where all particles are black to some intermediate time $t_1$ and then stop it.  We see a particular particle color configuration $\eta_{t_1}$.  Let us now turn the ring counter-clockwise for time $t_1$.  The dynamics is of course basically the same.  After that time $t_1$ of reversed motion we recover (by reversibility) the full black color.  In other words, $\eta_{t_1}$ is absolutely a ``bad'' configuration as it leads to a decrease in entropy and no relaxation to equilibrium greyness. However, the point is that such ``bad'' configurations are highly untypical for a given macroscopic condition.
They are not seen when randomly drawing a configuration corresponding to a particular macroscopic condition (here, some average color).
\item \underline{Periodicity} (or the Zermelo objection): The Poincar\'e recurrence time is very large (age of the observable universe) for $N\propto 10^{23}$. It is not relevant for the physical relaxation processes that we observe. Nevertheless, the presence of recurrence also explains why relaxation to equilibrium cannot be a statement about a closed isolated system containing only a few degrees of freedom. 
\item \underline{Reversibility} (or the Loschmidt objection): The seeming inconsistency between the microscopic reversibility
and the macroscopic irreversibility is known as the Loschmidt
paradox: each trajectory $\eta_t$ has a reversed one. How could it be that the entropy is typically increasing, as for the reversed motion the entropy is decreasing?  Equality~\eqref{eq: main-macro} in a sense solves this
paradox and put it in a correct perspective: those macroscopic
trajectories obtained by time-reverting the typical ones
$(\phi^t(m)$;\, $t = 0,1,\ldots)$ are indeed observable for finite
$N$, however they are exponentially damped.\\ Therefore, we could say that the answer is in the emphasis on macroscopic control.  We do not choose particular microscopic states; we fix the macroscopic condition and nature by itself draws from there the microscopic state.  Most are `good' (reproducing 2nd Law behavior), a minority is `bad.'
\begin{figure}[H]
\centering
\includegraphics[scale=0.4]{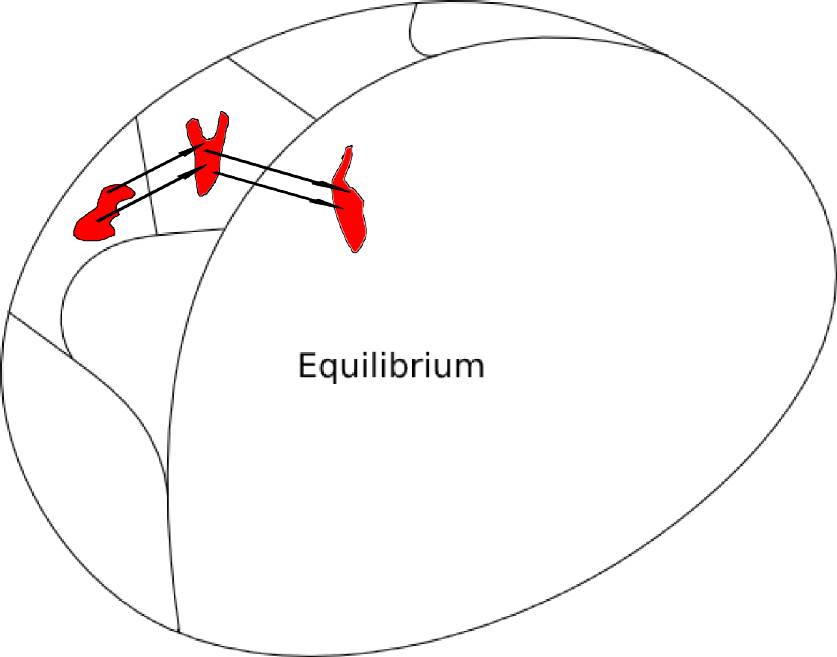}
\end{figure}
The red regions show micro states evolving with the microscopic time-evolution.  Think about classical mechanical dynamics and the physical partition of the constant energy surface. The Liouville theorem states that the volume of the region is conserved in time (but it can change shape). However this is not saying that the entropy is constant.  The relevant question is to determine the macro-state the system is now visiting. It is the volume of the regions corresponding to macro-state $m_{t}$ that increase  in time.   For large $N$ this dark region's volume is negligible compared to the rest of the macroscopic phase space volume. 
\end{itemize}

\section{H-theorems}\label{h1}
\lhead{C. Maes}
\rhead{H-theorem}

\begin{quote}
{\it Many important
phenomena, in physics and beyond, while they cannot be shown to hold
without exception, can be shown to hold with very rare exception, suitably
understood.}\\
Sheldon Goldstein in \cite{Goldstein2012Typicality}.
\end{quote}

\noindent\textbf{General argument:} Consider a dynamical system $(\Omega^{N},U^{N},\rho^{N})$, with $\Omega^{N}$ the state space, $U^{N}$ the dynamics and $\rho^{N}$ the invariant distribution, for example a mechanical system with $N$ degrees of freedom. Now we want to define a Boltzmann entropy using the many-to-one map (from micro to macro)
$$x\in\Omega^{N}\longmapsto m^{N}(x)\in\mathcal{M}$$
with $\mathcal{M}$ a metric space collecting the values of macroscopic variables. The Boltzmann entropy is then given by the fluctuation formula
\begin{equation}
\label{be}
S(m):=\lim\limits_{\delta\rightarrow 0}\lim\limits_{N\rightarrow\infty}\frac 1{N}\log \text{Prob}_{\rho^{N}}\left[m^{N}(x)\stackrel{\delta}{=}m\right]
\end{equation}
where the $\delta-equality$ refers to the maximal distance between $m^N(x)$ and $m$: $d(m^N(x),m)\leq \delta$ where $d$ is the metric on $\mathcal M$.  The H-theorem states that $S(m)$ must increase whenever there is macroscopic autonomy.  We can see that as follows:\\
Macroscopic autonomy means that there is a flow on $\mathcal{M}$, say $m_{t}=\phi(m_{0})$, such that
$$\lim\limits_{\delta,N}\frac{1}{N}\log \text{Prob}_{\rho^{N}}\left[m^{N}(x_{t})\stackrel{\delta}{=}m_{t}|m^{N}(x_{s})\stackrel{\delta}{=}m_{s}\right]=0\qquad\forall s\leqslant t$$  We claim that then indeed
$$S(m_{t})\geqslant S(m_{s})\qquad\text{for } s\leqslant t$$
which is the non-decreasing property of the Boltzmann entropy.\\

\noindent\textbf{Sketch of proof:}
\begin{eqnarray}
S(m_{t})&\simeq&\frac{1}{N} \log \text{Prob}\left[m^{N}(x)\simeq m_{t}\right]\nonumber\\
&\simeq&\frac{1}{N} \log \text{Prob}\left[m^{N}(x_{t})\simeq m_{t}\right]\nonumber\\
&\geqslant&\frac{1}{N} \log \text{Prob}\left[m^{N}(x_{t})\simeq m_{t}|m^{N}(x_{s})\simeq m_{s}\right|]\text{Prob}\left[m^{N}(x_{s})\simeq m_{s}\right]\nonumber\\
&=&\frac{1}{N} \log \text{Prob}\left[m^{N}(x_{t})\simeq m_{t}|m^{N}(x_{s})\simeq m_{s}\right|]+\frac{1}{N} \log \text{Prob}\left[m^{N}(x_{s})\simeq m_{s}\right]\nonumber\\
&=& 0 + S(m_{s})\nonumber
\end{eqnarray}
In the second equality we used the invariance of $\rho^{N}$, then the definition of conditional probability was used on the third line and the last line follows from our assumption of macro-autonomy and the definition of Boltzmann entropy. We can see that the entropy serves as a Lyapunov function on the macroscopic evolution.\\
Note that we have not shown the strict inequality $S(m_{t})>S(m_{s})$ for $s<t$. There could in fact be dissipationless hydrodynamics, like in the Euler equations.\\

\noindent\textbf{Examples:}
\begin{itemize}
\item Linear heat equation
$$\frac{\partial m_{t}(x)}{\partial t}=Dm_{t}''(x)$$
with diffusion constant $D>0$, $x\in[0,1]$, $m_t(0)=m_t(1)=1$ and  $m_{t}(x)\geqslant 0$.\\
Take $$S[m]=-\int_{0}^{1}\id x\, m(x) \log m(x) + \int_{0}^{1}\id x \left(m(x)-1\right)$$ and check that it is a Lyapunov function, {\it i.e.}, $\frac{\id S[m_t]}{\id t}\geq 0$ along the solution of the heat equation.\\
What if $m(0)=1$ and $m(1)=2$? Now the system is out of equilibrium and $S[m]$ no longer is a Lyapunov function. 
\item Nonlinear diffusion
$$\frac{\partial \rho_{t}(\vec{r})}{\partial t}+\vec{\nabla}\cdot\vec{J}_{t}(\vec{r})=0$$
with $\vec{r}\in V$ and $\rho_{t}(\vec{r})=\rho_\text{eq}$ for $\vec{r}\in\partial V$. For the current we take
$$\vec{J}_{t}(\vec{r})=\stackrel{\leftrightarrow}{\chi}\vec{F}_{t}(\vec{r})$$ where $\stackrel{\leftrightarrow}{\chi}$ is the susceptibility tensor and the force is 
\begin{equation}\label{hf}
\vec{F}_{t}(\vec{r})=-\vec{\nabla}\frac{\delta\mathcal{F}}{\delta\rho(\vec{r})}(\vec{r},t)
\end{equation}
 with free energy functional $\mathcal{F}$ of the field $\rho_{t}$. It is easily checked that $\mathcal{F}$ is a Lyapunov function.\\
The linear Fick equation is a special case and can be recovered by choosing
$$\mathcal{F}[\rho]=\int d\vec{r}\left[\rho(\vec{r}) \log\rho(\vec{r})-\left(\rho(\vec{r})-\rho_\text{eq}\right)\right]$$ such that $\frac{\delta\mathcal{F}}{\delta\rho(\vec{r})}= \log\rho(\vec{r})$ and $\stackrel{\leftrightarrow}{\chi}=D\rho$.
\end{itemize}
Other systematic arguments and examples have appeared, {\it e.g.} in \cite{villani}.

\section{Limitations and extensions}\label{limita}
\lhead{C. Maes}
\rhead{Limits to the Boltzmann picture}

Despite the simplicity and clarity of the Boltzmann picture outlined above, many questions appear to remain or continue to appear.  They cannot be ignored; they point to aspects that are less addressed in standard explanations.\\
We gather some questions and give them some thought.\\

\begin{itemize}
\item
What selects the correct or appropriate macroscopic variables?  Are we free to take no matter what as macroscopic variables?  Does that then downgrade the entropy and its monotonicity into something `subjective?'\\

That question is often combined with the conviction that we need a more fundamental dynamical understanding\footnote{and chaos is often quoted as essential.  Obviously, ergodicity or mixing properties of the microscopic dynamics is neither necessary nor sufficient for relaxation to equilibrium, \cite{bricmont2022making}.}.  It is sometimes phrased then by the saying that the Boltzmann picture makes us ``the parents instead of the children'' of irreversibility, which is not what physics should be doing.\\
The answer can take various directions, and there is no problem to acknowledge that there are indeed many ways of coarse-graining\footnote{For the purpose of modeling language, as a totally different class of problems, one can for instance think of `meaning' as a coarse-grained notion of `text,' where one should take into account that the map from `text' to `meaning' is often discontinuous. Such a coarse-graining can be achieved by introducing a dissipative dynamics between (mutually interacting) words that converges to `meaning.'} and reducing the description. Nevertheless, there is often a (more or less unique) meaningful ``physical'' coarse-graining, where the macroscopic description leads to a consistent thermodynamics with corresponding observable consequences.  For example, Carnot efficiency only depends on the temperature of the heat baths, not on their color or taste.  Arguably, a good choice of macroscopic variables is one in which their dynamics forms a first-order set of differential equation.  It means autonomy (or independence of microscopic details) and reproducibility for given initial conditions. In Section \ref{h1} we have given autonomy of the macroscopic variables as sufficient condition for having a corresponding entropy that is non-decreasing in time.  It is easy to find examples where the violation of autonomy makes the entropy non-monotone in time.\\
So indeed there may be more entropies as based on different sets of macroscopic observables, but not all are physically useful, far from.  The macroscopic observables of thermodynamics are not ``randomly'' chosen; they correspond to an increasing entropy and reproducible (thermodynamic) phenomena.  Once we have chosen the macroscopic framework, we must live with the objective consequences...

\item 
What determines the relaxation times?  After all, Boltzmann's picture seems to suggest that relaxation to equilibrium, if not immediate, is fast and ubiquitous in macroscopic systems; are there no exceptions?\\

Indeed, Boltzmann's picture as presented above lacks quantitative estimates if we get away from dilute gases.  There is work to be done for the non-dilute case...  However, there is quite some physics understanding of what can happen.  Indeed, physical systems can be kinetically constrained, both classical and quantum.  It leads to phenomena of localization and glassy dynamics, to mention two (even broadly related) classes of ``exceptional'' behavior.  What then is ``missing'' in Boltzmann's picture?\\
We can say that the picture above probably over-emphasizes the role of the size of the phase-space regions corresponding to the different macroscopic conditions. It seems to suggest that only size matters.  That is not true; the transitions between phase space regions specifying a macroscopic condition are not solely determined by their volume, {\it i.e.}, their entropy. What also matters, especially when still far from equilibrium, is the accessibility and escape rates (or life times).  In other words, there must be sufficient dynamical activity to leave a phase-space region and to enter another: the traffic between two adjoining rooms or cities is not only determined by their size, but also by the size of the exits and entrances.  
\begin{quote}
    {\it to determine whether under a given set of planetary conditions life is the preferred state or only a metastable state, we cannot just compare the lifeless state and the known biological state, but must consider the transitions between these states.}\\
(Charles Bennett, as quoted by Rolf Landauer in 1975)
\end{quote}
Lifetimes and accessibility, escape rates and dynamical activity come to the foreground in open nonequilibrium systems.\\
We have summarized that point via the notion of frenesy, or the frenetic contribution as we  encounter it naturally in response theory, \cite{fren}.  Here, it is a way of saying the obvious, that kinetics, and not only thermodynamics, matters for understanding relaxation (times).  The (im)possibility of time-symmetric dynamical activity is responsible for localization and glassy behavior.  That matters most when the sizes of the phase-space regions are still relatively small, much smaller than the overwhelming size of near-equilibrium regions. 

\item 
Is this Boltzmann picture restricted to closed and isolated classical macroscopic systems?  {\it E.g.}, what does it say or imply for small open systems? Is there actually a nonequilibrium entropy?\\

Entropy or other thermodynamic functionals are in the first place associated to the macroscopic equilibrium reservoirs that make the environment of the small open (nonequilibrium) system.  {\it E.g.}, we do not need to speak at all about the entropy or the temperature of our nonequilibrium system\footnote{when it satisfies a condition known as {\it local detailed balance}; see Section \ref{dlbb}.} to define its heat capacity. Heat capacity is related to the First Law of thermodynamics, where heat is defined with respect to the reservoirs.\\
Nevertheless, it remains true for large steady nonequilibrium systems that the Boltzmann picture retains its meaning for trying the correct coarse-graining; see {\it e.g.}, \cite{lebowitz_goldstein2004boltzmann}. We can still speak about the Boltzmann entropy as related to and governing static macroscopic fluctuations\footnote{They can also be obtained, at least in principle, as a contraction from the dynamical fluctuations (governing the plausibilities of trajectories). The pseudopotential (somewhat of a generalization of the Boltzmann entropy) is a certain solution of a Hamilton-Jacobi equation.} as we did in \eqref{be}. It need no longer be related to heat (and heat capacities) though, and it may be much less clear what are the relevant {\it working} macroscopic variables. A first guess is to stick to the ones for equilibrium, which is the usual strategy for steady state and irreversible thermodynamics. For example, under the assumption of local equilibrium, the balance equation of entropy includes fluxes and production in terms of the usual thermodynamic and hydrodynamic observables.\\
When dealing with a single small system, even when connected to equilibrium baths, there will of course not be a stone-wall character of the First and the Second Laws.  Fluctuations may be very important, and it makes little sense to speak about 'average' behavior.  Nevertheless, at least in some cases, the macroscopic limit can be replaced with certain asymptotic regimes where signal-to-noise ratios do get large.  We may think of very large driving or very low temperatures.\\
The Boltzmann picture is not restricted to classical physics.  Also for quantum systems, the same ideas continue except that the large deviation and fluctuation theory are less developed.\\
Another important gap is to apply the Boltzmann picture to gravity, or to problems of cosmology (via general relativity).   Fluctuation theory for geometric degrees of freedom comes, logically,  before we can develop a quantum theory of gravity.  We really have no statistical mechanics nor a good notion of entropy for fluctuating architectures, graphs or manifolds.  We need a good-working fluctuation theory of geometric degrees of freedom and we ought to understand notions such as geometric dissipation and activity, also to build geometric reservoirs. 

\item What does that teach us about biology?  What are the thermodynamic considerations in physics of life?  Does it make sense to ask for the entropy of an eye?  What is the entropy of a tiger?\\

Clearly, thermodynamics has an excellent record for understanding biological functioning as well.  At the least, the notions of energy, heat and work are relevant.  On the other hand, biology concerns (mostly) nonequilibrium phenomena and we should be aware that reasoning about {\it e.g.} reaction rates based on (free) energy landscapes has a limited validity.  It seems altogether not such a bad idea to introduce also there elements of nonequilibrium fluctuation and response theory.\\
In that same sense, understanding bio-realistic intelligence (or, associative memory to start) probably requires totally different ideas than used today in machine learning.  We cannot quite believe that the central principle of assembly, pattern formation and selection would be based on a cost function.  Variational evolution, such as (an even noisy) gradient flow, is probably not the mechanism to look for.\\
There is no urgency nor need to speak about the complexity of the eye or of the life of a tiger in terms of ``entropy.''  Complexity is different anyhow where it refers to the length of an algorithm or to the multitude of possible operational steps to achieve some result. In physics of life, kinetics and the time-symmetric fluctuation sector will be much more relevant than envisioned from the point of view of thermodynamics.  We cannot think of life as revealed by a state function or its change. 

\item What is next?\\

Even if not always explicit, much emphasis in
the study of nonequilibrium phenomena has gone to the study of the
entropy production, or to some nonequilibrium extension and
generalization of thermodynamic potentials. Nevertheless there are
reasons to doubt the unique relevance of the entropy production
concept, as traditionally understood, in far from equilibrium
setups. Similar thoughts have already been expressed half a century
ago, \cite{landauer}.
 The characterization of nonequilibrium could very well require to
consider observations that are somewhat foreign to equilibrium
thermodynamics.  Entropy production governs
 equilibrium fluctuations and remains useful for close-to-equilibrium
processes via the concept of heterogeneous equilibrium.  Yet, that
{\it hydrodynamic experience} is mostly related to the problem of
return to equilibrium. For all we can imagine, perhaps other
quantities must complement the entropy production to account for
other relevant nonequilibrium features that have to do not only
with dissipation but perhaps also with more constructive aspects
of the nonequilibrium kinetics and its
dynamical activity. 
\end{itemize}

\chapter{Onsager's picture}\label{onsp}
\lhead{C. Maes}
\rhead{Onsager's picture}

Lars Onsager has been a towering figure of statistical physics and belongs to a group of people (together with Einstein, Smoluchowski, Langevin, Kramers, Kirkwood, van Kampen and others) who emphasized the importance of considering a third level in the explanation of macroscopic behavior, the so-called mesoscopic scale.  The main point is to understand response of a macroscopic system from its dynamical fluctuations.  That embedding of macroscopic physics, especially response behavior, into stochastic dynamical processes on a mesoscopic level of description, is particularly relevant for nonequilibrium physics.

\section{Macroscopic equations}

While macroscopic equations describing the return to equilibrium have been conceived and applied even before the atomistic picture of matter was widely accepted, their derivation shows important mathematical and conceptual difficulties.  After all, hydrodynamic and thermodynamic behavior is described autonomously in only a few macroscopic variables and it must be understood how these variables get effectively decoupled from the many microscopic degrees of freedom. Moreover, in modifying the scale of description, the character of the dynamics could drastically change, from a unitary or a Hamiltonian to a dissipative evolution as possibly one of the most remarkable features\footnote{Perhaps the earliest example where that question became manifest is through d'Alembert's paradox (1752) for reconciling, what we call today, Euler's equation with that of Navier-Stokes. As d'Alembert wrote indeed, ``It seems to me that the theory (potential flow), developed in all possible rigor, gives, at least in several cases, a strictly vanishing resistance, a singular paradox which I leave to future Geometers to elucidate.''}. \\
In his microscopic derivation of irreversibility, Boltzmann already introduced a third scale of description in which the fluctuations are still visible, and he understood the macroscopic limit as a law of large numbers.  
A very famous next result was that of Lars Onsager in 1931 \cite{O1,O2}, showing that the linearized macroscopic evolution has a symmetry, called reciprocity and appearing in the matrix of the linear response coefficients as a consequence of microscopic reversibility. 
Again, fluctuations enter, as reversibility implies that the typical return path to equilibrium can be obtained from the small fluctuations away from that equilibrium.  That study was continued by Onsager and Machlup in 1953 on the same level of linearized hydrodynamics pioneering there the description in terms of probabilities on path-space \cite{OM}. In the 1950-60s, probably under the influence of or from the attraction by the quantum formalism, response theory was treated in terms of perturbation theory of operators.  The connection between response theory and  fluctuation theory on path-space was taken back up in the derivation of response relations around nonequilibrium, \cite{fdr}.

\section{Onsager reciprocity}\label{onsa}
\lhead{C. Maes}
\rhead{Reciprocity}

We examine a macroscopic system for which we have defined the relevant macroscopic variables, and their corresponding entropy $S$.  To fix the notation, while we think of the reduced variables in terms of
functions on phase space (of a very special and physical
character), possibly infinite-dimensional, we label the macroscopic states simply by $z\in \bbR^n$, and look at the concave function $S(z)$.  Assume that there is a unique sharp maximum
of the entropy, $S(0) = S_\text{eq}$ (for convenience) at
equilibrium $z_i=0$.  We then have a function $S(z) \leq
S_\text{eq} = S(0)$ of $n$ real variables. The first derivative of $S$ around its maximum (= equilibrium) vanishes. Therefore, we
take the entropy $S(z)$ as function of the macroscopic variables,
well-approximated around the equilibrium value $z=0$, by
\[
S(z) = S_\text{eq} + \frac
1{2}\sum_{i,j=1}^n\frac{\partial^2}{\partial z_i\partial
z_j}S(0)\,z_i\,z_j = S_\text{eq} - \frac 1{2}(z,{\bbS} z)
\]
where we have introduced the matrix ${\bbS}$, with matrix elements
\begin{equation}\label{cals}
{\bbS}_{ij} = -\frac{\id^2S}{\id z_i\id z_j}(0)
\end{equation}
The matrix ${\bbS}$ is a positive-definite symmetric matrix because of
the assumed concavity of the entropy.  \\
Now comes the Onsager idea (going back to Gibbs) of introducing thermodynamic forces. We define force as the first derivative
\begin{equation}\label{ofo}
F_i(z) = \frac{\partial}{\partial z_i}S(z) = -\sum_j {\bbS}_{ij} z_j
\end{equation}
hence linear in the displacements $z_j$ from the equilibrium values.
The role of the forces is to imitate Newton by writing
 \begin{equation}\label{linon}
\frac{\id}{\id t} z(t)  = \bbL \, F(t)
\end{equation}
a linear relation between the {\it currents} (or, rates of change or displacement) and the {\it forces},  where we define $\bbL$ as
the matrix of kinetic coefficients, or the matrix of linear
response coefficients.  We ignore the possible presence of an extra operator, like a spatial gradient, to connect ``displacement/change'' and ``current,'' since we do not want to specify here a spatial or chemical structure behind the states $z$. \\
 We assume that the above equations are approximately correct for times that exceed typical microscopic relaxation times, and close to equilibrium. We want to know more about the matrix $\bbL$ as it tells us about response, how forces create (currents or) displacements.\\

To start, we assume that the equation of motion of the reduced
variables is closed and linear with
\begin{equation}\label{go}
\frac{\id}{\id t} z_i(t)  = -\sum_j \bbG_{ij}\,z_j(t)
\end{equation}
for some matrix $\bbG$ (not necessarily symmetric).  On the other hand, by inverting \eqref{ofo}, we have $z = -{\bbS}^{-1}F$ as vector equality, and hence, \eqref{go} becomes
\[
\frac{\id}{\id t} z(t) = \bbG\,{\bbS}^{-1}
F(t)
\]
Hence, from \eqref{linon} we find that
\begin{equation}\label{conco}
\bbL = \bbG\,{\bbS}^{-1}
\end{equation}
for the linear response matrix.\\

Next, we set out to derive that $\bbL$ in \eqref{linon} is a symmetric matrix. Here is the brilliant idea of Onsager: the symmetry of $\bbL$ follows from statistical time-reversal invariance (borrowed from microscopic reversibility), by embedding the dynamics within (here linear) fluctuation theory.  (A nonlinear generalization is obtained in \cite{pons}.)\\
Suppose indeed a Markov process $X_t$ in $\bbR^n$ with transition kernel (density)
\[
\text{Prob}[X_t=y|X_0=x] = p_t(x,y)
\]
that describes fluctuations around the linear macroscopic evolution
\begin{equation}\label{mao}
\dot z_t = - \bbG z_t
\end{equation}
as we had it in \eqref{go}.
We assume the process $X_t$ is reversible with stationary distribution $\rho_\text{eq}$,
\begin{equation}\label{regr}
\rho_\text{eq}(x)\,p_t(x,y) = \rho_\text{eq}(y)\,p_t(y,x)
\end{equation}
and we also assume, in the spirit of small fluctuations around the law of large numbers, that this equilibrium distribution $\rho_\text{eq}(x)$ is Gaussian with covariance matrix ${\bbS}^{-1}$, having
\begin{equation}\label{gan}
\rho_\text{eq}(x) = \sqrt{\frac{\det \bbS}{(2\pi
k_B)^n}}\,e^{-(x,{\bbS}x)/2k_B}
\end{equation}
as a probability density with respect to $\id x_1\ldots \id x_n$.  We have the matrix $\bbS$ appearing as in \eqref{cals}.
Detailed balance \eqref{regr} (and microscopic reversibility) has also been called a regression
  hypothesis: it implies that the probability of a fluctuation $x\rightarrow y$ equals the probability of the relaxation $y\rightarrow x$, when $x$ is the typical (equilibrium) condition.\\
  We find back the macroscopic evolution \eqref{mao} as the evolution of the average $\langle X_t\rangle$ (which is the typical value in the macroscopic limit):
\[
z_t =  \langle X_t\,|\,X(0) =x\rangle,\quad \text{ or }z_t(x) = \big(\int\id y \text{ or }  \sum_y \big)\, y\,p_t(x,y)
\]
which relates the underlying stochastic process with the macroscopic variables.\\
We keep the  same conclusion as in \eqref{conco} of course, 
\begin{equation}\label{onsgr}
\dot z_t = \bbL \,\id S(z_t)
\end{equation}
with $\bbL = \bbG{\bbS}^{-1}$.  We proceed to show that $\bbL$ is symmetric. .\\

We consider a two-time correlation function under the fully time-reversible (equilibrium) dynamics:
\begin{equation}\label{eqt}
\langle X_i(0)\,X_j(t)\rangle_\text{eq} = \big\langle X_i(0)\;\langle X_j(t)\,|\,X(0)\rangle\big\rangle_\text{eq}
\end{equation}
by first conditioning on the initial value, which is then averaged out with the equilibrium distribution.
But $\langle X(t)\rangle_{X(0)=x} = [\exp - \bbG t]\,x$, or
for very small times $t$,
\[
\langle X_j(t)\,|\,X(0)=x\rangle = x_j - t\sum_k \bbG_{jk}x_k
\]
so that the  correlation function \eqref{eqt} to linear order in $t$ equals
\[
\langle X_i(0)\,X_j(t)\rangle_\text{eq} = \langle X_i(0)X_j(0)\rangle_\text{eq} - t\sum_k \bbG_{jk}\langle X_i(0) X_k(0)\rangle_\text{eq}
\]
The first term on the left-hand side is $({\bbS}^{-1})_{ij}$ and is symmetric.  The second term is $-t(\bbG {\bbS}^{-1})_{ji}$.  The proof is finished by imposing the time-reversal symmetry
\[
\langle X_i(0)\,X_j(t)\rangle_\text{eq} = \langle X_i(t)\,X_j(0)\rangle_\text{eq}
\]
 which allows switching the indices to get the symmetry of $\bbL = \bbG{\bbS}^{-1}$ as well.\\
The symmetry of the response matrix $\bbL$ is the Onsager reciprocity; it follows from time-reversal symmetry, in particular from the time-symmetry of correlation functions.    The evolution toward equilibrium is then linear gradient flow.  The symmetry of the matrix $\bbL$ is more generally (for nonlinear force-current relations) due to the convexity structure in the frenetic contribution to the pathspace action.  That was basically the result in \cite{pons}.\\

Remark that we have used that all the variables are even under time-reversal, and that the Hamiltonian is even in the momentum.  That restriction is not always satisfied (in the presence of magnetic fields or with Coriolis forces) and is indeed not needed.  Still, the Onsager relations change a bit into what is often called Onsager-Casimir relations, of the form
\[
\bbL_{ij}(B) = \varepsilon_i\varepsilon_j\,\bbL_{ji}(-B)
\]
where we indicate the dependence on the magnetic field $B$, and the $\varepsilon_i$'s are the parity ($\pm$) under kinematical time-reversal.

\begin{example}[Peltier {\it versus} Seebeck]
Currents interfere, as made explicit in the field of irreversible thermodynamics. More specifically, the Onsager symmetries connect currents that are  at first sight unrelated. We consider a well-known case.\\

We enter the physics of thermo-electric effects.  Think of a system consisting of two junctions through which two different metals are connected.   Each junction is characterized by an internal energy $U_1, U_2$, charges $e_1, e_2$, temperatures $T_1, T_2$ and electric potential $\varphi_1, \varphi_2$.  The total entropy, charge and energy is the sum for the two junctions:
\[
\id S = \frac 1{T_1}\, \id U_1 - \frac{\varphi_1}{T_1}\,\id e_1 +  \frac 1{T_2} \,\id U_2 - \frac{\varphi_2}{T_2}\,\id e_2
\]
Our system being closed, we have $\id e_1 + \id e_2 =0, \id U_1 + \id U_2 =0$ and so
\[
\id S = [\frac 1{T_1} - \frac 1{T_2}]\, \id U_1 + [\frac{\varphi_2}{T_2}- \frac{\varphi_1}{T_1}]\,\id e_1
\]
We have here the entropy $S=S(U_1,e_1)$ with $U_1,e_1$ as the reduced variables.  Taking the derivatives we obtain the thermodynamic forces:
 \[
 F_1 = \frac 1{T_1} - \frac 1{T_2},\quad F_2=\frac{\varphi_2}{T_2}- \frac{\varphi_1}{T_1}
\]
The linear current-force relations are
\[
\dot{U}_1=\bbL_{11}F_1 + \bbL_{12}F_2 = \bbL_{11}[\frac{1}{T_2}-
\frac{1}{T_1}] + \bbL_{12} [\frac{\varphi_2}{T_2}-
\frac{\varphi_1}{T_1}]
\]
and
\[
\dot{e}_1=\bbL_{21}F_1 + \bbL_{22}F_2 = \bbL_{21}[\frac{1}{T_2}-
\frac{1}{T_1}] + \bbL_{22} [\frac{\varphi_2}{T_2}-
\frac{\varphi_1}{T_1}]
\]
From Onsager reciprocity we know that $\bbL_{12}=\bbL_{21}$.
\end{example}

\subsubsection{Seebeck effect}
A difference in temperature between the junctions produces a difference in their electric potentials. 
To study it more closely we take the equations above and we put a small temperature difference $T_2 -  T_1=\delta$, making $F_1 \simeq \delta/T^2, F_2 \simeq E/T$ where $E$ is the difference in electric potential.  If there does not flow an electric current, then $\dot{e}_1=0$ and hence
\[
0 = \bbL_{21}\frac{\delta}{T^2} + \bbL_{22}\frac{E}{T}
\]
to find $E=a\delta$ for the Seebeck coefficient
\[
a =-\frac{\bbL_{21}}{\bbL_{22}}\frac 1{T}
\]
We read the above as $\delta$ creating $E$, which is the Seebeck effect.

\subsubsection{Peltier effect}
The Peltier effect shows that heat is generated by an electric current.  Imagine indeed that we produce an electric current $\dot{e}_1=j$ while we keep $T_1=T_2$.  Then,
\[
\frac{\dot{U}_1}{j}= \frac{\bbL_{12}}{\bbL_{22}}
\]
or
\[
\dot{U}_1 = b\, j,\qquad b= \frac{\bbL_{12}}{\bbL_{22}}
\]
with $b$ the Peltier coefficient.  We see that the Onsager reciprocity implies
the relation (called, Thomson relation because the earlier Lord Kelvin already tried to derive it in 1854),
\[
b=-aT
\]
A simple model dynamics for realizing these effects is contained in \cite{Maes2005}.

\section{Pathspace action}\label{psa}
\lhead{C. Maes}
\rhead{Action}

The work of Onsager-Machlup \cite{onsagermachlup,Onsager1953} introduced dynamical ensembles.  They were still restricted to Gaussian processes (as also used for the previous considerations) but they are the start of dynamical fluctuation theory.  Interestingly and quite different from static fluctuation theory, describing the probability of trajectories is not very different for nonequilibrium, nonlinear processes compared to the equilibrium and linear case.  In mathematical language, we speak about Girsanov transformations and Cameron-Martin expressions.  They are by now standard practise and we will not repeat here their derivation.  The upshot is that we can deal with a (sort of) Lagrangian (on spacetime configurations). That is taken up for its physical meaning in the next chapter.  We give here the (formal) example of Markov jump processes.

\begin{example}[Action for Markov jumps]
A continuous time Markov process on state space $K$ has piecewise constant trajectories $\omega=(x_t, t\geq 0)$ which are specified by giving the sequence $(x^{(0)},x^{(1)},x^{(2)},\ldots)$ of states together with the jump times
$(t_1,t_2,\ldots)$ at which the jumps, respectively, $x^{(0)}\rightarrow x^{(1)}, x^{(1)} \rightarrow x^{(2)}, \ldots$ occur; see Fig.~\ref{ju}.  We take the convention that $x_t$ is right-continuous, so that $x_{t_{i}+\varepsilon} = x_{t_i}$ for sufficiently small $\varepsilon$, while $x_{t_{i+1}-\varepsilon} = x_{t_i}$ for all $0<\varepsilon<t_{i+1}-t_i$, no matter how small.\\
To give the probability distribution over these possible paths means to give the distribution of waiting times $t_{i+1}-t_i$ when in state $x$, and to give the jump probabilities $x\rightarrow y$ when a jump actually occurs.  For the first (the waiting times) we take an exponential distribution with rate $\xi(x)$, {\it i.e.}, when in state $x$ at the jump time $t_i$, the time to the next jump is exponentially distributed as
\[
\mbox{Prob}[t_{i+1}-t_i \in [s,s+\id s]| x_{t_i}=x] = \xi(x) e^{-\xi(x)s}\,\id s
\]
Secondly, at the jump time $t_{i+1}$ the jump goes $x\rightarrow y$ with probability $p(x,y)$, for which we assume that $p(x,x)=0$.  We thus get
\[
 \xi(x_0)e^{-\xi(x_0)t_1} p(x_0,x_1) \,\xi(x_1)e^{-\xi(x_1)(t_2-t_1)} p(x_1,x_2)\ldots
\]
as probability density on path space $\Omega$. 
The products
\[
k(x,y) = \xi(x) p(x,y),\quad \mbox{with then }  \xi(x) = \sum_y k(x,y)
\]
are the {\bf transition rates} for the jumps $x\longrightarrow y$.  We can thus say that a path $\omega =(x^{(0)},t_1;,x^{(1)},t_2;\ldots,t_{n}; x^{(n)})$ over the time interval $[0,\tau]$ with $t_n< \tau\leq t_{n+1}$ has probability density
\begin{equation}\label{ppp}
k(x^{(0)},x^{(1)})k(x^{(1)},x^{(2)})\ldots k(x^{(n-1)},x^{(n)}) \,\exp \{- \int_0^\tau\xi(x_s)\id s\}
\end{equation}
The last integrand is
\[
\int_0^\tau\xi(x_s)\,\id s = \xi(x_0) t_1 + \xi(x_1)(t_2-t_1) + \ldots \xi(x_n) (\tau-t_n)
\]
for jump times $t_1<t_2<\ldots <t_n<\tau$.  These $\xi$'s are called {\bf escape rates}, quite appropriately.\\

The fact that the waiting times are exponentially distributed is equivalent with the Markov property, as we can see from the following argument.\\
Call $\cal T$ the waiting time between two jumps, say while the state is $x$.
For $s \geq 0$ the event ${\cal T > s}$ is equivalent to the event $\{x_u = x \text{ for } 0\leq u\leq s\}$. Similarly, for $s,t \geq 0$ the event ${\cal T >  s+t}$ is equivalent to the event $\{x_u = x \text{ for }
0\leq u \leq s+t\}$. Therefore,
\begin{eqnarray*}
\nonumber
 P[\cal T>s+t|\cal T>s]
&=& P[x_u = x \text{ for } 0\leq u \leq s+t| x_u = x \text{ for } 0\leq u \leq s]\\
\nonumber
&=& P[x_u = x \text{ for } s\leq u \leq s+t| x_u = x \text{ for } 0\leq u \leq s]\\
\nonumber
&=& P[x_u = x \text{ for } s\leq u \leq s+t| x_s = x]\\
\nonumber
&=& P[x_u = x \text{ for } 0\leq u \leq t| x_0 = x]\\
&=& P[\cal T>t]
\end{eqnarray*}
Thus, the distribution of $\cal T$  is memoryless, which means that it is exponential.\\

Finally, we observe that the path-probability \eqref{ppp} can be informally written as 
\[
\text{Prob}[\omega]\propto e^{-{\cal A}[\omega]}
\]
for ``action''
\begin{equation}\label{acj}
{\cal A}[\omega] := -\sum_{t_i} \log k(x^{(i-1)},x^{(i)}) + \int_0^T\xi(x_s)\id s
\end{equation}
where the first sum is over the jump-times.  Such formul{\ae} are useful for various physical considerations from comparing actions, hence probabilities, when parameters of the process are being changed.  
\end{example}

The above example plays on mesoscopic scales and can be repeated for diffusion processes.  There, white noise replaces the Poisson noise and serves as basic reference for building probabilities on pathspace.  However, also for fluctuations around macroscopic evolutions, there is a pathspace formulation.  That is the subject of macroscopic dynamical fluctuation theory, or how to express the asymptotics of the $\log-$probabilities of trajectories.  It turns out that, under the condition of local detailed balance, there is a canonical structure.  We postpone the discussion to the following chapter.

\chapter{Nonequilibrium structures}
\lhead{C. Maes}
\rhead{}
We turn to the study of stationary processes that are driven away from equilibrium following the prescription of local detailed balance.  We assume that the system is coupled to at least one heat bath, separated from the rest of the environment, in thermal equilibrium at fixed temperature $T$.\\

We elaborate by starting from the condition of local detailed balance. We enter then dynamical fluctuation theory and probabilities on pathspace (as continuation of the previous section).  Path-dependent entropy fluxes dominate the time-antisymmetric fluctuation sector.  However, as we discover by searching nonlinear structures beyond the linear regime around equilibrium, there appears another quantity, the frenesy, that characterizes the time-symmetric fluctuation sector, and becomes a crucial complement for the formulation of response theory.

\section{Local detailed balance}\label{dlbb}
\lhead{C. Maes}
\rhead{Local detailed balance}

What makes a good or well-motivated physical model on some reduced level of description? Those questions were first thoroughly discussed like seventy years ago, for instance in \cite{Lebowitz1956Thesis,leb55}. To obtain a self-consistent open system dynamics we need to enter into some limiting regimes, {\it e.g.}, of weak coupling where the interactions between system and reservoir are small and the reservoir has itself fast relaxation behavior to allow time-scale separation. In the end, we may wish to describe the thermodynamic changes (in energy, particle number and entropy) in the equilibrium reservoirs entirely in terms of the system trajectories. There are methods (such as weak coupling or van Hove limits) that take care of that and give rise to an effective reduced dynamics from more microscopic beginnings. However, we do not want to repeat such derivations in each case.  Rather we rely on and we apply a number of qualitative and general features of a reduced dynamics that give constraints on all our effective modeling.  That implies we need a brief discussion on the condition of local detailed balance.\\

One important feature of an open system dynamics is the relation of time-reversal with entropy changes, \cite{maes2002timereversalentropy,DeRoeck2006,Callens2004,JacobsMaes2005}.  As a matter of fact, already for contact with a single thermal bath, there is the condition of (global) detailed balance which arises from the microscopic time-reversibility underlying Hamiltonian dynamics, and it can be expressed in terms of the entropy change in the bath; recall Section \ref{onsa}.\\

The next simple idea is that when such equilibrium reservoirs are well-separated, we should impose a {\it local} detailed balance, meaning that we identify what changes in the system refer to entropy changes in what bath. Local has then a spatial meaning, using that the updating rules in the neighborhood of one bath depend on the thermodynamic parameters of that same bath, and not on the others.  There are derivations of that local detailed balance condition, and we refer to \cite{kls,leb55,leb57,xing2025,derrida,hal,chi,time,ldb} for various contexts.  

Also without spatially extended architecture, local detailed balance can be imposed by leaving the location of the entropy change unspecified.  
The hypothesis of ``local'' detailed balance then identifies the inherent asymmetry of Markov transition rates with entropy changes. In the case of Markov jump processes, local detailed balance states that for every single transition $x \rightarrow y$ between two configurations of the system, the antisymmetric component of the logarithm of the transition rate $k(x,y)$ corresponds to the increase of entropy in the environment,
\begin{equation}\label{forldb}
  k_B \log\frac{k(x,y)}{k(y,x)} = \Delta S_\text{env}(x \rightarrow y)\,.
\end{equation}
The environment may consist of several reservoirs not at mutual equilibrium; only one of them may assist in each transition. 
The entropy change in the reservoir is determined from the heat and the particle flow. {\it E.g.}, for a ``heat'' bath at temperature $T$ involved in the transition $x\rightarrow y$, $\Delta S_\text{env}(x \rightarrow y) = \Delta Q(x \rightarrow y) / T$ with $\Delta Q$ the heat ``leaving'' the system, where we apply the Clausius relation to that reservoir.  The condition \eqref{forldb} can be installed for diffusions and for systems with memory as well; see for instance \cite{gle1}.  A difficulty (especially for mathematicians) with \eqref{forldb} is that it does not appear to restrict the mathematics.  Indeed, it actually connects the mathematical formalism with a nonempty physical condition to start a sensible application of path-dependent thermodynamics on mesoscopic scales.  Speaking about entropy and heat flux is not for free, and the assumption of the corresponding physics is at the heart of conditions like \eqref{forldb}.\\

The by now well-known fluctuation symmetries
of the entropy production, be it transient or in the steady state,
are on a formal level nothing but expressions of that relation
between entropy production and time-reversal breaking\footnote{Those fluctuation relations are giving rise to identities between correlation functions and derivatives which, formally, follow the idea of Ward-Takahashi identities in quantum field theory, \cite{ward1950identity,jackiw1997evolution}.}.  That point
was especially emphasized in \cite{crooks,mn,poincare}.  The fundamental reason for local detailed balance is the
time-reversibility of an underlying microscopic dynamics over which
our effective stochastic model is presumably built. Hence, violating
such a condition reduces the physical interpretation of our
stochastic model. \\

If the system is coupled to a single heat reservoir and driven by conservative forces with potential $E(x)$ then $\Delta S_\text{env}(x \rightarrow y) = [E(x) - E(y)] / T$ as in the Clausius relation, yielding the usual ``(global) detailed balance'' condition of ``equilibrium'' dynamics,
\[ 
  e^{-\beta E(x)} k(x,y) = e^{-\beta E(y)} k(y,x)\,,\qquad \beta = \frac{1}{k_B T}\,.
\] 
The general ``global'' detailed balance condition requires that $\sum_{i=0}^n \Delta S_\text{env}(x_i \rightarrow x_{i+1}) = 0$ for any ''closed'' sequence of configurations $x_0 \rightarrow \ldots \rightarrow x_n \rightarrow x_{n+1} = x_0$.  Equivalently, $\Delta S_\text{env}(x \rightarrow y) = \psi(y) - \psi(x)$  for some function $\psi$ on configurations. ``Global'' detailed balance is violated under nonequilibrium conditions, {\it e.g.}, (i) if the system is driven by nonconservative forces, or (ii) when different transitions are assisted by different reservoirs which are not mutually at equilibrium.\\

Finally, note that the notion of ``local equilibrium'' (mostly used in macroscopic considerations such as for hydrodynamics) has not much in common with ``local detailed balance'' (which is used for mesoscopic dynamics, connecting time-reversal breaking with entropy production).  The latter allows strong spatially-local deviations from equilibrium. The dissipation length for a nonequilibrium system on which a driving force $F$ acts while coupled to a heat bath at inverse temperature $\beta$ can be defined as $\ell_\text{diss} = (\beta\, F)^{-1}$.  Interesting nonequilibrium effects are often associated with the regime where $r\lesssim \ell_\text{diss} \ll L$ where $L$ is a global length scale (like the system size), and $r$ is a local length (like the interaction range). For local detailed balance, there can be strong local gradients and/or strong external driving at short scales.  Local equilibrium covers the opposite regime\footnote{Local equilibrium is also different from close-to-equilibrium, and not only because local equilibrium presupposes a spatial extension and the close-to-equilibrium condition is global.  Local equilibrium is for instance compatible with large Reynolds number, where inertial forces are much greater than viscous forces. It allows chaotic motion and turbulence in high-speed or large-scale fluid motion. Close-to-equilibrium, a notion which is (only) well-defined under the condition of local detailed balance, the system's behavior is dominated by the linear approximation of the equations around an equilibrium point, while not avoiding long-range correlations.}. It fails {\it e.g.} when there are shock waves and ballistic transport plays a big role.   However, both local equilibrium and local detailed balance enable an uncomplicated way of speaking about entropy fluxes and entropy production, and there are regimes and regions where they do overlap: imagine {\it e.g.}, a metal rod with one end being heated: the overall rod is not in equilibrium, but a tiny section of it near the hot end is, and can be described locally by its own temperature.

\section{Decomposition based on time-reversal symmetry applied to response}\label{pif}
\lhead{C. Maes}
\rhead{Response decomposition}

Response theory is not merely a matter of writing Taylor expansions. It is straightforward to derive formal relations between a stimulus (small perturbations of a nonequilibrium condition) and the corresponding response, namely the induced change in the system’s evolution. Instead, what is required is an operational and physically transparent understanding, together with a systematic way of addressing different scenarios. In particular, we seek formulas expressed in terms of measurable quantities and a general framework that guides how response should be conceptualized. From this perspective, the most natural approach to response theory for nonequilibrium systems is provided by the theory of dynamical ensembles.\\ 
Dynamical ensembles are central tools in (mesoscopic) nonequilibrium statistical mechanics and are represented by weights Prob$[\omega]$ on pathspace, as we already had in Section \ref{psa}. The path $\omega$ is a trajectory of the system (on some reduced level of description) over a time-interval $[0,t]$.   This path-probability or -density, or path-weight, depends on the parameters of driving and reservoirs and would generally change when a perturbation is added to the system.  We compare
the path-weights with a reference process and associate an action ${\cal A}(\omega)$ to each trajectory $\omega$ via 
\begin{equation}\label{sqa}
\text{Prob}[\omega]=e^{-{\cal A}(\omega)}\,\text{Prob}_0[\omega]
\end{equation}
with Prob$_0(\omega)$ is the weight of the same path for the reference process\footnote{Mathematically speaking, the two path-probabilities, perturbed and reference, must be absolutely continuous, {\it i.e.}, compatible in their designation of what events have 0/1 probabilities.}. Let us think of a generic perturbation $h \to h+\id h$ which changes Prob$^h[\omega]$ to Prob$^{h+\id h}[\omega].$   This is how these days we start response theory around nonequilibrium conditions, \cite{resp,prlresponse}.

The change in expectations for a path-observable $O$ due to the perturbation is now conveniently expressed from \eqref{sqa} as,
\begin{eqnarray}
&& \langle O(\omega) \ra^{h+\id h} -\langle O(\omega) \ra^{h}  \cr 
&&=  \int \id\omega \,\text{Prob}[\omega] 
\left(e^{-A_{h+\id h}(\omega)}-e^{-A_{h}(\omega)}\right) O(\omega) 
\end{eqnarray}
where  the notation is purely formal (but hopefully suggestive).
For small perturbations $\id h$ this leads to a general differential response formula,
\begin{equation}
\frac \id{\id h} \langle O(\omega)\rangle^h = - \left \langle O(\omega)\frac {\id}{\id h} {\cal A}_h(\omega)  \right\rangle ^h \label{eq:dQA}
\end{equation}
where the right-hand side is a correlation function in the unperturbed process. 

Ignorant in the given generality of other symmetries, it is useful to decompose the action into two components by using the involution $\theta$ on pathspace that time-reverses the trajectory.  We then write\footnote{Beware --- here, $\cal F$ does not denote a free energy.}, 
\[
\cal A_h(\omega) = {\cal F}_h(\omega) - \frac 12 {\cal S}_h(\omega)
\]
where ${\cal S}_h(\omega)= -{\cal S}_h(\theta\omega)$ is the time-antisymmetric excess in entropy flux associated with the trajectory $\omega$, and the time-symmetric part is the excess frenesy ${\cal F}_h(\omega)= {\cal F}_h(\theta\omega)$.  The reason for ${\cal A}_h(\theta\omega) - {\cal A}_h(\omega) = $ entropy flux (per $k_B$)  is local detailed balance (of the previous section), 
\[
\log \frac{\text{Prob}[\omega]}{\text{Prob}[\theta\omega]} = \text{entropy flux over } \omega  \text{ per } k_B
\]
After the decomposition, the response \eqref{eq:dQA}  takes the form 
\begin{equation}
\frac \id{\id h} \langle O(\omega)\rangle^h = \frac 12 \left \langle O(\omega)\frac {\id }{\id h}{\cal S}_h(\omega)  \right\rangle^h - \left \langle O(\omega)\frac {\id }{\id h}{\cal F}_h(\omega)\right\rangle^h  \label{eq:dQSD}
\end{equation}
That is a very general relation.\\

Let us check what that becomes when the reference is an equilibrium process.  Note first that taking $O\equiv 1$ in \eqref{eq:dQSD}
implies that 
\begin{equation}
\frac 12 \left \langle \frac {\id }{\id h} {\cal S}_h(\omega)  \right\rangle^h_x  = \left \langle \frac {\id }{\id h} {\cal F}_h(\omega)\right\rangle^h_x  \label{eq1:dQSD}
\end{equation}
for all initial conditions, {\it e.g.},  fixing the initial state of the system to be $x$.  To work our way towards the Kubo formula  (linear response formula around equilibrium) we take the observable $O(\omega) = O(x_t)$ to depend only on the state at the final time $t$.  In the equilibrium process (say at $h=0$),
\begin{eqnarray}\label{linma}
\left\langle O(x_t)\frac {\id }{\id h}{\cal F}_h(\omega)\right\rangle_{\text{eq}} &=& \left \langle O(x_0)\frac {\id }{\id h}{\cal F}_h(\omega)\right\rangle_{\text{eq}} = \frac 12 \left \langle O(x_0) \frac {\id }{\id h}{\cal S}_h(\omega)  \right\rangle_{\text{eq}} \nonumber\\
&=& -\frac 12 \left \langle O(x_t) \frac {\id }{\id h}{\cal S}_h(\omega)  \right\rangle_{\text{eq}}
\end{eqnarray}
where the first and last equalities follow from the time-reversal invariance of the equilibrium process, and the middle equality follows from \eqref{eq1:dQSD}.  The last identity can be substituted in \eqref{eq:dQSD} to conclude that, with ${\cal S}'_0 = \id {\cal S}_h/\id h (h=0)$,
\begin{equation}
\frac \id{\id h} \langle O(\omega)\ra^h|_{h=0} = \left \langle O(x_t)\,{\cal S}'_0(\omega)  \right\rangle_{\text{eq}}  \label{eq2:dQSD}
\end{equation}
which, we claim, is exactly the classical Kubo formula\footnote{Since we are dealing here with open systems and mesoscopic dynamics, we do not try here to make the bridge with the correct Kubo formulas that relate, in the end, to correlation functions in the equilibrium microcanonical ensemble.}.    Indeed, it is but we need to be more specific, as we do next.
The fact that for the response there appears a universal correlation in the equilibrium process between the observable and the entropy flux ${\cal S}'_0(\omega)$ makes this relation also known under the heading of a fluctuation--dissipation theorem.\\ 

To apply the above we need to specify ${\cal S}_h(\omega)$ and ${\cal F}_h(\omega).$ Let us take the example of Markov jump processes (as in the example above), with transition rates  $k(x,y)$.  Escape rates are $\xi(x) = \sum_y k(x,y)$.  Paths $\omega$ are piece-wise constant\footnote{They are, by convention, continuous to the right and having left limits, which reverses under time-reversal. That is of little or no relevance.} with jumps at times $s_i$ and have weight (as in \eqref{ppp}),
\begin{equation}
\text{Prob}[\omega] \propto \mu_0(x_0)\prod_{s_i} k(x_{s_i},x_{s_{i+1}}) e^{-\int_0^t \xi(x_s) \id s}
\end{equation}
for initial distribution $\mu_0(x_0).$ To write the action $\cal A_h(\omega)$, we need to choose a reference process. It is easy to show that the final response formula does not depend on this choice. So, for our purpose we take the simplest reference process defined by $k_0(x,y)=1$ iff $k(x,y) \ne 0.$  Then, as in \eqref{acj},
\begin{equation}
{\cal A}(\omega) = -\sum_{s_i} \log k(x_{s_i},x_{s_{i+1}}) + \int_0^t ds[\xi(x_s) -\xi_0(x_s)] \label{eq:A}
\end{equation}
 The entropy flux and frenesy associated with trajectories can be identified as the time antisymmetric and symmetric components of ${\cal A}(\omega).$ Denoting the time--reversed trajectory as $\theta \omega$,
\begin{eqnarray}
{\cal S}_h(\omega) &=& {\cal A}(\theta \omega) - {\cal A}(\omega)  \cr
{\cal F}_h(\omega) &=& \frac 12 [{\cal A}(\theta \omega) + {\cal A}(\omega)] \label{eq:ShDhbis}
\end{eqnarray}
from which follows, 
\begin{eqnarray}
{\cal S}_h(\omega) &=&  \sum_i s(x_{s_i},x_{s_{i+1}}) \cr
{\cal F}_h(\omega) &=& \int_0^t \id s \, \xi(x_s) - \sum_{i} \log \psi(x_{s_i},x_{s_{i+1}})  
\end{eqnarray}
 Note that $\int \id s\, \xi_0(x_s)$ in \eqref{eq:A} can be ignored for differential response, as it does not depend on $h$.\\

For diffusion processes we take the example of a colloid with position ${\vec r}_t\in {\mathbb R}^3$ and motion following
 \begin{equation}
 \dot{\vec r}_s = \chi \, \vec F({\vec r}_s) + \sqrt{2k_BT\,\chi}\,\vec \xi_s,\qquad {\vec \xi}_s=\mbox{ standard white noise vector}\label{overd}
 \end{equation}
The mobility $\chi$ is a positive $3\times3-$matrix that for simplicity we choose not to depend on position here. It implies that in the frenesy, only the escape rates change when we change $F$ with respect to a reference choice, {\it e.g.}, by putting
\[
\vec F(\vec r) = h\,\vec f(\vec r)  + \vec g(\vec r)
\]
where $\vec f$ and $\vec g$ are vector functions.  The constant $h$ is a parameter and we let $h=0$ to correspond to the reference dynamics.
We want the excess frenesy and entropy flux per $k_B$ for $h\neq 0$, as defined above. Recall that ${\vec \xi}_s,s\in [0,t],$ is (formally) a stationary Gaussian process whose weights carry over to the trajectory via the quadractic form
 \begin{equation}\label{stc}
\frac 1{2}{\vec \xi}_s \cdot {\vec \xi}_s =  [\dot{\vec r}_s - \chi \, \vec F({\vec r}_s)]\cdot\frac 1{4k_BT\,\chi}\,[\dot{\vec r}_s - \chi \, \vec F({\vec r}_s)]
	\end{equation}
To obtain the action ${\cal A}$, that must be integrated over time $s\in [0,t]$ after which we must take the difference between the expressions for $\vec F = \vec g$ and for $\vec F=\vec g + h\vec f$.\\
At the same time, we must be clear about the stochastic integration.  In \eqref{stc}, It\^o-integration must follow, which is not symmetric under time-reversal. It is useful to change to Stratonovich-integration therefore.  That rewriting uses the relation
\begin{equation}\label{is}
\int_0^t \vec G({\vec r}_s)\circ \id {\vec r}_s = \int_0^t \vec G({\vec r}_s)\,\id {\vec r}_s + k_BT\int_0^t(\chi \nabla) \cdot \vec G({\vec r}_s)\,\id s
\end{equation}
for general smooth functions $G$, 
that connects for \eqref{overd} the Stratonovich-integral (left-hand side) to the It\^o-integral (first term on the right-hand side).\\
Using that the Stratonovich-integral $\int_0^t \vec f({\vec r}_s)\circ \id {\vec r}_s$ is antisymmetric under time-reversal, the result for \eqref{overd} is
\begin{eqnarray}\label{frda}
  {\cal F}(\omega) &=& \frac{h^2\beta}{4}\int_0^t \id s\, \vec f\cdot \chi \vec f + \frac{h\beta}{2}\int_0^t \id s \,\vec f\cdot \chi\,\vec g + \frac{h}{2} \int_0^t \id s\, \chi\nabla\cdot \vec f\label{fres}\\
  \cal 
S(\omega) &=& h\,\beta\int_0^t \id {\vec r}_s \circ \vec f({\vec r}_s)\label{se}
\end{eqnarray}
The entropy flux $S(\omega)$ per $k_B$ due to the perturbation is time-extensive, being $\beta$ times the work done by the nonconservative
 force as given in \eqref{se}.  It is the Joule-heat divided by $k_BT$.  The highest order in the excess parameter $h$ appears in the frenetic part.  Indeed, frenesy matters more and more at larger excesses, farther from equilibrium.\\
 
  When ${\vec f}=\nabla V$ is conservative, then the second and third term in $\cal F$ (the linear part of the frenesy \eqref{fres}) add up to become proportional to the time-integral of the backward generator $L$ acting on $V$:
\begin{eqnarray}\label{kons}
\text{for } \vec f &=& \nabla V,\qquad \vec f\cdot \chi\,\vec g +  k_BT\chi\nabla\cdot \vec f=  LV\notag\\
{\cal F}(\omega) &=& \frac{h^2\beta}{4}\int_0^t \id s\, \vec f\cdot \chi \vec f + \frac{h\beta}{2}\int_0^t \id s \,LV(\vec r_s)\\
S(\omega) &=& h\beta[V({\vec r}_t)-V({\vec r}_0)]\notag
\end{eqnarray}
for the backward generator $Lu = \nabla u\cdot \chi \vec g + k_BT(\chi\nabla) \cdot \nabla u$ (on a function $u$) of the reference dynamics.  On the other hand, the entropy flux \eqref{se} becomes a time-difference.\\
    We can also specify to the case where $\vec g = -\nabla U$ and $\vec f$ being the nonconservative (or rotational) part of the force $\vec F$.  The reference dynamics ($h = 0$) satisfies the condition of detailed balance (time-reversibility). 
The excess frenesy \eqref{frda} now equals
\begin{equation}\label{fred}
  \cal F(\omega) = \frac{h^2\beta}{4}\int_0^t \id s\, \vec f\cdot \chi\vec f - \frac{h\beta}{2}\int_0^t \id s \,\vec f\cdot \chi\nabla U + \frac{h}{2} \int_0^t \id s\, \chi\nabla\cdot \vec f
\end{equation}
 For more examples and theory, we refer to \cite{upd,Maes2020} with some typical results highlighted in Section \ref{sue}.  We see that trajectory-based analysis as a proper complement to more (functional) analytic approaches that probably and to some extent appear to have their origin in the strange quantum mood of the 1950-70s where it was celebrated that quantum mechanics is not to be formulated in terms of real trajectories.\\
 Finally, we note that the above response formulas are essential ingredients to derive (not only fluctuation-dissipation relations like giving mobilities in terms of diffusion and dynamical activity, \cite{proS,neg,fren1}, but also) extensions of fluctuation-dissipation relations of the second kind; see {\it e.g.} \cite{maes2014,stef,maes2020fluctuating,Tanogami2022,pei2026transferActiveMotion,beyen2025couplingelasticstringactive}.

\section{Thermokinetic aspects of driven diffusions}\label{therk}

Heat and entropy production for an ensemble of independent colloids subject to a Markov diffusion process can be described for a class of
$d-$dimensional inhomogeneous diffusions \begin{equation}\label{gsd1}
  \dot x_t = \bigl\{\chi(x_t) \bigl[ F(x_t) -
   \nabla U(x_t) \bigr] + \nabla\cdot D(x_t)\bigr\}
  + \sqrt{2 D(x_t)}\, \xi_t
\end{equation}
to be interpreted in the It\^o way for independent standard Gaussian white noise $\xi_t$.
The mobility $\chi(x) = \be D(x)$ is a positive $d
\times d-$matrix which is Einstein-related to the diffusion matrix $D(x)$ via the inverse
temperature $\be > 0$. $\nabla \cdot D$ is a vector with components
$\sum_\ga \partial_\ga D_{\ga \al}$.\\
We may think of either periodic (the particle moves on the unit torus
$[0,1)^d$ and the fields $U$, $F$,
and $\chi$ are smooth functions on the torus), or on the infinite space when the potential $U$ grows fast enough at infinity so that the
particle is essentially confined to a bounded
 region, and the density and its derivatives vanish at infinity.

The evolution of the density profile $\mu_t$ of a
macroscopic amount of independent colloids each moving according
to~\eqref{gsd1} is given by the
Fokker-Planck equation,
\begin{equation}\label{fk1}
  \frac{\partial\mu_t}{\partial t} + \nabla\cdot j_{\mu_t} = 0,\qquad
  j_\mu = \chi \mu \,(F - \nabla U) - D\nabla\mu
\end{equation}
The stationary density $\rho$ satisfies
$\nabla \cdot j_\rho = 0$.  At time $t\geq 0$ and for all functions $f$,
\begin{equation}\label{str-identity0}
  \Bigl\langle  f(x_t) \Bigr\rangle_{\mu_0}
  = \int f(x)\,\mu_t(x)\,\id x
\end{equation}
where $\langle\cdot\rangle_{\mu_0}$ is the average over the
process started with density profile $\mu_0$.\\
Moreover, using a Stratonovich-stochastic integral,
\begin{equation} \label{str-identity}
  \Bigl\langle \int_0^\tau f(x_t) \circ \id x_t \Bigr\rangle_{\mu_0}
  = \int_0^\tau \id t \int f(x)\, j_{\mu_t}(x)\,\id x
\end{equation}
which means that $j_\mu$ gives the profile of the
particle current at given density $\mu$.  Taking $f(\,\cdot\,) = \de(\,\cdot\,-x)$, we indeed find that
$\mu_t$ and $j_{\mu_t}$ are the time-dependent (or transient) local
density, respectively the local current for that density\footnote{For computational reasons, we may be interested in the score $\nabla \log \rho$ which can be used to abbreviate some formulas here and below.  However, the stationary distribution $\rho$ is in general unknown and with little operational meaning; it does play a role in the game of time-reversing but the conceptual framework of relating time-reversal with entropy production is thermodynamically supported by local detailed balance only, as is illustrated next.} 

\subsection{Entropy production}

The entropy of a closed macroscopic system is defined via counting
microstates, giving the entropy as the fluctuation functional governing static fluctuations and verifying
a Second Law inequality, discussed in Chapter \ref{bpic}. For our mesoscopic system, we think of an
ensemble of $N \to \infty$ independent copies. A density $\mu$
then becomes a macro-observable telling us the relative occupation
of the space, and the associated counting entropy equals the
relative entropy (with respect to the flat distribution):
\begin{equation}\label{static-ent}
  s(\mu) = -\int \mu(x) \log \mu(x)\,\id x
\end{equation}
By construction it is the (static) fluctuation functional in the
probability law for observing the empirical density $\mu$ when
sampling the particles from the flat distribution. For $F = \nabla
U = 0$ (i.e.\ for a thermodynamically closed system) the entropy
rate satisfies $\id s / \id t \geq 0$ along the solution of the
Fokker-Planck equation.\\
Yet, our system is open and dissipates heat that results in an extra entropy
production in the environment.\\

To start, we avoid notational problems and treat
\begin{equation}\label{m2}
	\gamma \dot{x}_t = \cal G(x_t) + \sqrt{2\gamma k_{B} T}\, \xi_t,
\end{equation}
where the force on $x$ is $\cal G(x) = F(x)-U'(x)$. 
From  It\^o's Lemma,
\begin{eqnarray}\label{dup}
U(x_t+\id x_t) - U(x_t) &=& U'(x_t)\id x_t + (1/2) U''(x_t) (\id x_t)^2
\nonumber\\ &=&\mu[(F(x_t) -  U'(x_t))U'(x_t)+ k_B T U''(x_t)]\id t \nonumber\\ &&+\sqrt{2\gamma k_B T}\,U'(x_t)\, \xi_t\id t \label{eq:itoU}
\end{eqnarray}
with $\mu=1/\gamma$, and  $\xi_t\id t =\id B_t$ for the standard Wiener process $B_t$. As a consequence, the expected energy change $\dot{\cal{U}}(x;t)$ per unit time at time $t$, when $x_t = x$, equals
\begin{equation}
    \dot{\cal{U}}(x) = \mu[(F(x) -  U'(x))U'(x)+ k_{B} T U''(x)].
\end{equation}
On the other hand, the power  $ \dot{\cal{W}}(x;t)$ exerted on the system at time $t$ when  $x_t = x$, is 
\begin{align}\label{heateq}
\dot{\cal{W}}(x) =& \langle F (x)\circ \dot x\rangle \nonumber\\ 
=&\langle F (x) \dot x\rangle +\mu k_B T F' (x) \nonumber \\
=&\mu[ {F}^2 (x) -  F (x)\,U'(x) + k_{B}T F'(x)] .
\end{align}
Finally, the expected heat flux to the bath $ \dot{\cal{Q}}(x)$ is obtained from applying the First Law, $\dot{\cal{Q}}(x) =\dot{\cal{W}}(x)-\dot{\cal{U}}(x) $:
\begin{align}\label{heateq}
\dot{\cal{Q}}(x) 
 = \mu \Big[  ( F(x)-U'(x))^2 + k_B T( F'(x) -U''(x)) \Big].
\end{align}
or 
\[\dot{\cal{Q}}(x;t) = \mu [ {\cal G}^2(x) + k_{B}T {\cal G}'(x)]
\]
For getting it as function of the density profile, the rate $\dot{\caQ}(\mu)$ of mean heat
dissipation comes from the work performed by the nongradient force
and from the change of the energy of the system:
\begin{equation}
\begin{split}
\dot{\caQ}(\mu) &= \int F \cdot j_\mu\,\id x
  - \frac{\id}{\id t} \int U\,\mu\, \id x
\\
  &= \int (F - \nabla U) \cdot j_\mu\,\id x + \int \nabla \cdot (U j_\mu)
  \,\id x
\end{split}
\end{equation}
where we have used the Fokker-Planck equation~\eqref{fk1}. By the boundary conditions, the last integral equals zero.\\ 
The environment is a heat reservoir at inverse temperature $\be$
that remains itself at equilibrium during the whole process (the
weak coupling assumption). Hence, the mean entropy flux equals
$\be \dot{\caQ}(\mu)$ and the total entropy production rate reads
\begin{equation}\label{ent-prod}
\begin{split}
  \si(\mu) &= \be \dot{\caQ}(\mu) + \frac{\id s}{\id t}(\mu)
\\
  &= \int (\be F - \be\nabla U - \nabla \log \mu) \cdot j_\mu\,\id x
\\
  &= \int j_\mu \cdot (\mu D)^{-1} j_\mu\, \id x
\end{split}
\end{equation}
where we have used again the Fokker-Planck equation~\eqref{fk1}
and the boundary conditions. In the language of irreversible
thermodynamics, $\be F - \nabla \log (\mu\, e^{\be U})$ is a
generalized thermodynamic force conjugated to the current $j_\mu$.
Clearly, $\si(\mu) \geq 0$ in agreement with the second law,
proving the thermodynamic consistency of our diffusion model.  It is thanks to the (local) Einstein relation that we get local detailed balance.  We explain:\\

The density of the path-distribution
$\bsP_{\mu_0}$ of the process with respect to the reference
process $\bsP^0$ where $F =
\nabla U = 0$, equals
\begin{equation}\label{path-distribution}
  \id\bsP_{\mu_0}(\om) = \mu_0(x_0)\,e^{-\int_0^T \caL^+(x_t)\,\id t
  - \int_0^T \caL^-(x_t,\id x_t)}\id\bsP^0(\om)
\end{equation}
where $\om= (x_t)_{t=0}^T$ is a trajectory and $\mu_0$ is an
initial density.\\
Following \eqref{pif}, we split the action in a
time-symmetric part,
\begin{align}\label{lagr-plus}
  \caL^+(x) &= \frac{\be^2}{4} [ (F - \nabla U) \cdot D(F - \nabla U) ]
  + \frac{\be}{2} \nabla \cdot [ D(F - \nabla U) ]
\\\intertext{and a time-antisymmetric part}
\label{lagr-minus}
  \caL^-(x,\id x) &= -\frac{\be}{2} (F - \nabla U) \circ \id x
\end{align}
Applying the identity~\eqref{str-identity},
\begin{equation}
  -\Bigl\langle \int_0^T \caL^-(x_t,\id x_t) \Bigr\rangle_{\mu_0}
  = \frac{\be}{2} \int_0^T \id t\int (F - \nabla U) j_{\mu_t}\,\id x
  = \frac{\be}{2} \int_0^T \dot\caQ(\mu_t)\,\id t
\end{equation}
given in terms of the entropy flux. This is again an
instance of a fluctuation symmetry, \cite{gibbsian,mn,jmp}, which, finally, is always based on local detailed balance\footnote{With conclusion that for Markov diffusions, local detailed balance is directly obtained by imposing the traditional Einstein relation between friction and noise amplitude, as done in \eqref{gsd1}.}\\

\subsection{Traffic}

 The
time-symmetric contribution is the frenesy with expectation computed analogously
via~\eqref{str-identity0} to obtain
\begin{equation}
  \Bigl\langle \int_0^T \caL^+(x_t)\,\id t \Bigr\rangle_{\mu_0}
  = \int_0^T \caT(\mu_t)\,\id t
\end{equation}
in terms of the traffic $\caT(\mu)$, which equals
\begin{equation}\label{traffic-def}
  \caT(\mu) = \frac{\be^2}{4} \int \mu(F - \nabla U) \cdot D(F - \nabla U)\,\id x
  + \frac{\be}{2} \int \mu \nabla \cdot [ D(F - \nabla U)]\,\id x
\end{equation}
A crucial feature of diffusion systems, not valid in general
beyond the diffusion approximation, is that the
traffic~\eqref{traffic-def} and the entropy
production~\eqref{ent-prod} are not independent.
The reason is the relation
\begin{equation}\label{ent-traffic}
  \caT(\mu) = \frac{\si(\mu)}{4}
  - \int \frac{\nabla\mu \cdot D \nabla\mu}{4\mu}\,\id x
\end{equation}
where the last term only depends on the distribution $\mu$ and
not on the dynamics. That is a simplification\footnote{For jump processes, \eqref{ent-traffic} only holds close to equilibrium.} in the structure
of fluctuations that is characteristic and restricted to
diffusions, \cite{Maes2008Steady}. A more interesting nonequilibrium theory should reach beyond the
Langevin or Markov diffusion approximation.

\section{Macroscopic fluctuation theory}\label{maft}
\lhead{C. Maes}
\rhead{Fluctuation theory}
With the required amount of exaggeration, the two previous chapters (Boltzmann and Onsager pictures) can be summarized by saying that Boltzmann characterized the static fluctuations while Onsager considered the current fluctuations, always at or near equilibrium.  These two approaches merge in a so-called dynamical fluctuation theory for macroscopic (nonequilibrium) systems.\\
Before we start, it is essential to keep in mind that we should not be content with a formal discussion where fluctuation theory and large deviations are presented in terms of mathematical terms that have no physical or operational meaning. The successes of fluctuation theory should be measured, literally.

\subsection{Structure of dynamical fluctuations}\label{strdy}
An autonomous equation for the state $z$ of a macroscopic physical system is obtained by specifying how the time-derivative or displacement $\dot z$ depends on $z$. The latter may refer to spatial profiles or to the densities of various species in the limit where the number of particles $N$ tends to infinity.\\
Before taking the macroscopic limit, there are in general many possible paths and we write them as $(z(s),j(s)), s\in [0,t],$ for time-dependent states and currents related via $\dot z(s) = D j(s)$ for some operator $D$ (such as minus the divergence).  Which paths are more plausible depends on the initial data and on the type of kinetics, interactions and driving. All that gets summarized in the Lagrangian $\cal L(j,z)\geq 0$ governing the dynamical large deviations in the sense that for $N\uparrow\infty$,
\begin{equation}\label{prob}
\text{Prob}\big[(z(s),j(s)), s\in [0,t]\big] \sim e^{-N\cal S(z_0)} \,e^{-N\,\int_0^t \id s \,\cal L(j(s),z(s))\,}
\end{equation}
renders the probability of possible paths.  We write $\cal S$ for the pseudopotential governing the occupation statistics of the initial data. We call $\cal L$ a Lagrangian, but (yet) without ``mechanical'' motivation.  To obtain an explicit \eqref{prob} from more microscopic considerations requires understanding the fluctuations around a law of large numbers, which we ignore here.  Instead, we take the road (in the form of path-space considerations) as pioneered in the Onsager-Machlup theory, \cite{Onsager1951,Onsager1953}, but generalized to nonequilibrium nonlinear evolutions.\\
In the macroscopic limit $N\uparrow \infty $, the ``true'' or deterministic updating rule {\it emerges} in the form $\dot z = D j_z$, and $j_z$ is found by minimizing $\cal L(j,z)$ over $j$ (zero-cost limit), that is, maximizing probability \eqref{prob}.  Once we know the structure of the Lagrangian in \eqref{prob}, we find  $j_z$, {\it i.e.}, determining the macroscopic equation $\dot z = D j_z$.  That is similar but not identical to characterizing motion from the Principle of Least Action; see for instance \cite{mecos,Beyengradientflow,beyen2025noether_GENERIC} for a comparison.\\

To derive the form of the Lagrangian and hence the structure of its zero-cost limit, we assume local detailed balance as discussed in Section \ref{dlbb}, \cite{time,ldb}.  That means more generally that we suppose a ``Hamiltonian'' flow $J^H$ and ``force'' $F$, functions of the state $z$ with $J^H(z)\cdot F(z) = 0$, so that for all $(j,z)$,
\begin{equation}\label{lbd}
\cal L\left(J^H(z)- j, z \right) - \cal L\left(J^H(z) + j, z \right) = j\cdot F(z)
\end{equation}
In other words, the antisymmetric contribution to the Lagrangian is given by the entropy production (current times force), as first pronounced in \cite{maes1999fluctuation,maes2000entropy} for $J^H=0$.  The nonequilibrium aspect is in the fact that the force $F$ does not need to be the gradient/derivative of a thermodynamic potential.  A typical situation is that $F$ sums over multiple thermodynamic forces arising from coupling the system with different equilibrium baths (in terms of temperature or chemical potential and/or rotational forces).  We do not always indicate the functional dependence of the Lagrangian via $\cal L = 
\cal L_F$ on the force $F$, but it is of course important to remember.\\
Perhaps surprising,  under the sole condition of local detailed balance \eqref{lbd}, the Lagrangian $\cal L$ in \eqref{prob} can always be written as 
\begin{eqnarray}
\cal L(j,z) &=& \psi(j-J^H(z),z)+\psi^\star(\frac{F(z)}{2},z)-(j-J^H(z))\cdot \frac{F(z)}{2}\nonumber\\
\cal L(j+ J^H(z),z) &=& \psi(j,z)+\psi^\star(\frac{F(z)}{2},z)-j\cdot \frac{F(z)}{2}\label{LGEN}
\end{eqnarray}
where $(\psi,\psi^*)$ are a Legendre pair (which typically can depend on $F$).  We emphasize that this structure of dynamical fluctuations is completely general, and has various (many known) realizations, including the macroscopic fluctuation theory for either jump or diffusion processes, \cite{M,F,T,Maes_2008,Maes2008Steady,maes2007entropy}.\\

Aspects of the canonical structure \eqref{LGEN} for nonequilibrium dynamical fluctuations have already been observed in \cite{Maes_2008}.  We refer to Lemma 2.1 in \cite{MPR13} for the case $J_H=0$ and in the context of reversible Markov processes.  In \cite{Kaiser2018}, \eqref{lbd}--\eqref{LGEN} was studied without $J^H$, but with nongradient $F$, following up on the canonical decomposition in \cite{Maes_2008,maes2007entropy,Maes2008Steady}. Similarly, the local detailed balance condition \eqref{lbd} (with $J^H=0$) has appeared in Chapter 10.5.3 of  \cite{Hoeksema2023}.\\

The canonical structure \eqref{LGEN} generalizes the case of boundary driven diffusions, \cite{M,F,T}, where we have the structure
   \begin{eqnarray}
{\cal L} (j,z) &=& \frac 1{4} \int \id r [j- j_z]\cdot\chi(z)^{-1}[j- j_z]\notag\\
&=& \frac 1{4} \int \id r \frac{|j+{\cal D}(\rho) \nabla \rho|^2}{\chi(\rho)}\notag
\end{eqnarray}
Here,
$j_z $ is the actual zero-cost flow (the hydrodynamic current), with $\dot z = Dj$ which means $\partial_t \rho + \text{ div }j =0$, when
 $z=\rho$ is macroscopic density profile of particles and $j$ the associated current. We write
 ${\cal D}(\rho) =$ for the hydrodynamic diffusion, and $\chi(\rho)$ is the mobility; they are related via the Einstein relation ${\cal D}(\rho) = f''\,\chi(\rho)$, where $f$ is the free energy density.\\
For example, in the case of the weakly bulk-driven simple exclusion on a ring, which is still diffusive with small external field $\cal E$, we have
 \begin{eqnarray}
\psi^*(\rho,f) &=& \frac 1{2} (f,\chi(\rho)f), \qquad
\chi(\rho)= \rho(1-\rho)\notag\\
F(\rho) &=& {\cal E} - \nabla \frac{\delta{\cal F}}{\delta \rho(r)},\; \quad {\cal F}(\rho) = T \int \id r\,[\rho(r)\log \rho(r) + (1-\rho(r))\log(1-\rho(r))]\notag
\end{eqnarray}
and the zero-cost flow, discussed in the next section, is the viscous Burgers equation.  All of that is also generalizable to dynamics on Riemann manifolds (if wished).

\subsection{Structure of relaxation equation}

From minimizing \eqref{LGEN}, {\it i.e.}, putting $\cal L(j,z) = 0$, as a simple consequence of Legendre duality, the relaxation equation must have the form
\begin{equation}\label{rel}
\dot z=D\,J^H(z) + D\,\partial_f\psi^\star(\frac{F(z)}{2},z)
\end{equation}
where $\partial_f\psi^\star$ is the derivative with respect to the first argument in the convex function $f\mapsto \psi^*(f,z)$.    We repeat that the Lagrangian and hence $\psi^*$ can also depend on the force $F$, in particular on its rotational part.  In other words, a functional dependence of $\psi^*(f,j) = \psi^*_F(f,j)$  makes a double dependence in \eqref{rel} of the current on the driving:
\begin{equation}\label{relf}
\dot z=D\,J^H(z) + D\,\partial_f\psi^\star_F(\frac{F(z)}{2},z)
\end{equation}
That structure \eqref{relf} unifies a wide variety of equations that describe relaxation to nonequilibrium.  It also extends the framework of {\tt GENERIC} to relaxation toward nonequilibrium steady conditions.  When the force $F = D^\dagger \id S$ in \eqref{relf} can be derived from a potential $S$, we recover the relaxation to a stable equilibrium as for {\tt GENERIC}, \cite{gradientflowmetricspace,ottinger2005beyond,grm1,grmela2025rheologicalmodelinggenericonsager,GEN}.   Indeed, the case of relaxation to equilibrium is the special case where in \eqref{relf}, $F = D^\dagger \id S$ and $D\,J^H \cdot \id S=0$.  Then
\[
\dot S = \id S\cdot \dot z = 2D^\dagger\id S/2\cdot\partial \psi^*(D^\dagger \id S/2) \geq 0
\]
where the last inequality uses $\psi^*(f,z) \geq \psi^*(0,z) =0$.  In other words, $S$ never decreases in time, a scenario that realizes the H-theorems of Section \ref{h1}.  That is of course (and luckily) lost for relaxation to (stationary) nonequilibrium\footnote{The relaxation to nonequilibrium can be much richer than to equilibrium, including the convergence to limit cycles and turbulent or chaotic behavior. Moreover, for a given macroscopic system various observables could be considered, each with their own relaxational behavior.}.\\

\noindent{\bf Remarks:}\\

{\bf 1:} Note that the input for \eqref{relf} is threefold, apart from the choice of state space: we need (1) to identify the Hamiltonian flow $J^H$ and the thermodynamic force $F$, (2) we need to understand the operator $D$, mostly just minus the divergence, the unit operator or some stocheometric matrix, and (3) we need the convex function $\psi^*$.  {\it Ad} (1), $J^H$ and $F$ follow from local detailed balance.  {\it Ad} (2), the operator $D$ gets specified from the kinematic or geometric relation between the temporal change in the state, aka the displacement, and the current. Finally,  {\it ad} (3), $\psi^* = \psi^*_F$ is determined by the frenetic part in the Lagrangian.
  Note indeed that $\psi^*$ appears in the frenetic part \cite{fren} of the Lagrangian \eqref{LGEN}: that is, complementary to \eqref{lbd}, we have the symmetric contribution
\[
\frac 1{2}\left[\cal L\left(J^H(z)- j, z \right) + \cal L\left(J^H(z) + j, z \right) \right]= \psi(j,z) + \frac 1{2}\psi^\star(\frac{F(z)}{2},z)
\]
where $\psi(j,z) = \psi(-j,z)$ is the Legendre transform of the function $\psi^*(f,z)$.  In particular, the universal structure \eqref{relf} highlights the role and importance of the time-symmetric part in the Lagrangian, and how it gets influenced by the forcing $F$.\\

{\bf 2:} About every first-order dynamical system can be written as in \eqref{relf}, for example, by taking $\psi^*$ quadratic in the force, but that is not the point.  The real interest is that there appears to be a narrow link between the physics of dynamical fluctuations and the physics of dynamical response.  When one identifies the function $\psi^*_F$, one is able reconstruct the Lagrangian and hence the dynamical fluctuations.  When one knows the frenetic contribution to the fluctuations, one knows the function $\psi^*_F$, and hence the relaxational behavior.  That is the strongest instance of a fluctuation-response correspondence. In other words, moving between \eqref{prob} and \eqref{relf}, one connects fluctuations and response, \cite{pons,onsager1931,kirkwood1946}.\\

{\bf 3:}  Where symplectic geometry describes the mathematics of Hamiltonian mechanics, contact geometry describes the geometry of systems with friction, thermodynamics, optics, and nonconservative dynamics.  It gives a visualization of the trajectory of a macroscopic variable, following steepest descent of some free energy.  That evolution is {\it variational}, often called {\it gradient flow}.  It corresponds to the image of a particle moving in a landscape; the landscape often stands for some profile as function of the variable.  It would of course be very nice to also have such visualizations for nonequilibrium dynamics.  One possibility is offered by motion on a Finsler manifold, which can be thought of as a Riemann manifold equipped with extra arrows.  It has a natural version in so-called Zermelo navigation, where minimizing the time-span (for moving a sailing boat subject to wind or flow drift) is expressed as geodesic motion for some Finsler metric (the Randars metric).  It creates a direction-dependent distance measurement that is nonRiemannian but easier to analyze.

\subsection{Static fluctuations}\label{bent}
Suppose we want the probability of an event at a fixed time, say  very late after relaxation. We are then dealing with macroscopic fluctuations in the stationary distribution, asking {\it e.g.} about the plausibility of a density profile in the stationary nonequilibrium condition (imagining the system shows spatial extension).  That is governed by the pseudopotential $\cal S$, as a contraction from dynamical fluctuations.  More precisely, we need to integrate over all paths that end at the required target.  In the usual
situation, the integral over all paths $\omega$ can be approximated by the
contribution of a single dominant path:
\begin{equation}\label{res}
\begin{split}
  \text{Prob}[ z(0)=b\,|\,z(-\tau)=a] &\simeq \int_{a\rightarrow b} \id\omega\,
  \,\exp\Bigl[-N \int_{-\tau}^0 \id s\,\cal L(z(s),j(s))\Bigr]
\\
  &= \exp\Bigl[-N \inf_{a\rightarrow b} \int_{-\tau}^0 \id s\,\cal L(z(s),j(s))\Bigr] \sim e^{-N\cal S(b)}
\end{split}
\end{equation}
Hence,
\begin{equation}\label{SfromHam}
  \cal S(b) = \inf_{a,a\rightarrow b}\int_{-\infty}^0
  \id s\,\cal L(z(s),j(s))
\end{equation}
or, $\cal S$ is Hamilton's principal function for the Hamiltonian
$\cal H$ associated to the Lagrangian $\cal L$.  There is therefore, skipping the details, a Hamilton-Jacobi equation 
\begin{equation}\label{hj}
\cal H\Bigl({\cal D}^\dagger\frac{\id\cal S}{\id z},z\Bigr)=0
\end{equation}
to find that $\cal S$, which is the functional for static fluctuations. \\

The Hamiltonian $\cal H$ is obtained from the Lagrangian $\cal L$ of \eqref{prob} through a Legendre transformation
\begin{equation}\label{hmd}
    \cal H(f,z) = \sup_j \Big\{j \cdot f - \cal L(j,z) \Big\}, \qquad   \cal L(j;z) = \sup_f \Big\{j \cdot f - \cal H(f;z) \Big\}
\end{equation}
where $f$ is dual to the current $j$ and represents a variable thermodynamic force. As in Hamiltonian mechanics, the thermodynamic force $f$ and the macroscopic variable $z$ are independent variables.  Ignoring the Hamiltonian flow, $J^H=0$, the expression is obtained from taking the Legendre transformation of \eqref{LGEN},
\begin{equation}
    \cal H (f,z) = \psi^*(f + \frac 1{2}F(z),z) - \psi^*(\frac 1{2}F(z),z)
\end{equation}
which we need to plug into \eqref{hj} to find the maximal solution $\cal S$ of the Hamilton-Jacobi equation.  The result would give the stationary probability of things like a density profile.   General aspects include the nonlocal nature of the pseudopotential, reflecting the  long-range spatial correlations that are so common in steady macroscopic nonequilibrium systems, \cite{Spohn1983,DorfmanKirkpatrickSengers1994}. However, in general it is hard to solve \eqref{hj} (except in equilibrium).  Not too many people have attempted perturbation theory here, as is the tradition in celestial mechanics (for finding approximate solutions for systems that are ``nearly integrable'' such as the three-body problem).  Nevertheless, once we know the frenetic contribution to the Lagrangian in \eqref{prob}, doing as outlined above reduces finding the pseudopotential for static fluctuations to a specific computational problem.\\

For the study of (time-integrated) current fluctuations, it is intuitively clear that we need information about the occupation times; those are the residence times of some macroscopic state $z$ over a long time duration. That refers to  the question how we can change the forcing so that the typical residence times for that new dynamics coincide with what is abnormal for the original process.  It refers to a so-called {\it tilted} process, which is often discussed in the literature but is not taken up here; see for instance \cite{derrida}.

\section{Frenetic steering and selection}
\lhead{C. Maes}
\rhead{Blowtorch theorem}

The word dissipation has become synonymous with loss by heat, where energy degrades and efforts get wasted.  It has of course played an important role in the discussion of efficiencies and of labor ethics and organization.  The 19th century industrial revolution and the more recent attempts (here and there) for clean-up and increased sustainability continue to turn around thermodynamic efficiency, albeit with a different emphasis and under different constraints.  From that shift (in what is allowed, beneficial and what is efficient) we already realize that what we consider as ``dissipation'' (in the sense of waste or losses) depends on what we see as useful, or on what parameters we are able to change or manipulate.\\
More recently, there has been quite a change in the theoretical meaning or in the practical role of dissipation as it has become increasingly clear that many nonequilibrium systems (in particular, involving life processes) need and take advantage of dissipation. Without dissipation, birds cannot fly (and life or evolution appears impossible all together). The very fact of dissipating is so to speak creating possibilities, and unleaches a freedom of action that only gets tempered for and around equilibrium processes (where there is no or much less dissipation).  That idea is not entirely new\footnote{and it is probably in that sense that Ilya Prigogine wrote about ``dissipative structures,'' \cite{ds1,ds2}.   In those poetic ways, nature had to invent time to allow dissipation to take advantage of a huge freedom in kinetics.   Even for heat engines, only dissipation allows work to be done in finite time.}.  Presently, that is being taken to another scientific level, where we think about the constructive ways of dissipation, \cite{nondiss}. Paradoxically, thanks to time-reversal asymmetry, the time-symmetric fluctuation sector becomes much more important and it allows frenesy to take control \cite{frenesy}.\\

There is no intention here to describe in detail some of the nonequilibrium scenario's where we see the possibly constructive role of dissipation via nondissipative actors as summarized by the notion of frenesy. It concerns, always via kinetic effects, the stabilization (of assemblies, of patterns \cite{eplu,Maeskarel2020}, or of bound states \cite{prithasimon}), proofreading \cite{Hopfield1974}, error-correction and the selection of stationary distributions \cite{land} and of currents and their directions \cite{Maes2015}, the control and frenetic steering of probes coupled to nonequilibrium media via the mean force \cite{LefebvreMaes2024,MaesNetočný2019} and friction \cite{pei2026transferActiveMotion} on them, and of evolution. Here, we only give a quick idea by recalling the blowtorch theorem \cite{Landauer1993Machinery,heatb}.

\begin{example}[Blowtorch theorem, \cite{Landauer1993Machinery,heatb}]
Let us consider a random walk on the ring $\{1,2,\ldots,N\}$ where we identify $N+1=1$.  The transition rates are taken
\[
k(x,x+1) = p_x\,e^{\frac{\beta}{2} q_x},\quad k(x+1,x) = p_x\,e^{-\frac{\beta}{2} q_x},\qquad x\in \bbZ_N
\]
for $p_x>0$ and, following detailed balance, with total heat $Q = \sum_{x=1}^N q_x$ along the ring.  Global detailed balance implies $Q=0$; here we assume $Q>0$.
Between any $x,y\in \bbZ_N$ there is a positively (clockwise) and a negatively (anti-clockwise) oriented path, $\caD^+(x,y)$ respectively $\caD^-(x,y)$, with corresponding heat along these paths satisfying
\[
q(\caD^+(x,y)) = -q(\caD^-(y,x)),\qquad q(\caD^+(x,y))+ q(\caD^+(y,x)) =  Q
\]
In particular, $q(\caD^+(x,x+1))=q_x$ and
$q(\caD^-(x,x+1)) = q_x - Q$.
Hence, the sites $x$ and $x+1$ can be ``heat-ordered'' provided
both heat quanta $q(\caD^\pm(x,x+1))$ have the same sign, {\it i.e.}, for
$q_x \not\in (0,Q)$. If this condition is verified for all sites $x$ then the heat-order becomes \emph{complete}, extending the detailed balance case
($Q = 0$) where the completeness is obvious. Remark that whenever $q_x \geq 0$ for \emph{all} sites $x$ then the heat-order can only be \emph{partial} since our complete-order condition is then in contradiction with the constraint
$\sum_x q_x = Q > 0$. Furthermore, in a diffusion regime where $q_x = \caO(1/N)$ while keeping $Q = \caO(1)$, the heat-order only applies for the neighboring sites $(x,x+1)$ such that $q_x \leq 0$.\\
In the specific case $N = 3$, the stationary occupations follow~
\[
\frac{\rho(1)}{\rho(2)} = \frac{e^{-\beta q_1} \,p_1\, p_2 \,e^{\frac{\beta}{2} (q_1-q_2)} +
e^{\beta (q_2+q_3)} \,p_2\, p_3 \,e^{-\frac{\beta}{2} (q_2+q_3)} + e^{-\beta q_1} \,p_1\, p_3 \,e^{\frac{\beta}{2} (q_1+q_3)}}{p_1\, p_2 \,e^{\frac{\beta}{2} (q_1-q_2)} +  p_2\, p_3 \,e^{-\frac{\beta}{2} (q_2+q_3)} + p_1\, p_3 \,e^{\frac{\beta}{2} (q_1+q_3)}}
\]
\emph{Non-ambiguous case.}
It is immediate that when $q_2 + q_3 = -q_1$ (detailed balance), then
$\rho(1) = e^{-\beta q_1}\,\rho(2)$.  But we also see that if
$q(\caD^+(1,2)) = q_1\geq 0$ and $q(\caD^-(1,2)) = -q_3-q_2 \geq 0$, then still $\rho(1) \leq \rho(2)$, independent of the $p_{1,2,3}$. Here the heat released along both paths from $1$ to $2$ is always nonnegative and that is why the occupation of $1$ cannot exceed the occupation of $2$.\\
\emph{Ambiguous case.}\footnote{to be compared with M.C. Escher's lithograph prints `Ascending and Descending' (1960) or `Waterfall' (1961).}
Take now both $q_1 > 0$ and $q_2 + q_3 > 0$ as well.  Then, by taking $p_1$ very small we have $\rho(1) > \rho(2)$ while for $p_3$ very small we find $\rho(1) < \rho(2)$.
In this case the heats along the two paths from $1$ to $2$ have a different sign, and hence the higher-occupied state cannot be determined without knowing the detailed kinetics (and both possibilities can indeed be designed).\\

If there are some \emph{a priori} energy levels
$E(1) < E(2) < \ldots$ assigned to the states so that the heat functions take the form
$q(x,y) = E(x) - E(y) + F(x,y)$ where $F$ accounts for the nonequilibrium driving, then the situation where, {\it e.g.}, $\rho(1) < \rho(2)$ is usually called a population inversion. We have seen that a necessary condition for this effect to occur is the absence of heat-order between the states $1$ and $2$.
\end{example}

Of course, the fact that the time-symmetric reactivities (part of the frenesy) \emph{also} determine the stationary distribution is rather obvious from the mathematical point of view.
What is actually more surprising is that at detailed balance (and in the neighborhood of it) heat and heat alone plays such a dominant role, and determines on its own the stationary occupations.  It is then the challenge of nonequilibrium physics to identify a complement to heat and entropy for characterizing the steady behavior. In other words, to understand exactly what aspect of the kinetics matter.  Dynamical fluctuation theory, where the concepts of activity, traffic or frenesy have been used, indeed specify the relevant kinetic features, {\it e.g.} in \cite{nondiss,frenesy}.

\section{Long-range correlations}
\lhead{C. Maes}
\rhead{Long-range spatial decay}
 There are, even in the presence of dissipation, well-documented ({\it i.e.}, experimental) situations where relaxation may be very slow and/or (standard) linear response formula (also around equilibrium) show divergences or singularities\footnote{Most pronounced in one dimension, and depending on disorder and momentum conservation, the conductivity may diverge as function of the system size (and separation between the temperature sources); see {\it e.g.}, \cite{Livi2003}. Indeed, in low-dimensional systems, heat conduction can deviate from the predictions of standard diffusion (Fourier's law), exhibiting phenomena like superdiffusion or size-dependent thermal conductivity.  On the theory side, (nonlinear) fluctuating hydrodynamics is employed to analyze how microscopic interactions and fluctuations in low-dimensional systems lead to complex and sometimes anomalous heat transport behavior; see \cite{Das2014, Spohn2014}  An inspiring model here remains the one of Fermi-Pasta-Ulam-Tsingou, where the complexity of energy localization and anomalous transport are studied, \cite{Dauxois2008}.}.  As one but important class of examples, we mention the case of ``fat tails.'' \\ 
When a system at equilibrium is perturbed slightly (say by an external field or a forced but small displacement of particles), its relaxation is governed by equilibrium time-correlation functions.
However, not all equilibrium systems relax exponentially or with a single relaxation time: the precise decay depends on the structure of correlations, conservation laws, hydrodynamic modes and/or disorder and frustraton.  As a matter of fact and again, even in equilibrium, relaxation can be algebraic (fat-tailed) rather than exponential. {\it E.g.}, in fluids or magnets, conserved densities (momentum, energy, particle number) relax diffusively. The result is that correlations decay as power laws, aka ``long-time tails''.  A classic example is the velocity autocorrelation in a fluid, which decays like $t^{-d/2}$ in
$d$ dimensions, \cite{AlderWainwright1967}.\\  
Another phenomenon is critical slowing down; near a critical point, relaxation times diverge and correlation functions can decay as power laws in both time and space.\\ But even below a certain temperature in equilibrium, there can be generically stretched exponential relaxation or power-law relaxation. It happens in glasses and in many situations of disorder, {\it e.g.}, even in the Griffiths regime of spin glasses.  For other references, we include \cite{AlderWainwright1967,DorfmanCohen1970,vanBeijeren1982,KuboTodaHashitsume1991}.\\

The possibility of long-time tails in the approach to equilibrium need not lead to long-range correlations for the stationary (then, equilibrium) condition.  A diffusive dynamics typically show power-law relaxation, and yet the equilibrium distributions have correlations that decay like the (static) interactions, except for some special (critical) points in the phase diagram.  However, for nonequilibrium systems, that symmetry of time-reversal invariance is gone, and there is {\it a priori} good reason to see long-range correlations.  Indeed, power-law decay is typical in nonequilibrium stationary conditions, and it also reflects the nature of the spatial dependencies in the pseudopotential and static fluctuations.  Depending on other symmetries, such as spatial symmetries, the decay in a two-points function will be of the order $r^{-d}$ in $d$ dimensions, which is related to the quadrupole nature of the source-term in the relevant  Poisson equation.  We take some space to formulate briefly the main structure. 

A first pioneering result is contained in \cite{Spohn1983}.  It treats an infinite-temperature lattice gas subject to a gradient in chemical potential making it boundary-driven. The system is infinite in the other (orthogonal) directions.  The rigorous result showed a correlation decay as the inverse Laplacian, {\it i.e.}, $r^{d-2}$ for dimensions $d\geq 3$.  It is fully compatible with fluctuating hydrodynamics for nontranslation-invariant stationary nonequilibria with macroscopic gradients maintained by boundary reservoirs.\\
In the case of translation-invariant systems, for diffusive systems where the temporal decay goes like $t^{-d/2}$ and spacetime rescales as $t\sim r^2$, we have a spatial power-law decay $r^{-d}$ of the two-point correlation functions.  We refer to the review \cite{Spohn1983} and to some specific models \cite{Basu_deBuyl_Maes_Netocny_2016,Baek2018,Maes1990}.\\
Also for the stationary distribution itself, long-range effects may dominate and even in first order around equilibrium.  The relation between (interaction potential) and correlation is more complicated when the system gets driven and, not surprisingly, also there, long-range or nonlocal effects appear; see {\it e.g.}, \cite{Maes20099}.

\chapter{Active matter}
\lhead{C. Maes}
\rhead{active}

Active matter refers to systems of (mostly soft) condensed matter whose individual components are (often internally) agitated or supplied with energy. By now, the literature is vast, easy to access, and we do not even attempt to start giving references to reviews or meta-reviews. Typical examples arise in biological contexts, such as suspensions of bacteria, collections of cells, or assemblies of tissues and membranes. The consequences of this internal driving or agitation are diverse; in the simplest cases, it leads to self-propelled motion, for example, swimming driven by flagellar activity, or, for synthetic particles such as Janus particles, driven by local chemistry and/or optics.  Another area of application is materials science, in particular for targeting specific functionalities (as in biomimetics) or for manufacturing (self-reproducing or self-assembling) structural properties achieved by active processes.  What follows is a considerably drier account, exploring corners of the subject that are rarely highlighted in the literature.

\section{Run-and-tumble equations of motion}

In the spirit of the present lecture notes, we do not delve into the detailed and varied mechanisms that convert fuel into motion.
Instead, still restricting ourselves to the case of a single point particle (or a dilute gas thereof), we focus on one realization where locomotion is described via a diffusion process on a spatial landscape\footnote{The agitated molecules of Section \ref{afmp} can also be considered as active matter, where the landscape is energetic though, and the delivered work may not be in the form of translational locomotion.}, coupled to and agitated by other (hidden/internal) dynamical degrees and switches.\\   To be specific and restricting to one-dimensional notation, we follow \cite{Demaerel2018} and we consider 
the internal degrees of freedom to be discrete, calling them {\it spin} for short. 
Particle position is denoted by $x\in\mathbb{T}$ where $\mathbb{T}\subset \mathbb{R}$ stands for the circle $S^1$, an interval $[-\ell,\ell]$, the entire line or some discrete (lattice) version thereof, and $\sigma\in K$ is the spin taking $n$ possible values.  The spin value gets updated in a Markov jump process with transition rates $k_x(\sigma,\sigma')$ for $\sigma\rightarrow \sigma'$ at fixed position $x$.  The coupled spin-orbit dynamics in the form of an overdamped diffusion is
\begin{eqnarray}\label{mode1}
\dot{x}(t) - v(\sigma(t))= - \chi(\sigma(t))\, \frac{\id U}{\id x}(x(t),\sigma(t)) + \sqrt{2D(\sigma(t))}\;\xi_t\\
\log\frac{k_x(\sigma,\sigma')}{k_x(\sigma',\sigma)} =  [U(x,\sigma) - U(x,\sigma') + F(x;\sigma,\sigma')]/T_\text{eff}\nonumber\\
\sqrt{ k_x(\sigma,\sigma')\;k_x(\sigma',\sigma)} = \psi(x;\sigma,\sigma')\nonumber
\end{eqnarray}
where additional boundary conditions on $x$ may depend on the geometry.
The first driving is in terms of a fixed function $v(\sigma)$ of the spin $\sigma$, called propulsion velocity.  There may be a  conservative coupling between position $x$ and spin $\sigma$  via the potential $U(x,\sigma)$, but that is often not detailed.  The mobility coefficient is $\chi(\sigma)\geq 0$ and $T= D(\sigma)/\chi(\sigma) \geq 0$ (independent of $\sigma$) is the temperature of a thermal environment represented by standard white noise $\xi_t$. (Boltzmann's constant is set to one.)  There is the antisymmetric $F(x;\sigma,\sigma')=-F(x;\sigma',\sigma)$ as a possible extra source of nonequilibrium driving in spin-space.  Another nonequilibrium driving comes through the effective temperature $T_\text{eff}$ that may be much higher than $T$. Often, the limiting regime $T_\text{eff}\uparrow \infty$ is taken, also to replace the detailed implementation of the potential $U$ and the forcing $F$. The reactivity $\psi(x;\sigma',\sigma)=\psi(x;\sigma,\sigma')$ represents the time-symmetric agitation and becomes most important for characterizing the persistence of the locomotion at high $T_\text{eff}$.\\

A standard case  considers a particle with position $x$ pushed with velocity field $c_\sigma(x)$ depending on spin $\sigma=\pm 1 $ in a bath at temperature $T$.  In other words, we take \eqref{mode1} with $\left[v(\sigma) - \chi(\sigma)\,U'(x,\sigma)\right] = c_x(\sigma)$ and $k_x(\sigma,-\sigma) = k_\sigma(x)$:
\begin{equation}\label{zerot}
\dot{x}(t) = c_{\sigma(t)}(x(t)) + \sqrt{2T}\;\xi_t,\; \,\qquad \text{and jumps  } \sigma \longrightarrow -\sigma \text{ at rate } k_\sigma(x)
\end{equation}
where the $c_\sigma(x)$ are smooth in $x$ with $c_+> 0 > c_-$.  The model is called athermal when $T=0$ in \eqref{zerot}.\\
The joint probability density $(\rho_+(x,t),\,\rho_-(x,t))$ to be in $x$ with spin $\pm 1$ satisfies
\begin{eqnarray} \label{dyn}
\partial_t \rho_+ & =& -\partial_x(c_+ \rho_+) + k_-\rho_--k_+\rho_+ + T\partial^2_{x}\rho_+\notag\\ 
\partial_t \rho_- &=& -\partial_x(c_- \rho_-) + k_+\rho_+-k_-\rho_-+ T\partial^2_{x}\rho_-
\end{eqnarray}
Making it even simpler, taking $c_{+} = c= -c_{-}$ and $k_\sigma(x)\equiv a>0$, we get
\begin{equation}\label{nzerot}
\dot{x}(t) = c\,\sigma(t) + \sqrt{2T} \,\xi_t,\; \,\qquad \text{and jumps  } \sigma \longrightarrow -\sigma \text{ at rate } a
\end{equation}
The corresponding evolution equation for the spatial density $\rho= \rho_+ + \rho_-$ verifies
\begin{equation} \label{T-tel}
(\partial_t -T\partial_x^2)^2\,\rho - c^2\partial_x^2\rho = -2a\,(\partial_t-T\partial_x^2 )\rho
\end{equation}
The derivation is given in \cite{Demaerel2018}.  There is another instructive way to write that thermal telegraph equation \eqref{T-tel}; it is equivalent to the following system for functions $\pi(x,t), \rho(x,t)$,
\begin{equation}\label{hal}
\begin{cases}
& (\partial_t -T\partial_x^2)\rho = \pi \\
& (\partial_t -T\partial_x^2)\pi = c^2\partial_x^2\rho -2a\,\pi
\end{cases}
\end{equation}
Defining the energy-like functional
\begin{equation}
{\cal H}[\pi,\rho]:=\frac{1}{2}\int \id x\left(\pi^2+c^2(\partial_x\rho)^2\right)
\end{equation}
one can rewrite \eqref{hal} and thus also \eqref{T-tel} as 
\begin{equation}
\begin{cases}
& \partial_t\rho = \frac{\delta {\cal H}}{\delta \pi}-\frac{T}{c^2}\frac{\delta {\cal H}}{\delta \rho} \\
& \partial_t \pi = -\frac{\delta {\cal H}}{\delta \rho}-(2a-T\partial_x^2)\frac{\delta {\cal H}}{\delta \pi}
\end{cases}
\end{equation}
with a Hamiltonian plus a dissipative part.
One easily verifies that ${\cal H}$ is a Lyapunov functional, and  the athermal telegraph equation appears in the limit $T \downarrow 0$. \\

We also mention the pleasant and useful relation with the Dirac electron, \cite{struyve2012zigzag,struyve2012jphta,Krekels2024ZigZag,Maes2022RunAndTumble}.   The (free) Dirac equation for a single electron in four spacetime dimensions has a wave function $\psi(\mathbf{x}, t) \in \bbC^4$ satisfying 
 \begin{equation}\label{die}
(i\hbar\gamma^\mu \partial_\mu - mc)\psi =0
 \end{equation}
where we use the gamma-matrices 
$\gamma^\mu$. In the Weyl (or chiral) basis for the gamma matrices, the 
wave function is a bispinor, a pair of two-component Weyl spinors 
$\psi = (\psi_-,\psi_+)^T$, describing respectively a left-handed and a 
right-handed spinor. The Dirac equation \eqref{die} is then equivalent with the 
coupled equations
\begin{equation}\label{wey}
  i\hbar\,\sigma^\mu \partial_\mu \psi_+ = mc\psi_-,\qquad   
  i\hbar\,\overline\sigma^\mu \partial_\mu \psi_- = mc\psi_+
\end{equation}
where $\sigma^\mu = (1,\vb*{\sigma})$ and $\overline\sigma^\mu = 
(1,-\vb*{\sigma})$ for Pauli matrices $\vb*{\sigma}$.
The Dirac current
\[
j^\mu = \overline{\psi}\gamma^\mu\psi\quad \text{with }\,\; \overline{\psi} = 
\psi^\dagger \gamma^0,\;\;\text{satisfying }\;\; \partial_\mu j^\mu=0
\]
follows the decomposition $j^\mu = j^\mu_+ + j^\mu_-$ for
\[
j^\mu_+ = \psi_+^\dagger \sigma^\mu\psi_+,\qquad  j^\mu_- = \psi_-^\dagger 
\overline\sigma^\mu\psi_-
\]
As a result of the coupling, there appears a source term in their continuity equation:
\[
\partial_\mu j^\mu_{\pm} = \pm F, \quad \text{with }\;\; F= 
2\frac{mc}{\hbar}\Im\psi_+^\dagger\psi_-
 \]
which is
\begin{eqnarray}\label{rt}
\partial_t\rho_+ + {\mathbf\nabla}\cdot({\mathbf v}_+\rho_+) &=& a\rho_- - b\rho_+\nonumber\\ 
\partial_t\rho_- + {\mathbf\nabla}\cdot({\mathbf v}_-\rho_-) &=& b\rho_+ - a\rho_-
\end{eqnarray} for the densities $\rho_+ = j^0_+ = \psi^\dagger_+\psi_+$ and 
$\rho_- = j^0_- = \psi^\dagger_-\psi_-$,
where, equations (22)--(23) in \cite{struyve2012zigzag},
\begin{eqnarray}\label{rte}
\vb{v}_+ = c\,\frac{\psi_+^\dagger \,\vb*{\sigma}\, 
\psi_+}{\psi^\dagger_+\psi_+}, &&  \vb{v}_- = -c\, \frac{\psi_-^\dagger 
\,\vb*{\sigma}\,\psi_-}{\psi^\dagger_-\psi_-}\nonumber\\
a  = 2\frac{mc^2}{\hbar}\,\frac{\left( \Im\psi_+^\dagger 
\psi_-\right)^+}{\psi^\dagger_-\psi_-},&& b= 2\frac{mc^2}{\hbar}\,\frac{\left( 
\Im\psi_-^\dagger \psi_+\right)^+}{\psi^\dagger_+\psi_+}
\end{eqnarray} with  $h^+ = \max\{h,0\}$. \\
There have been no approximations and \eqref{rt} reproduce the equations of run-and-tumble motion \eqref{dyn} at $T=0$, making it easy to visualize the electron trajectories, \cite{Maes2022RunAndTumble,Krekels2024ZigZag}.  Probabilities for trajectories arise from the initial condition where the positions are drawn according to the Born rule, and from the random tumbling events. 
The Dirac electron shows its mass in the zig-zag motion, due to the chirality coupling - much like the Higgs mechanism.  The speed $|\mathbf v_{\pm}| = c$ is invariably equal to the speed of light, $j^\mu_+j_{+\mu} = j^\mu_-j_{-\mu} = 0$.  The velocity jumps from $\mathbf{v}_\chi$ to $\mathbf{v}_{-\chi}$ at every transition $\chi\rightarrow -\chi$ in the direction of the particle.  That tumbling is a consequence of the coupling between left- and right-chirality and occurs at a rate proportional to the mass,  $2mc^2/\hbar$, also known as the angular frequency of the {\it Zitterbewegung}. The resulting dynamics is a time-dependent Markov process $(\mathbf x_t,\chi_t)$ and that dynamics at every moment reproduces the correct densities $\rho_{\pm}$. The dynamics is also time-reversal invariant, in contrast to models of 
run-and-tumble particles where there is dissipation from fuel consumption, \cite{Maes2022RunAndTumble}.  

\section{Stationary behavior}
The stationary density  $(\rho_+,\rho_-)$ for \eqref{dyn} at $T=0$ solves
\begin{equation} \label{stat}
\begin{cases}
& 0 = -(c_+ \rho_+)' + k_-\rho_--k_+\rho_+\\
& 0 = -(c_- \rho_-)' + k_+\rho_+-k_-\rho_-
\end{cases} 
\end{equation}
On the line, by adding the two equations in  \eqref{stat} and integrating over $x$, we get
\begin{equation} \label{res}
c_+\rho_++c_- \rho_- = \text{constant  } \,J
\end{equation}
where $J$ is the total current in the $x$-direction.\\
Combining \eqref{res} with \eqref{stat} we get
\[
-(c_\sigma\rho_\sigma)'+\frac{k_{-\sigma}}{c_{-\sigma}} \left(J - c_\sigma \rho_\sigma\right)-k_\sigma \rho_\sigma=0
\]
with general solution 
\begin{equation} \label{expr1}
\rho_\sigma(x)= \frac{1}{c_\sigma(x)}e^{\phi(x)}\left[\frac{J}{2}+\sigma\,A+ J\,\int_0^x \frac{k_{-\sigma}(x')}{c_{-\sigma}(x')}e^{-\phi(x')}\id x'\right]
\end{equation}
for integration constant $A$, and where $\phi$ is defined through 
\begin{equation} \label{def}
\phi(0)=0,\qquad  \phi'=-\frac{k_+}{c_+}-\frac{k_-}{c_-}
\end{equation}
Boundary conditions matter of course.\\

When capturing the run-and-tumble particle in a harmonic potential, the Master equations become
\begin{equation} \label{EOMstat}
\begin{cases}
& 0 = \partial_x \left(-[c-\kappa\,x]\rho_++T \partial_x \rho_+\right) + a\,(\rho_--\rho_+)\\
& 0 = \partial_x \left([c+\kappa\,x]\rho_-+T \partial_x \rho_-\right) + a\,(\rho_+-\rho_-)
\end{cases}
\end{equation}
Three pairs of equations follow by multiplying \eqref{EOMstat} with $1, x$ and $x^2$, and integrating over $x\in \mathbb{R}$, to find
\begin{equation} \label{temp}
\kappa\,\langle x^2 \rangle = T + \frac{c^2}{\kappa+2a} =: T_\text{eff}(\kappa)
\end{equation}
where the effective temperature $T_\text{eff}\geq T$.
For the first moments, for fixed $\sigma=+1$ and $\sigma=-1$ respectively,
\begin{equation}
\langle x \rangle_+ = \frac 1{2}\frac{c}{\kappa+2a} = -\langle x\rangle_-
\end{equation}
which signals that the stationary density  acquires a bimodal character as a result of the activity, showing nonequilibrium steady population statistics.\\ 

Moving to two or higher dimensions, there appears the possibility of Motility Induced Phase Separation, \cite{Cates_2015}.  When the particles mutually repel each other, for large enough density and persistence, we get the formation of particle clusters separated by voids.   We can view it as the result of a dynamical activity that depends on the local density.  The feedback consists in lower escape rates when there are a greater number of persistent particles in the neighborhood.  That mechanism is not restricted to run-and-tumble particles but is a general feature of active particles, including active Brownian and active Ornstein-Uhlenbeck processes.\\
More of that and richer behavior emerges when the propulsion speeds are not all equal, and other interaction schemes are tried.  Such a dispersion in propulsion strength for active multi-particle systems with 
short-range attractive interaction, delivers novel phases that
combine spatial and orientational ordering; see \cite{DuttaBasu2026}.  

\section{Response behavior}
To obtain the diffusion constant $D$ we release the process with initial data concentrated at $x=0$ and track $\langle x^2\rangle(t)$ for $t$ large. The relative probability of starting with $\sigma =1$ or $\sigma=-1$ is not important. 
Multiplying equation \eqref{T-tel} by $x^2$ and integrating, one gets
\[
\ddot{\langle x^2\rangle}-2c^2=-2a\dot{\langle x^2\rangle}+4aT
\]
For $t\to +\infty$, one then obtains that $\langle x^2\rangle-(2T+\frac{c^2}{a})t$ converges and the diffusion constant is
\begin{equation}\label{diffu}
D = T + \frac{c^2}{2a}
\end{equation}
There is already diffusion at zero temperature $T=0$, making the situation reminiscent of the one in quantum mechanics.
Furthermore,  comparing with \eqref{temp}, \eqref{diffu} implies
\begin{equation}\label{ineq}
	T\leq T_\text{eff}(\kappa)\leq T_\text{eff}(0)= D
	\end{equation}

To see the mean velocity (or current) $\nu = \lim_{t \to \infty} \frac{\langle x\rangle(t)}{t}$ resulting from the application of an extra electric field $\cal E$, we modify the drift $c_{\sigma}=\sigma\,c$ to $c_{\sigma}= \sigma\,c + \cal E$.  A trivial calculation then yields that  $\nu = \cal E$; the mobility defined as $\mu := \frac{\nu}{\cal E}$  thus equals $1$.
Plugging in the calculated values, one obtains
\begin{equation} \label{Suth}
\frac{D}{\mu T}=\frac{D}{T} = 1 +\frac{c^2}{2a \,T} > 1
\end{equation}  
so that the Sutherland-Einstein relation is broken, as in Section \ref{sue}. 
Active particles are relatively more diffusive than their passive counterparts, compatible with the Taylor-Aris effect \cite{taylor1953dispersion,aris1956dispersion}.\\

Response theory also plays an important role in derivations of extended fluctuation-dissipation relations of the second kind.  Interestingly, for a bath of active particles, an immersed probe can become active as well, \cite{pei2026transferActiveMotion}.

\section{Biological functioning}
There is something truly special about our universe: it harbours life.  At its core, physics of life seeks to uncover the fundamental principles of life processes: how the laws of physics shape life, how biological phenomena can be unified and understood (even in extra-terrestrial environments), and what new physics might emerge from these inquiries.\\
At any rate, life does not answer Hamlet’s question, “To be or not to be?”  Life is a process, a dynamics, and is not a substance; it fills time. A single snapshot is not enough to define life; we need to watch the full film.\\   
The engines in living organisms do not function by maintaining real temperature differences. Instead, they require nutrients, such as glucose, which generates ATP (adenosine triphosphate) through cellular respiration. Plants use light and photosynthesis to produce their food, relying on low-entropy energy sources. The key here is that molecules are driven out of equilibrium by so to say chewing chemical food.  They get local bursts of energy. As a result of that agitation, the equilibrium distributions that describe the population level of chemo-mechanical configurations in the physics of Boltzmann and Gibbs no longer apply. Instead, we enter the realm of active matter, a vast and rapidly evolving field in physics. One of the earliest models, the Vicsek model, was introduced only in 1995. Through the physics of active matter and active particles, fresh insights into locomotion and self-organization start to emerge. Cooperative phenomena in active particles—such as phase transitions or collective flows—open still unfamiliar chapters in the study of states of aggregation.\\
The stimulus-response relationship in living organisms is entirely different from that in non-living matter. Homeostasis, for example, is not a static state but rather a steady, dynamic condition in which organs continuously adapt to the changes they encounter.  The physics of active matter may help to understand how stresses and mechanical perturbations affect the biochemical functioning of cells.\\  
To claim that life is an extraordinary stroke of luck, one must first establish a meaningful reference point—what {\it a priori} possibilities were truly equivalent? Any of the traditional claims that life is an improbable accident rely on probability distributions associated with equilibrium states. However, the maximum-entropy probability distributions of inert matter may not be the most appropriate framework for understanding the origin and evolution of life. The universe has always existed in a state far from equilibrium, driven by the low-entropy condition of an early homogeneous cosmos—an atypical state for gravity.  Ever since, matter has been agitated by gravity.

\vspace{0.5cm}
\center
\includegraphics[width=0.8\textwidth]{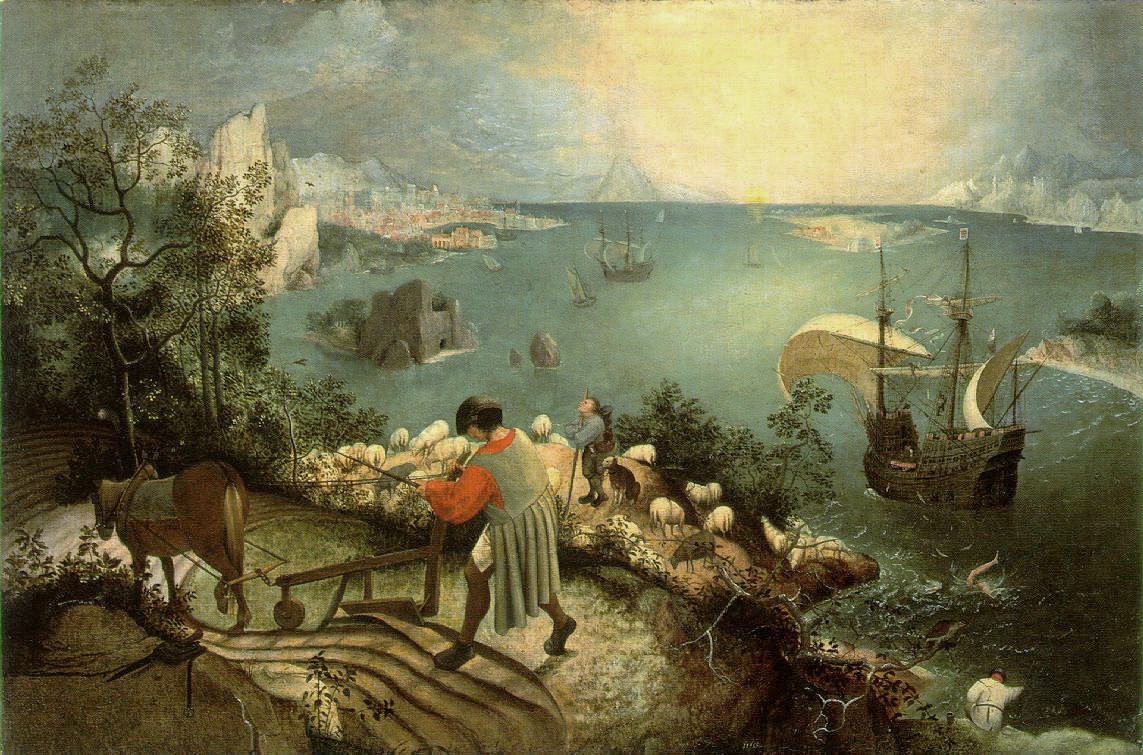}\par
\vspace{1.5cm}

\newpage
\lhead{C. Maes}
\rhead{Bibliography}

\bibliographystyle{unsrt}  
\bibliography{cleaned_references}

\end{document}